\documentclass[11pt,a4paper]{article}
\usepackage{jcappub}
\usepackage[utf8]{inputenc}
\usepackage[table]{xcolor}
\usepackage{verbatim}
\usepackage{amsmath,amsfonts,amssymb}
\usepackage{subfloat}
\usepackage{float}
\usepackage{xcolor}
\usepackage{array, booktabs, tabularx, makecell, multirow}
\usepackage{multicol}
\usepackage{tikz}
\usetikzlibrary{shapes.geometric, arrows}
\usepackage{tabularx}
\usepackage{multirow}
\newcolumntype{s}{>{\hsize=.5\hsize}X}

\usepackage[misc]{ifsym} 

\tikzstyle{startstop} = [rectangle, rounded corners, minimum width=3cm, minimum height=1cm,text centered, draw=black, fill=red!30]
\tikzstyle{io} = [trapezium, trapezium left angle=70, trapezium right angle=110, minimum width=3cm, minimum height=1cm, text centered, draw=black, fill=blue!30]
\tikzstyle{process} = [rectangle, minimum width=3cm, minimum height=1cm, text centered, text width=3cm, draw=black, fill=orange!30]
\tikzstyle{decision} = [diamond, minimum width=3cm, minimum height=1cm, text centered, draw=black, fill=green!30]
\tikzstyle{arrow} = [thick,->,>=stealth]

\tikzstyle{lcdmparam} = [rectangle, rounded corners, minimum width=3cm, minimum height=1cm,text centered, draw=black, fill=yellow!30]
\tikzstyle{stockcode} = [rectangle, rounded corners, minimum width=3cm, minimum height=1cm,text centered, draw=black, fill=red!30]
\tikzstyle{modcode} = [rectangle, rounded corners, minimum width=3cm, minimum height=1cm,text centered, draw=black, fill=blue!30]
\tikzstyle{modparam} = [rectangle, rounded corners, minimum width=3cm, minimum height=1cm,text centered, draw=black, fill=orange!30]
\tikzstyle{goal} = [rectangle, rounded corners, minimum width=3cm, minimum height=1cm,text centered, draw=black, fill=green!50]

\usetikzlibrary{mindmap,backgrounds}
\usepackage{adjustbox}
\usepackage{tabularx}
\hypersetup{%
,urlcolor=red
,citecolor=red
,linkcolor=blue
}

\def\mnras{Monthly Notices of the Royal Astronomical Society}
\def\aap{Astronomy and Astrophysics}

\def\jcap{Journal of Cosmology and Astroparticle Physics}
\def\apj{Astrophysical Journal}
\def\physrep{Physical Reports}
\def\prd{Physical Review D}

\def\apjl{Astrophysical Journal, Letters}

\title{Cosmological gravity on all scales III: non-linear matter power spectrum in phenomenological modified gravity}

\author[a]{Sankarshana Srinivasan,}
\author[b,a]{Daniel B Thomas,}
\author[a]{and Richard Battye}

\affiliation[a]{Jodrell Bank Centre for Astrophysics, School of Natural Sciences, University of Manchester, \\ Alan Turing Building, Oxford Road, Manchester, M13 9PL, United Kingdom}
\affiliation[b]{School of Physics and Astronomy, 
Queen Mary University of London \\
G O Jones Building, 327 Mile End Road, London, E1 4NS, UK}

\emailAdd{sankarshana.srinivasan.manchester.ac.uk}
\emailAdd{dan.b.thomas1@gmail.com}
\emailAdd{richard.battye@manchester.ac.uk}

\date{April 2020}

\abstract{
Model-independent tests of gravity with cosmology are important when testing extensions to the standard cosmological model. To maximise the impact of these tests one requires predictions for the matter power spectrum on non-linear scales. In this work we validate the \texttt{ReACT} approach to the non-linear matter power spectrum against a suite of phenomenological modified gravity N-body simulations with a time-varying gravitational constant, covering a wider range of parameter space than previously examined. This vanilla application of \texttt{ReACT} has limited range and precision due to the different concentration-mass relation $c(M)$ that occurs when gravity is modified. We extend this approach with a fitting function for a modified concentration-mass relation, allowing for accurate (1$\%$) computation of the matter power spectrum up $k=2\,h\,{\rm Mpc}^{-1}$ across a substantial range of parameter space. This fitting function allows precision model-independent tests of modified gravity to be carried out using the data from upcoming large scale structure surveys.
 }

\begin{document}

\maketitle

\section{Introduction}
Over the last two decades, the field of cosmology has entered an era of precision measurements. The current cosmological paradigm, the $\Lambda$CDM model contains Cold Dark Matter (CDM) and dark energy (sourced by the cosmological constant $\Lambda$) as its main components, but the underlying nature of both remains unknown. The validity of General Relativity (GR) is a crucial assumption in this picture since both CDM and $\Lambda$ are inferred from solely gravitational observations. In addition, efforts to detect dark matter directly or indirectly have not yet been successful, while quantum field theory predicts a $\Lambda$ value that is many orders of magnitude larger than that inferred from cosmological measurements. These issues form part of the motivation for modifying the law of gravity and the establishment of a vast space of modified gravity models \cite{ref:CliftonReview, Nojiri_2017} in order to explain the physics attributed to the dark sector. Efficiently testing this model space is a major challenge in cosmology.

The assumption that GR is the correct theory of gravity involves an extrapolation of several orders of magnitude from small scales (compact object mergers and solar-system tests) to the cosmological scales probed by measurements of the Cosmic Microwave Background (CMB). Upcoming experiments such as the \textit{Euclid} satellite \cite{ref:Euclid} \footnote{\href{https://www.euclid-ec.org}{https://www.euclid-ec.org}}, the Nancy Roman telescope (NRT) \cite{NancyRoman} \footnote{\href{https://roman.gsfc.nasa.gov}{https://roman.gsfc.nasa.gov}}, the Vera Rubin Observatory (VRO) \cite{LSST} \footnote{\href{https://www.lsst.org}{https://www.lsst.org}} and Square Kilometre Array (SKA) \footnote{\href{https://www.skatelescope.org}{https://www.skatelescope.org}}\cite{ref:SKACosmo} will generate an unprecedented volume of data in the so-called non-linear regime, where the density contrast $\delta \gg 1$, i.e., the fluctuation away from the average density of the Universe is large where GR has not been adequately tested. In the context of modified gravity, one requires a parameterisation of modified gravity models \cite{ref:BattyePearson, ref:Gleyzes} that  encompasses a large region of the parameter space. The lack of a robust tool that takes advantage of the additional constraining power from including non-linear scales to predict cosmological observables across the range of scales for general models of modified gravity results in strict data-cuts that excise non-linear scales from the data analysis, severely limiting constraining power (see, for example  \cite{DES_Y1_2018}).

Previous studies of modified gravity in the non-linear regime have primarily been $N$-body simulations of specific models such as $f(R)$ or Dvali-Gabadadze-Porrati (DGP) gravity (see e.g. \cite{Arnold2019, Arnold2022}). Such studies are useful to understand structure formation as a function of model-parameters, and explore observational signatures in specific theories \cite{ref:HassaniNBodyMG}. However, it is not practical to run dedicated $N$-body simulations for every modified gravity model in order to predict cosmological observables, as the associated computational cost would be too large. In addition, more model-independent approaches allow for a more comprehensive null test of $\Lambda$CDM and don't rely on us having already come up with the correct alternative.

One approach for testing gravity in a model-independent way on all scales is described in \cite{ref:DanPF}. In this approach, the standard parameterisation in linear theory involving two parameters is extended to all cosmological scales: $\mu$ is a change in the strength of gravity, and $\eta$ is the ``slip'', a possible difference between the two Newtonian gauge scalar metric potentials. Both of these can in principle be almost arbitrary functions of time and space. In our simulation work we follow ``the maximally-phenomenological pixels" approach described in \cite{ref:DanPF} in which the  modified gravity parameters consist of independent piece-wise constant functions composed of independent bins in time \footnote{The extension for $\mu$ and $\eta$ to be functions of space as well is not conceptually difficult, however we focus on time dependence as a first step and to ensure we understand this simpler case well before generalising further. We leave the generalisation of our N-body simulations to scale dependence to future work.} as this is most suited to our philosophy of model-independence.  

In \cite{Srinivasan2021} we studied the phenomenology of $N$-body simulations with binned time-dependent modified gravity parameters. We improved on the previous attempt to study such models \cite{ref:Cui1, ref:Cui2} in which $\mu$ was set to be a constant value at $z<50$, i.e., constant throughout the simulation redshift range. We focused on the matter power spectrum $P(k)$ measured from the simulations and the weak-lensing observables computed from $P(k)$. We found that the \texttt{ReACT} formalism \cite{Bose_2021, BoseReACT, ref:reactionCataneo} is most successful among those available in the literature in reproducing the cosmological observables computed in or from our simulations. We designed a pipeline to compute  cosmological observables in the non-linear regime using \texttt{ReACT}. Here, we validate this pipeline in detail across a much wider region of the (phenomenological) modified gravity parameter-space. In particular, we quantify the range of scales any cosmological observable can reliably be computed using \texttt{ReACT}. In addition, there were indications in \cite{Srinivasan2021} that this implementation would not be precise enough over a wide enough range of scales to maximise the gain from upcoming surveys, and also that this issue could be alleviated by modifying the concentration-mass relation. We investigate this issue in detail and provide an extension to the fit of \cite{Srinivasan2021} that substantially improves the range of precision of the fitting function. Our goal is to validate \texttt{ReACT} in order to run modified gravity forecasts for binned $\mu(z)$ and $\eta(z)$.  

The structure of this paper is as follows. In section \ref{sec:pipeline} we review our parameterisation, define the terms we will use to quantify the results from our simulations and present our adapted  implementation of \texttt{ReACT} with binned modified gravity functions. We define our pipeline for computing $P(k)$ and the weak-lensing convergence spectrum using \texttt{ReACT}. We also discuss how one can improve the accuracy of \texttt{ReACT} by modifying halo concentrations. We then present the suite of $N$-body simulations that we have run to validate our \texttt{ReACT} pipeline in section \ref{sec:sims}. In section \ref{sec:results}, we quantify the accuracy and precision of \texttt{ReACT} in predicting $P(k)$ and weak-lensing observables against our simulations both with and without the concentration modification. We discuss the relevance of these results in the literature and the impact they may have for cosmological parameter forecasts in the context of modified gravity for surveys such as \textit{Euclid}, NRT, VRO and SKA in section and conclude in section \ref{sec:conclusion}. \\

\section{Modified gravity framework}
\label{sec:pipeline}
As described in \cite{Srinivasan2021}, our parameterisation may be formulated as 
\begin{eqnarray}
\frac{1}{c^2}k^2 \tilde{\phi}_{\rm P} & = & -\frac{1}{c^2}4\pi a^2 \bar{\rho} G_{\rm N}\mu(a,k)\tilde{\Delta} \,, \label{eq:MGParam} \\
\tilde{\psi}_{\rm P} & = & \eta(a,k)\tilde{\phi}_{\rm P} \, ,
 \end{eqnarray}
where $\bar{\rho}$ is the background density, $\tilde{\Delta} = \tilde{\delta} - \frac{\dot{a}}{a}\frac{3}{c^2k^2}ik_i \tilde{v}_i$ is the gauge-invariant density contrast in Fourier space, $G_{\rm N}$ is Newton's constant, $\phi_P$ and $\psi_P$ are the standard Newtonian gravitational potentials\footnote{$g_{00}=-1-\frac{2\phi_P}{c^2}$; $g_{ij}=a^2\delta_{ij}\left(1-\frac{2\psi_P}{c^2} \right)$.} (normally found to be equal in GR) and $\mu(a, k)$ is a dimensionless function of space and time representing a change to the strength of gravity. These equations can be derived from the FLRW metric expanded up the first Post-Friedmann order (which is one step beyond leading order) at which the equations describe structure formation on all scales (see \cite{ref:DanPF} for details). An important consequence of this parameterisation is that the dynamics of massive particles are purely governed by \eqref{eq:MGParam}, i.e., one can build and run dark matter only $N$-body simulations where the structure formation is completely specified by a single modified gravity parameter, $\mu$. The second parameter $\eta$, the so-called slip parameter purely affects photon geodesics and, therefore, can be modelled in post-processing. 

\subsection{Applying the \texttt{ReACT} framework to modified gravity with binned $\mu$}
We showed in \cite{Srinivasan2021} that the \texttt{ReACT} framework performed significantly better than other candidates from the literature including the standard \texttt{halofit} prescription and the standard halo model in predicting $P(k)$ measured from our simulations. We showed how we can take the publicly available implementation of the \texttt{ReACT} code and apply it to our binned models. We call this implementation the `vanilla' implementation throughout this paper. 

 The fundamental idea at the centre of \texttt{ReACT} is that the accuracy of the halo model can be significantly improved by measuring the so-called `non-linear response of the power spectrum' \cite{Mead2017} i.e., the ratio of matter power spectra between two cosmologies that share the same linear power spectrum. The halo model reaction formalism computes the non-linear matter power spectrum for beyond $\Lambda$CDM cosmologies from the so-called `pseudo spectrum', which is the $\Lambda$CDM power spectrum with the the initial conditions modified to match the growth in the modified gravity model.  The use of the pseudo spectrum ensures that the transition from the 2-halo term to the 1-halo term is smooth. This transition was known to be a problematic region to model accurately for the standard halo model prescription. In this case, the similarities in the mass functions for the two cosmologies, i.e., since the two cosmologies have the same linear clustering, allows the inaccuracy in the transition region to be overcome. This was first shown in \cite{Mead2017} and later validated for $f(R)$ and DGP gravity in \cite{ref:ReactTheory}. This idea was implemented into a numerical fitting function dubbed the ``halo model reaction'' or \texttt{ReACT} \cite{ref:reactionCataneo, Bose_2021, Carrilho_2022}. 

 A key point to note is that in the publicly available implementation of \texttt{ReACT}, two \textit{separate} parameters, $\mu$ and $\mathcal{F}$ are used to model the linear and non-linear gravitational clustering, respectively. The linear modifications to the Poisson equation are encoded in $\mu(a, k)$, while $\mathcal{F}$ is computed by assuming a specific model and applying the spherical collapse formalism to specific modified gravity models with specified screening mechanisms. Given this prescription, the \texttt{ReACT} code computes the reaction from the following input information:
\begin{itemize}
    \item The linear power spectrum at the redshift(s) of interest for the target cosmology.
    \item The function $\mu$ which encodes the \textit{linear} modification to gravity via the Poisson equation applied to scales over which perturbation theory may be applied. 
    \item The function $\mathcal{F}$ that encodes the modification to the non-linear Poisson equation that is relevant for $N$-body simulations.  
    \item Additional input parameters connected to specific models of modified gravity/established linear theory paramterisations such as the Effective Field Theory (EFT) of dark energy.
\end{itemize}

However, in our parameterisation the Poisson equation is derived making no distinction between the clustering of matter due to gravity on linear vs non-linear scales (see sec. IV A in \cite{ref:DanPF}). Therefore, in order to make contact with our simulations within the the \texttt{ReACT} framework, one sets $\mathcal{F} \equiv \mu(a, k)-1$. In other words, we ensure that the parameter $\mu(a, k)$ specifies both the linear and non-linear clustering in our simulations, which in general is not true for many models considered in the basic \texttt{ReACT} framework. We also switch off/set to their respective $\Lambda$CDM values any additional model-specific parameters. 

The corrections from the modifications to the Poisson equation are propagated through to the halo model quantities such as the mass function via the spherical collapse model implemented within \texttt{ReACT}. This introduces a slightly different source term in the differential equation for the radius of the halo within spherical collapse. We describe this procedure in appendix \ref{app:spherical_collapse}. 

We quantify the performance of \texttt{ReACT} in later sections by comparing the matter power spectrum $P(k)$ measured from our simulations against that predicted by the fitting function. In practice, we will compute the so-called halo model `reaction' term, a dimensionless ratio of halo model quantities given by
\begin{equation}\label{eq:Reaction}
    R(z, k) = \frac{P (z, k)}{P_{\rm pseudo}(z, k)}
\end{equation}
where $P(z, k)$ is the matter power spectrum in the specified cosmology and $P_{\rm pseudo }(z,k)$ is the $\Lambda$CDM matter power spectrum with the initial conditions modified such that the linear growth factor is identical to the the numerator at the redshift of interest. Using eq.~\ref{eq:Reaction}, we define the $k_{\rm fail}$ parameter as the smallest wavenumber at which \texttt{ReACT} failed to predict $R(k)$ in our simulations at a specified accuracy threshold of 1\%.

\subsection{Modifying halo model ingredients in \texttt{ReACT}}\label{sec:conc_mod}
We found evidence in \cite{Srinivasan2021} to suggest that the accuracy of \texttt{ReACT} relative to our simulations may be improved by modifying the halo model parameters within the code. In particular, we modified the halo concentration-mass amplitude parameter. In this work we systematically develop this finding into a modification to ReACT that substantially improves the `vanilla' ReACT approach. In this subsection we describe this procedure to model $c(M)$ as a function of $\mu(z)$ to maximise the accuracy of \texttt{ReACT} relative to our simulations. We call this implementation the `extended' implementation of \texttt{ReACT}.

A key ingredient in the halo-model approach in \texttt{ReACT} is the density profile of dark matter halos. The standard profile is the Navarro-Frenk-White (NFW) density profile \cite{ref:NFW} 
\begin{equation}
    \rho(r) = \frac{\rho_{\rm s}}{\frac{r}{r_{\rm s}}\left(1 + \frac{r}{r_{\rm s}}\right)}\, ,
\end{equation}
where $r_{\rm s}$ and $\rho_{\rm s}$ are the scale radius and density, respectively. These are both free parameters within the simple parameterization.  In general, the NFW profile is divergent, but one can impose an upper limit on the size of the halo given by the virial radius $r = r_{\rm vir}$. Integrating the density profile gives the halo mass given by
\begin{equation}
    M = \frac{4\pi \rho_{\rm s}r_{\rm s}^3}{3}\left(\ln(1+c) - \frac{c}{1+c}\right)\, ,
\end{equation}
where the NFW concentration parameter $c = r_{\rm vir}/r_{\rm s}$. The parameterization is usually complemented by a mass-concentration relation, $c(M)$. This relation is measured from simulations, and it reduces the number of free parameters to one that is defined by the total mass of the halo given a specific background density.

Within the vanilla implementation of the \texttt{ReACT} algorithm, the $c(M)$ relation is specified by 
\begin{equation}
    c(M) = \frac{c_0}{1+z}\left(\frac{M}{M_*}\right)^{-\alpha} \, ,
    \label{conreact}
\end{equation}
where $c_0 = 9.0$, $M_*$ is the characteristic mass computed by imposing a condition on the peak height $\nu(M_*)=1$ of the Sheth-Tormen mass function and $\alpha = 0.13$. 

\begin{figure}
    \centering
    \includegraphics[width = 0.45\textwidth]{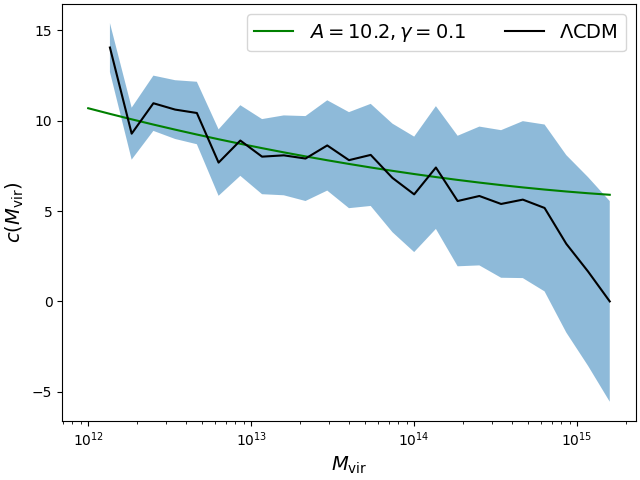}
    \includegraphics[width = 0.45\textwidth]{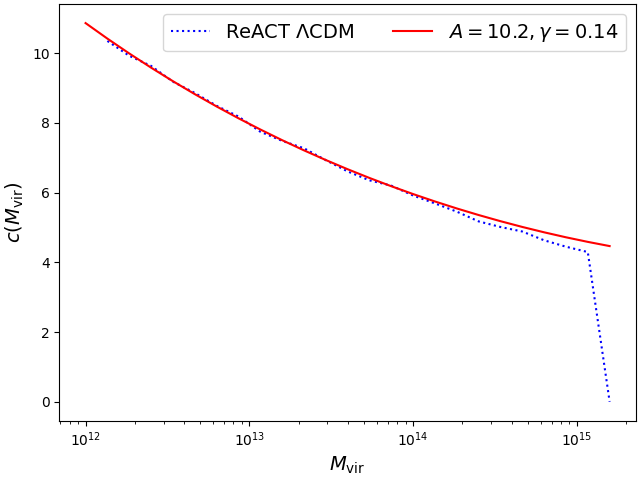}
    \caption[$\Lambda$CDM mass-concentration relation in our simulations and used in the vanilla implementation of \texttt{ReACT}]{In the left panel, we present the concentration-mass relation measured from our $\Lambda$CDM simulations and a fit for the {\it Planck} 2018 (TT+TE+low P+BAO) \cite{ref:Planck2018} parameters from (\cite{ref:MultiDark}). We note that there is significant scatter as our simulation suite has not been optimised for measuring $c(M)$.
    In the right panel, we illustrate that the \texttt{ReACT} $c(M)$ using the expression in \eqref{eq:Multidark} agrees with that from \eqref{conreact}. This allows us to express the concentration-mass relation in our modified gravity simulations as a ratio relative to the $\Lambda$CDM \texttt{ReACT} case.}
    \label{fig:conc_lcdm}
\end{figure}

The $c(M)$ relation for the $\Lambda$CDM model, has been measured from a suite of $N$-body simulations~\cite{ref:MultiDark} and is parameterized by
\begin{equation}\label{eq:Multidark}
    c(M) = A \left(\frac{M}{10^{12}h^{-1}M_{\odot}}\right)^{-\gamma}\left[1 + \left(\frac{M}{M_0}\right)^{0.4}\right]\, ,
\end{equation}
where $A_{\rm \Lambda CDM} = 10.2, \gamma_{\rm \Lambda CDM} = 0.1, M_0 = 10^{17} h^{-1}\,{\rm  M_{\odot}}$. We compare the concentration-mass relation above for {\it Planck} 2018 (TT+TE+low P+BAO) \cite{ref:Planck2018} parameters with the $c(M)$ measured from our simulations and the $c(M)$ relation in \texttt{ReACT} in fig.~\ref{fig:conc_lcdm}. Moreover, we compare \eqref{conreact} and (\ref{eq:Multidark}). Although we find that our simulations are quite noisy - they have not been optimised to measure the $c(M)$ relation - all are in agreement with each other. In what follows we will use (\ref{eq:Multidark}) to modify the $c(M)$ relation within \texttt{ReACT}, keeping $\gamma$ and $M_0$ constant and varying $A$.

 In general, the concentration scales with mass as a weak power law  at low redshifts. This trend can be traced back to the hierarchical nature of structure formation. At late times, when smaller halos have already formed and are merging to form larger halos, the inner regions of these large halos are relatively undisturbed by mass accretion. Therefore, as halos become larger, their virial radii become bigger while the core radius remains relatively unchanged. This then leads to more massive halos having lower concentrations in comparison to less massive halos \cite{ref:MultiDark}. As such, hierarchical structure formation leads one to expect that increasing $\mu$ at late redshifts will enhance structure formation at late times, affecting massive halos and thus \textit{reducing} the ratio of $A/A_{\rm \Lambda CDM}$ and vice-versa for when $\mu$ is increased at early times. We found evidence for this behaviour in \cite{Srinivasan2021} and in this work. In addition, we also find additional complexity arising at very low $z_{\rm mp}$ that we couldn't have resolved earlier due to the larger width of our $\mu$ bins relative to this work. 

In this work, we show that the computation of the full non-linear $P(k)$ with high accuracy in \texttt{ReACT} for generic modified gravity requires modification of the concentration-mass relation in section \ref{subsec:results_conc}. We do this by finding the value of $A$ in (\ref{eq:Multidark}) that gives the best representation of the non-linear power spectrum from the simulation, where this is defined as the representation that has the largest $k_{\rm fail}$. By performing a range of simulations with different modifications to gravity,  we provide a fitting function for the amplitude $A$ of the $c(M)$ relation as a function of $\mu(z)$ validated against the simulations.

We primarily use an accuracy threshold of 1\% to compute $k_{\rm fail}$. However, we note that baryonic feedback has been studied in a recent release of \texttt{ReACT} \cite{BoseReACT} quoting an accuracy of 3\%. Therefore, we also present, both in our abstract and conclusions, the 1\% and 3\% $k_{\rm fail}$ values. This is particularly relevant for our weak lensing results, where we also compute the corresponding values of $\ell_{\rm fail}$ for the weak-lensing convergence power spectrum (again for 1\% and 3\% accuracy thresholds), and we reflect on how robust these numbers are across our parameter space (see sec.~\ref{sec:lensing}).

We have optimised our simulation suite to investigate and fit the non-linear matter power spectrum, which is sub optimal for measuring the $c(M)$ relation from the simulations. As such,  we cannot perform a consistency check between the $c(M)$ in our modified gravity simulations and our fitting function prediction. Since we are only interested in a accurate way of creating the non-linear power spectrum from the linear one, this test is not essential, and we leave it to future work to examine this in detail. 

We note that modifications to gravity have been shown to affect the $\Lambda$CDM $c(M)$ relation. It has been shown that in cosmologies with modified growth histories such as smooth dark energy models that the $c(M)$ relation is modified in order to accurately predict $P(k)$ using the halo model \cite{Dolag2004}. In $f(R)$ gravity, it has been shown that the $c(M)$ is modified from a simple power law due to the fact that $M_*$ contains implicit environmental dependence on halo mass \cite{Lombriser2014, Shi2015, Ruan2023}. These findings further justify our choice to improve the fit from ``vanilla'' ReACT by modifying the $c(M)$ relation.

\section{
N-body simulation suite \& coverage of parameter space
}\label{sec:sims}

In \cite{Srinivasan2021} we studied the phenomenology of models with a time-dependent $\mu$ by implementing piece-wise constant bins\footnote{Up until section \ref{sec:lensing}, all references to ``bins'' refer to the piecewise constant parts of the $\mu$ function, not to tomographic weak-lensing analysis bins.}  in redshift for $\mu$ in our simulations. In particular, we focused on the matter power spectrum $P(k)$ in our simulations. We showed that non-linear clustering statistics can exhibit significant variation as a function of $\mu(z)$ for the same linear $P(k)$. This analysis revealed that the \texttt{ReACT} formalism was able to accurately predict $P(k)$ in our simulations, albeit for a small region of parameter space. In this work we quantitatively validate the \texttt{ReACT} formalism to understand the range of scales over which it can be used to predict $P(k)$ with specified accuracy and precision. To do this, we need to explore a wide range of modified gravity parameter space. Since we bin $\mu$ in redshift, this involves sampling a three dimensional parameter space where the degrees of freedom are the centre of the bin, the bin width, and the value of $\mu$.

As shown in \cite{Srinivasan2021}, the linear growth factor $D(z)$ between the start and end of a redshift bin is a good quantity to use to characterise the bin width (rather than working directly with a difference in $a$ or $z$). As such, we continue to work with redshift bins of constant $D(z)$ in $\Lambda$CDM (this can be computed analytically in linear theory).  We now describe the procedure we followed to fix $\mu(z)$ across our simulations.

 We designed two sets of simulations to extensively test the performance of \texttt{ReACT}. In simulation set $\alpha$ (presented in table \ref{tab:binning}) we fixed $D(z)$ and obtained 8 redshift bins. All the bins in these simulations are chosen such that the $\Lambda$CDM growth factor associated to them is constant and is $D(z_{\rm max})/D(z_{\rm min}) = 1.26$. We choose the initial value of $D(z)$ to be similar to that associated with the binning in \cite{ref:Casas2017} so that we may contextualise our results (and any future results derived from the data products in this work) with parameterised {\it Euclid} forecasts.

 In set $\alpha$ we pick two different values of $\mu$ on each side of $\mu=1$, with the expectation from our findings in \cite{Srinivasan2021} that our results are likely to be symmetric about the line $\mu=1$. In other words, the matter power spectrum exhibited equal and opposite effects when $\mu = 1.1$ and $\mu=0.9$, respectively in the same redshift bin. We found both in our previous work and in this analysis that this behaviour persists and is independent of the choice of $\mu(z)$, (see fig.~\ref{fig:pk} for more info).
 
 We also noted in \cite{Srinivasan2021} that the performance of \texttt{ReACT} varied with position and width of the bin in question. In particular, we found that the `maximal' departure from $\Lambda$CDM was in redshift bins at the extreme end of the redshift range associated to our simulations. This motivates a higher resolution analysis at low $z$ in set $\beta$ \footnote{Ideally, one would also like to sample high redshifts in the simulations. However, we find that the \texttt{ReACT} formalism fails to accurately reproduce $P(k)$ at $z>7$ across the range of values of $D(z)$ that we explored in our simulation suite. Exploration of \texttt{ReACT} at $z>7$ is left for future work. It is pertinent to point out that the range of scales that are non-linear at $z>7$ are too small to be relevant in this analysis.} 
 
\begin{figure}
    \centering
    \includegraphics[width = 0.9\textwidth]{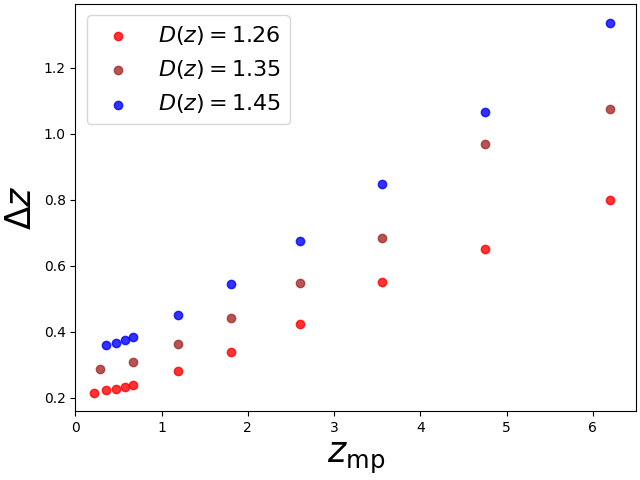}
    \caption{Locations of the set $\beta$ simulations in the $\Delta z-z_{\rm mp}$ plane: the width of each redshift bin $\Delta z$ as a function of the bin midpoint $z_{\rm mp}$ for the different values of $D(z)$ we used in our simulations. We present the redshift bins across all of our simulations and we see that for $z>1$ all the bin midpoints coincide. This allows us to use $z_{\rm mp}$ as a reference point to compare results across different values of $D(z)$. This figure contains both set $\alpha$ (red data-points) (see table \ref{tab:binning}) and set $\beta$ (blue and brown data points). In set $\beta$, we varied $D(z)$ and increased our redshift resolution at $z<1$ to sample the resulting change in shape in $P(k)$ for these low-$z_{\rm mp}$ bins. Note that we run simulations with $\mu = \{1.1, 1.15, 1.2\}$ for all these bins.   }
    \label{fig:mu_Dz}
\end{figure}

In simulation set $\beta$, we vary of $\mu, D(z)$ and over a range values to further validate \texttt{ReACT} in our approach. We restrict ourselves to $\mu>1$ (taking advantage of the aforementioned symmetry to save on computational resources) and introduce additional bins at low redshift (see fig.~\ref{fig:mu_Dz}). We implement finer sampling of $\mu$ at the lower range of $z$ in set $\beta$, which reveals that the shape of the $P(k)$ undergoes a dramatic change as the midpoint of the redshift bin approaches $z=0$ (see figs.~\ref{fig:pk}, \ref{fig:pk_transition} and associated discussion).  In order to develop and validate our fitting function for the concentration-mass amplitude parameter $A$, we ran additional simulations in which we vary the linear growth factor associated with our redshift bins; again, we restrict ourselves to $\mu>1$. We illustrate the full range of bins we have run in fig.~\ref{fig:mu_Dz}. 

We compare simulations across different values of $\mu$ and $D(z)$ by keeping the bin midpoints constant across the parameter-space. In other words, the endpoints of each redshift bin are varied in order to keep the midpoint the same for all $D(z)$. We will return to this point in the next subsection where we show our fitting function. This procedure allows us to systematically investigate the performance of \texttt{ReACT} for varying bin-widths, epochs and $\mu(z)$ in as orthogonal a manner as feasibly possible. Furthermore, as we will see in section \ref{sec:results}, keeping the midpoint constant allows a simple interpretation of the variation of $A(\mu, D(z))$. Our full suite of simulations represents an unprecedented exploration of phenomenological modified gravity parameter space on non-linear scales.

\begin{table}
    \centering
    \begin{tabular}{|c|c|c|c|c|c|c|c|c|}
    \hline
      $\Delta z_{\rm bin}$   &  $\mu_1$ & $\mu_2$ & $\mu_3$ & $\mu_4$ & $\sigma^1_8$ & $\sigma^2_8$ & $\sigma^3_8$ & $\sigma^4_8$ \\
      \hline
       $7 \geq z \geq 5.4$ & 0.8 & 0.9 & 1.1 & 1.2 & 0.7896 & 0.8002 & 0.8215 & 0.8321 \\
       $5.4 \geq z \geq 4.1$ & 0.8 & 0.9 & 1.1 & 1.2 & 0.7896 & 0.8002 & 0.8216 & 0.8322 \\
       $4.1 \geq z \geq 3.0$ & 0.8 & 0.9 & 1.1 & 1.2 & 0.7889 & 0.7999 & 0.8218 & 0.8329 \\
       $3.0 \geq z \geq 2.1$ & 0.8 & 0.9 & 1.1 & 1.2 & 0.7900 & 0.8004 & 0.8213 & 0.8318 \\
       $2.1 \geq z \geq 1.47$ & 0.8 & 0.9 & 1.1 & 1.2 & 0.7915 & 0.8012 & 0.8205 & 0.8302 \\
       $1.47 \geq z \geq 0.91$ & 0.8 & 0.9 & 1.1 & 1.2 & 0.7937 & 0.8023 & 0.8194 & 0.8281 \\
       $0.91 \geq z \geq 0.43$ & 0.8 & 0.9 & 1.1 & 1.2 & 0.7977 & 0.8043 & 0.8175 & 0.8241 \\
       $0.43 \geq z \geq 0$ & 0.8 & 0.9 & 1.1 & 1.2 & 0.8052 & 0.8081 & 0.8136 & 0.8165 \\
       \hline
    \end{tabular}
    \caption{The simulation parameters for set $\alpha$ of our simulation suite in order to validate \texttt{ReACT}. This choice was designed to mimic the binning in \cite{ref:Casas2017} which is likely to be representative of the bin-width typically chosen for a \textit{Euclid}-like survey. For this set of simulations, we varied $\mu$ symmetrically about $\mu=1$, to test if the symmetry we found in \cite{Srinivasan2021} persists. Note that these are all bins with equal $\Lambda$CDM growth, i.e., $D(z_{i-1})/D(z_i) = D(z_i)/D(z_{i+1}) = \text{constant} =  1.26$. This set does not represent the full suite of simulations we run (see fig.~\ref{fig:mu_Dz}). Due to the symmetry in the shape of the power spectrum, we only use $\mu>1$ for set $\beta$, to make optimal use of computational resources.  }
    \label{tab:binning}
\end{table}

\section{Results}\label{sec:results}
\subsection{Vanilla Implementation}
\label{subsec:results_stock}
\begin{figure}
    \centering
    \includegraphics[width=0.95\textwidth]{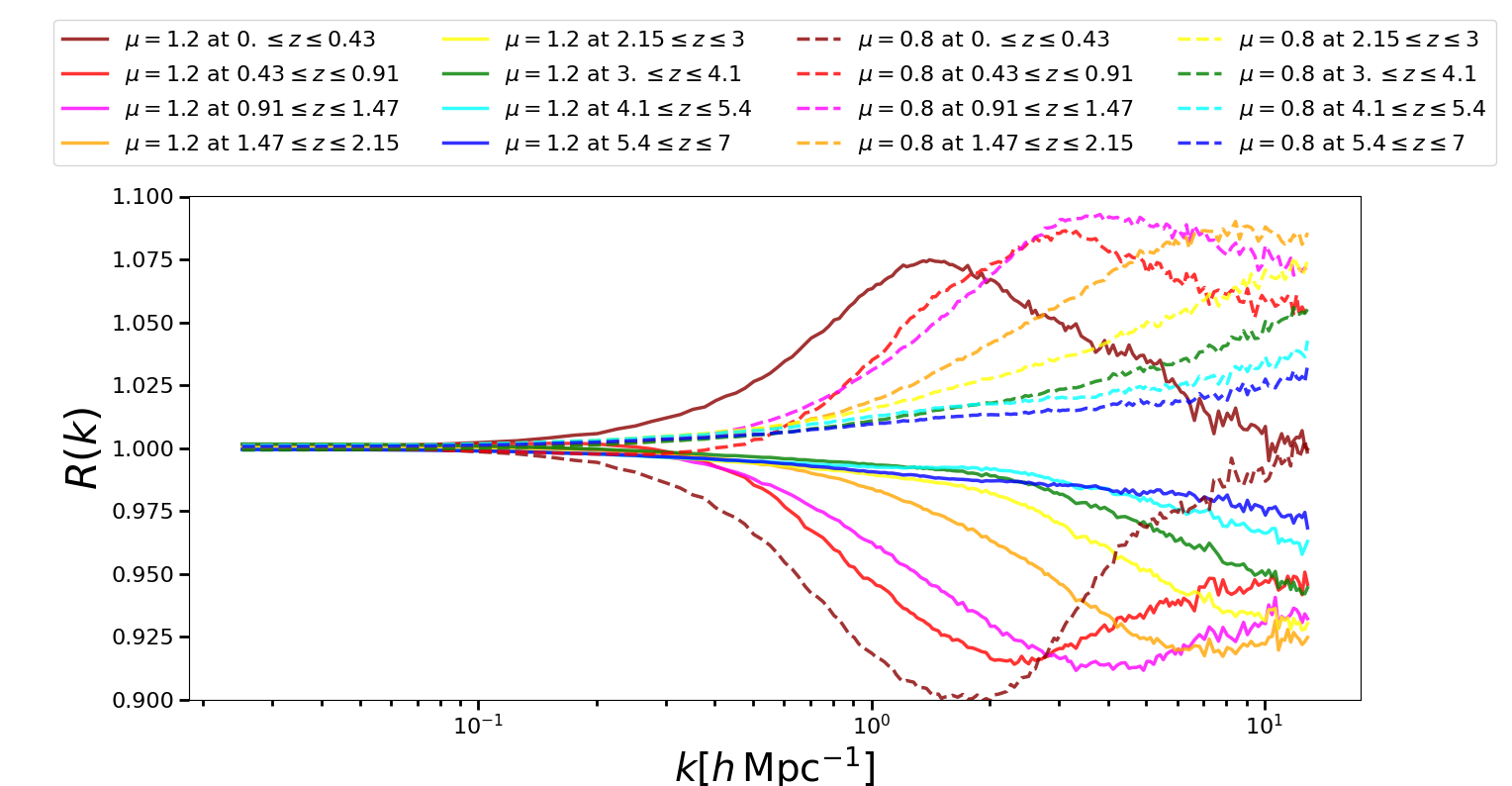}
    \includegraphics[width=0.95\textwidth]{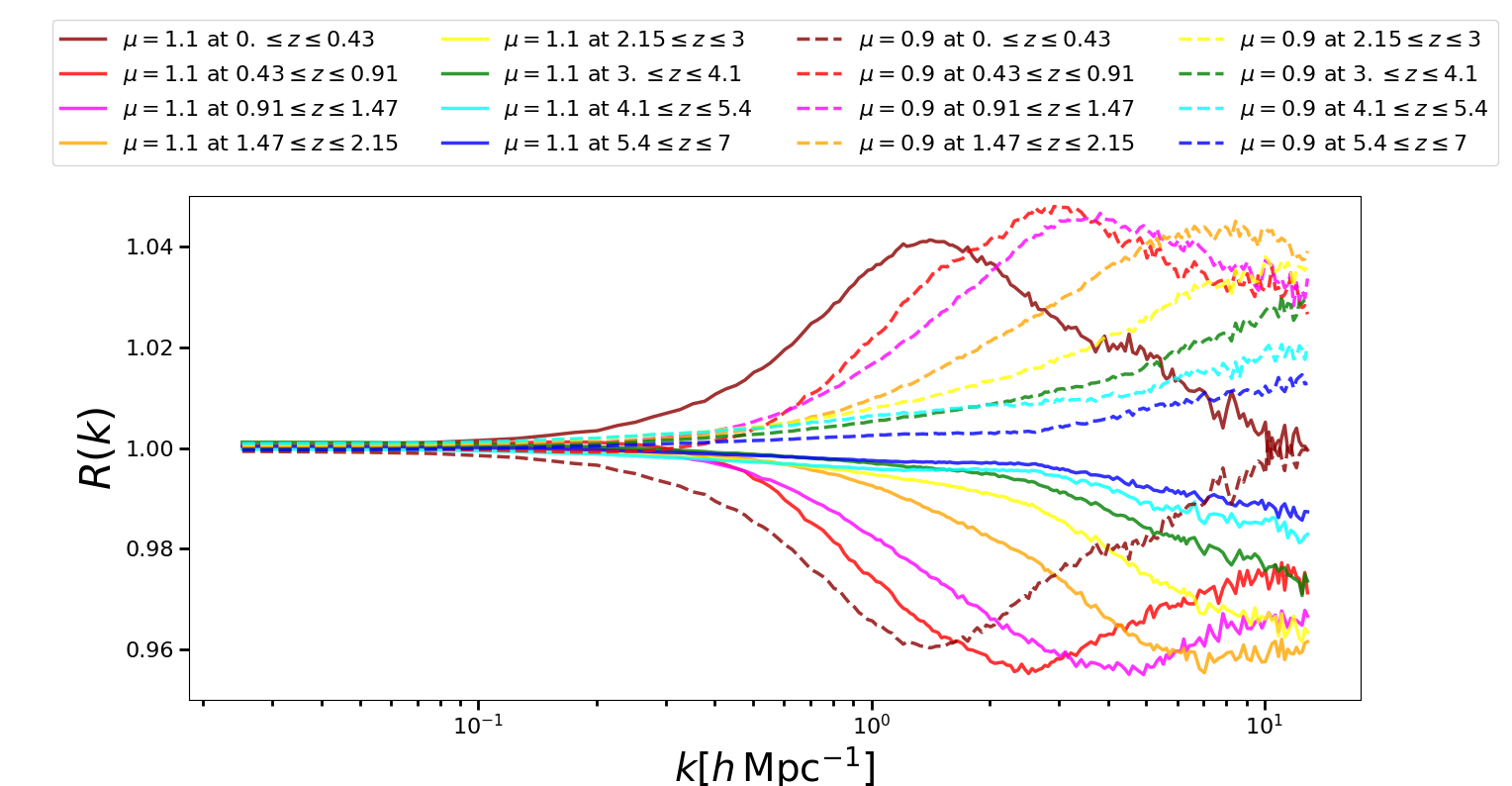}
    \caption{$R(k)$ at $z=0$ from all of our simulations with redshift bins associated with $\Lambda$CDM linear growth $D=1.26$. We observe symmetry in the power spectra about $\mu=1$. We also note that the extent to which the power spectra differ from $\Lambda$CDM increases as the redshift at which modified gravity is switched on decreases. Consistently with this, we see later that the accuracy of \texttt{ReACT} is better for the power spectra closer to $\Lambda$CDM, i.e. for smaller $|\mu-1|$ and higher $z_{\rm mp}$.}
    \label{fig:pk}
\end{figure}

In the previous sections, we have explained the reasons that motivated our choice of using the $\mu-D(z)$ parameter space to explore the modified gravity phenomenology in our simulations. In this section, we show the results of measuring the matter power spectrum from our suite of simulations and quantify the performance of \texttt{ReACT}, both without any modifications and with modifications to the halo concentrations. We present $R(k)$ for the different $\mu(z)$ presented in table \ref{tab:binning} in fig.~\ref{fig:pk}. We note that the symmetry about $\mu=1$ is preserved irrespective of $D(z)$, $\mu(z)$ and the midpoint of the redshift bin. As a consequence of this symmetry, we restrict ourselves to $\mu>1$ for the rest of our analysis. 

\begin{figure}
    \centering
    \includegraphics[width = 0.95\textwidth]{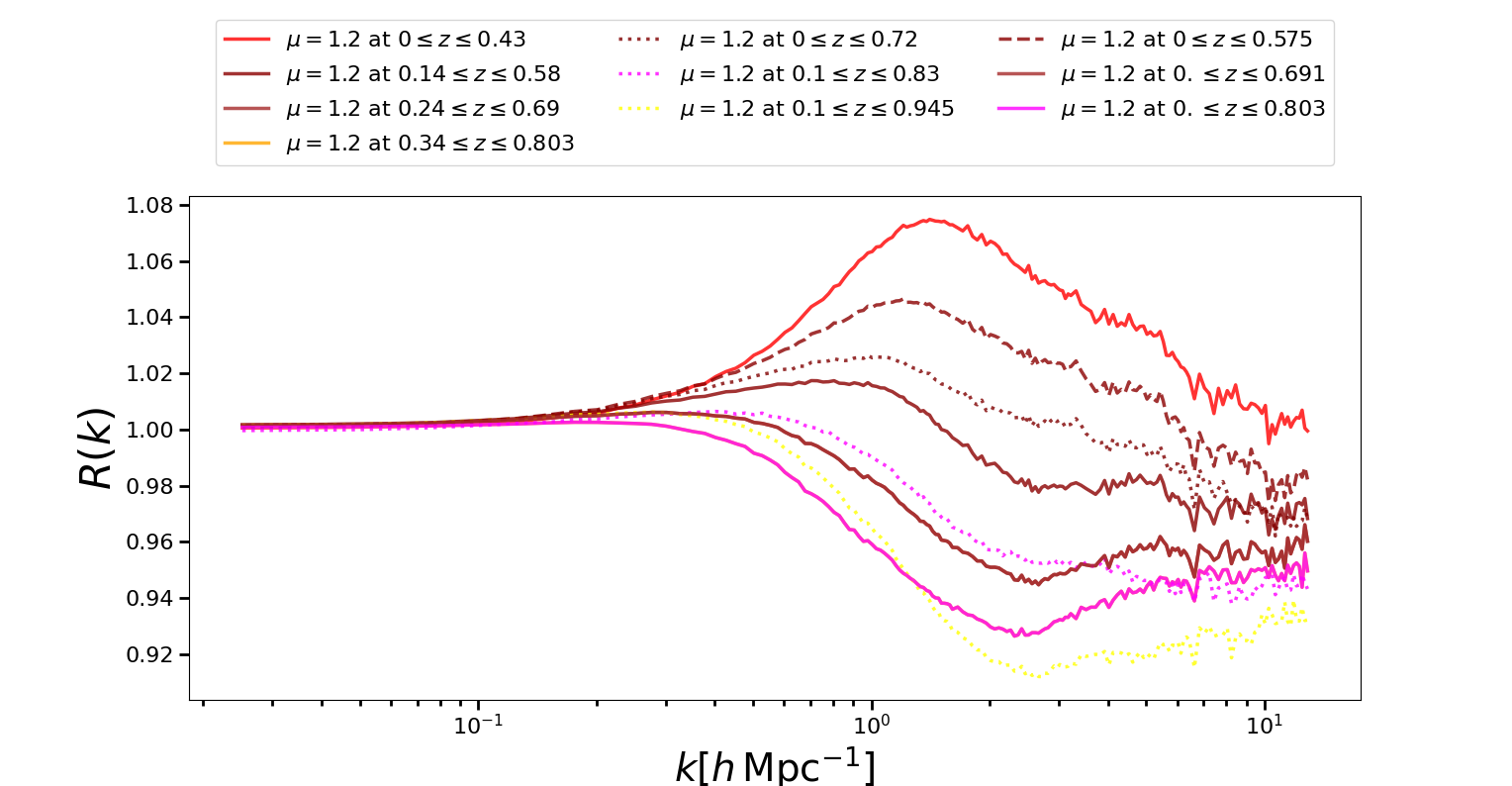}
    \caption{$R(k)$ from the simulations with the $\mu$ modified in the redshift range containing the transition from matter to dark energy domination. The general trend indicated in fig.~\ref{fig:pk} is clearly broken once the Universe enters dark energy domination. This fact will influence the performance of our fitting function (see fig.~\ref{fig:fit}). In the context of maximising the constraining power of future surveys, ensuring that $k_{\rm fail}>1\,h\,{\rm Mpc}^{-1}$ thus requires modifying the halo concentrations consistently with $\mu(z)$ (see sec.~\ref{sec:pipeline}).  }
    \label{fig:pk_transition}
\end{figure}

We also observe in fig.~\ref{fig:pk} that the `departure' from $\Lambda$CDM seems to be more dramatic the later $\mu$ is switched on. This behaviour is consistent with our earlier results in \cite{Srinivasan2021} (but now with greater redshift resolution in the transition regime), where the relative difference to $\Lambda$CDM decreases with increase in the redshift at which the modification occurs. Furthermore, the shape of the matter power spectrum changes drastically when $\mu$ is switched on in the bin $0 \leq z \leq 0.43$ compared to the general trend seen in the other bins. This regime is examined in greater detail in set $\beta$  see fig.~\ref{fig:mu_Dz}). We present the matter power spectra from this set of simulations in fig.~\ref{fig:pk_transition}, where it is clear that this transition seems to occur around the epoch where the transition to dark energy domination takes place. We will return to this point shortly. 

In \cite{Srinivasan2021} we hypothesised that the shape of the power spectra can be traced back to halo formation time as a function of its mass. In other words, due to the hierarchical nature of structure formation, the average mass of the halos forming in our simulations is large at low redshift. Therefore, the range of scales affected relative to the opposite case are  when $\mu$ is modified at low redshift is larger. In other words, there is a transfer of power (from small scales to large scales) that takes place when $\mu$ is switched on at late times. We will come back to this point when we discuss our fitting function for halo concentrations, since this tends to reduce the typical concentration of the halos. 

\begin{figure}
    \centering
    \includegraphics[width = 0.7\textwidth]{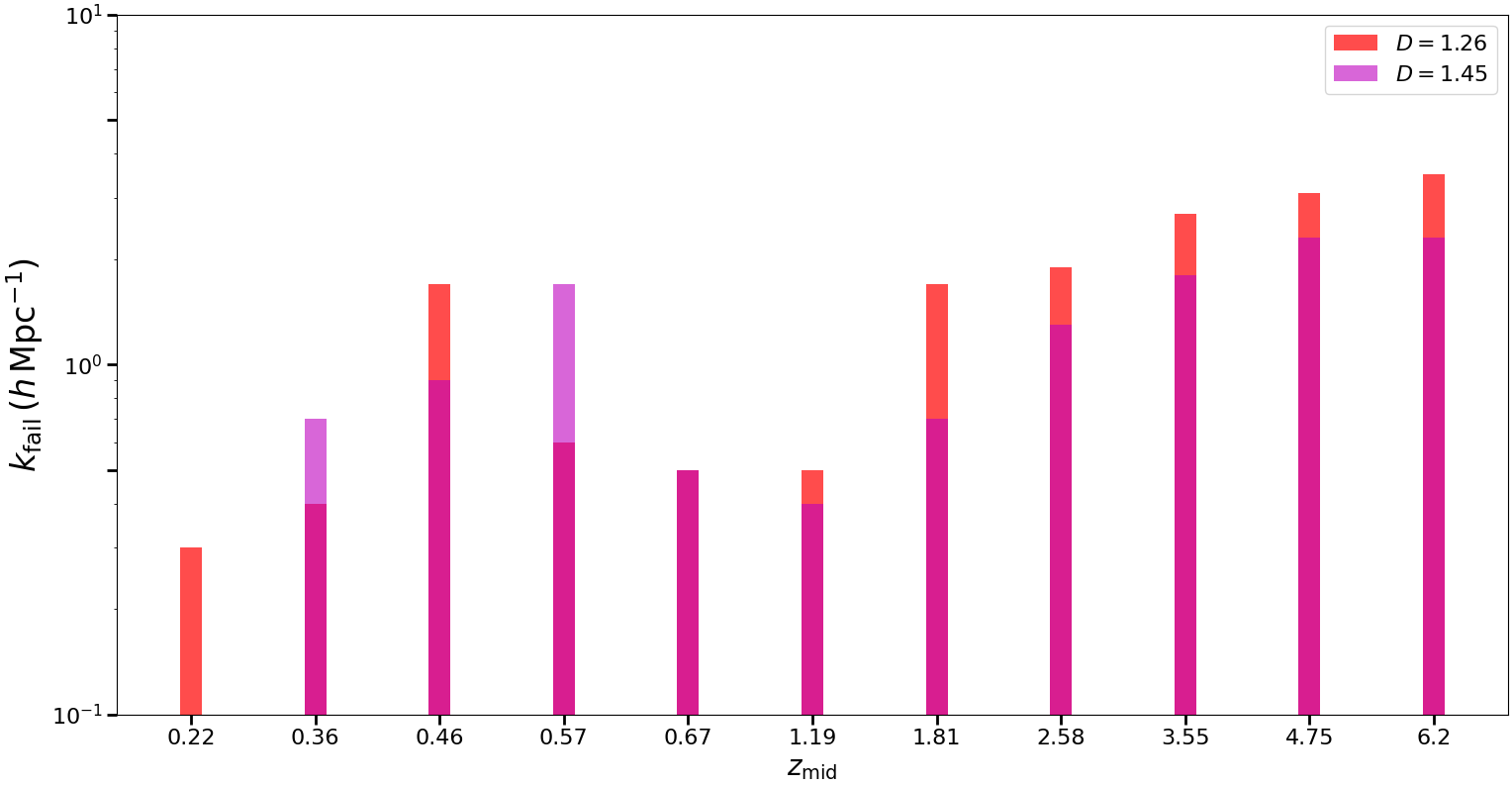}
    \includegraphics[width = 0.7\textwidth]{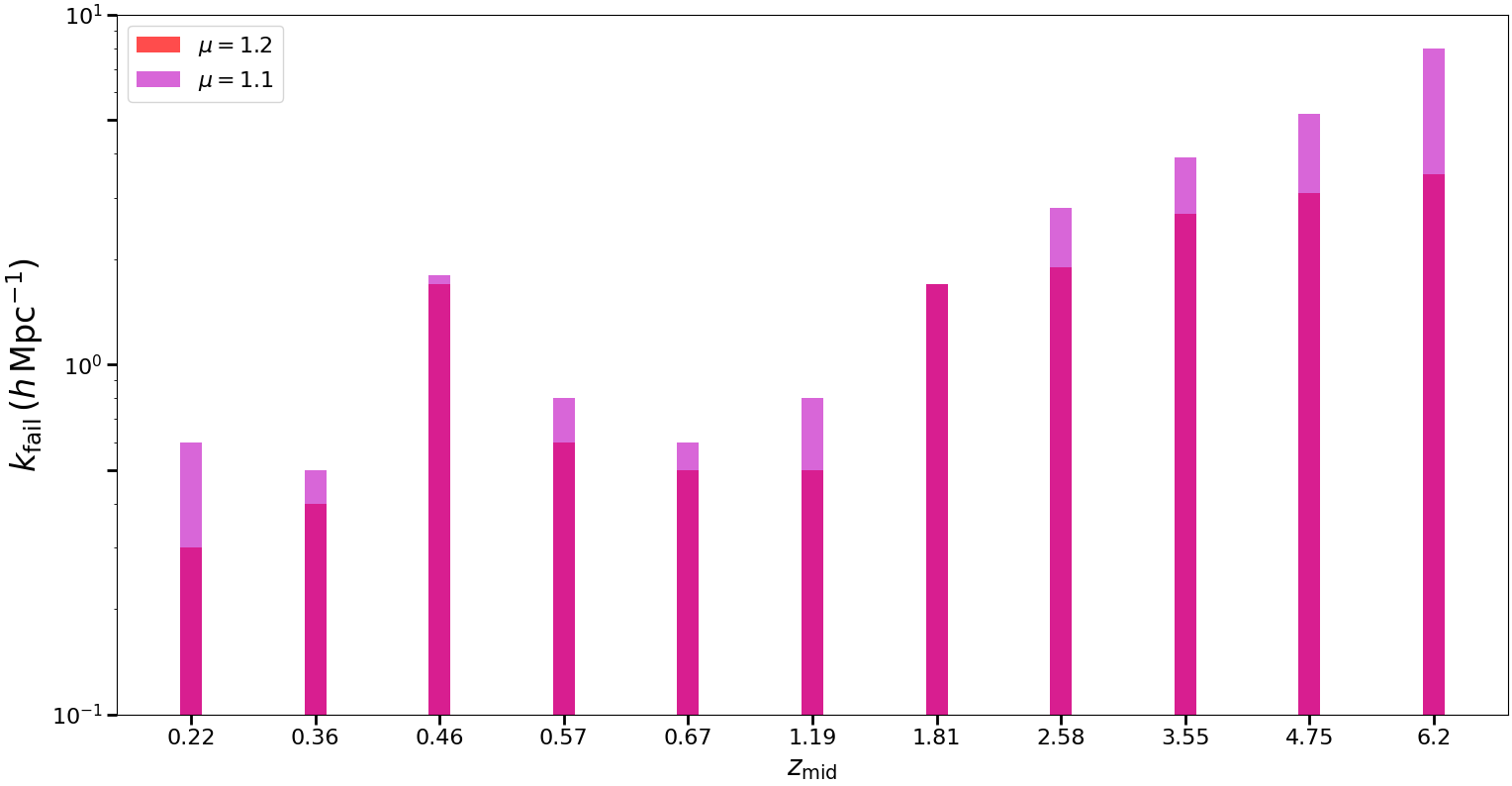}
    \caption{The $k_{\rm fail}$ values as a function of $z_{\rm mp}$ i.e., the midpoint of the redshift bin at which $\mu$ is `switched on' for our vanilla implementation of \texttt{ReACT}. We see that, as indicated in fig.~\ref{fig:pk}, the accuracy of \texttt{ReACT} when $\mu$ is closer to GR, i.e., $k_{\rm fail}$ monotonically decreases as $\mu$ increases. We note that $k_{\rm fail}$ tends to increase with $z_{\rm mp}$, but this trend is broken when $z_{\rm mp}$ is in the transition regime from matter domination to dark energy domination.  }
    \label{fig:kfail_standard}
\end{figure}

In fig.~\ref{fig:kfail_standard}, we quantify this point by computing the value of $k_{\rm fail}$ for our simulations in set $\beta$. We find that the departure from $\Lambda$CDM and, therefore, the accuracy of \texttt{ReACT} is monotonic with $\mu$. We don't see such a clear trend for $k_{\rm fail}$ as a function of $D(z)$. Interestingly, \texttt{ReACT} does much better for the simulations in which $\mu$ is modified at $2 \leq z \leq 7$, across parameter space. As mentioned earlier (see sec.~\ref{sec:conc_mod}, the additional complexity in structure formation arising from mergers of halos drives down $k_{\rm fail}$ for the lower redshift bins, with exceptions arising in cases where the bin-midpoint is such that the bin straddles the transition period between the two separate regimes of structure formation.\footnote{This is due to an artificial cancellation of errors}.

We note that the lowest $k_{\rm fail}$ is associated to the smallest redshift bin, i.e., $0\leq z \leq 0.43$. As mentioned before, the transition into dark energy domination seems to coincide with the change in shape of the power spectra at $z<0.5$ as we previously mentioned (see fig.~\ref{fig:pk_transition}), we see that the trend of how $P(k)$ is modified as a function of $\mu(z)$ is broken and indeed reversed in the range $0\leq z \leq 0.9$. Due to the accelerated expansion at these late times, structure formation appears to be `frozen in', and increasing $\mu$ simply adds power much more uniformly across a range of quasi-linear to non-linear scales relative to the pseudo-$\Lambda$CDM case as seen in fig.~\ref{fig:pk_transition}. 

From our analysis, we see that the main challenge in predicting $P(k)$ across our simulations is ensuring a consistent value of $k_{\rm fail}$ across all our simulations. Particularly in the context of maximising the constraining power of future surveys, ensuring that $k_{\rm fail}>1\,h\,{\rm Mpc}^{-1}$ would require a factor $>2$ improvement in $k_{\rm fail}$ for the lowest redshift bins. We have already mentioned in sec.~\ref{sec:pipeline} that the \texttt{ReACT} $P(k)$ prediction can be improved by modifying the halo concentrations consistently with $\mu(z)$. We will now describe how this can be achieved. 

\subsection{Extended Implementation}
\label{subsec:results_conc}

We perform a fit for the ratio of amplitude of the $c(M)$ parameter relative to the $\Lambda$CDM case, that is $A/A_{\Lambda{\rm CDM}}$, required to maximise the $k_{\rm fail}$ for all $\mu(z)$. This allows us to capture the varying phenomenology of our simulations more accurately using \texttt{ReACT}. Note that while we are fitting for a single parameter, i.e., the ratio of the amplitudes of $c(M)$, the actual fit contains multiple parameters (see eq.~\eqref{eq:Cfit}). Following the \texttt{ReACT} philosophy of computing quantities as a ratio relative to $\Lambda$CDM, we compute our fitting function for the ratio $A/A_{\rm \Lambda CDM}$, thus ensuring our analysis is insensitive to realisation-dependent effects \cite{ref:McDonaldRatio2006}. We remind the reader that in order to compute $R(k)$, one needs the modified gravity $P(k)$ as well as the $\Lambda$CDM $P(k)$ with equalised linear growth.

As earlier, we use the $k_{\rm fail}$ parameter to quantify the accuracy with which \texttt{ReACT} captures the simulation results. We vary the amplitude $A$ of the $c(M)$ relation to maximise $k_{\rm fail}$. We then calculate the ratio $A/A_{\rm \Lambda CDM}$ as a function of the bin midpoint for each set of simulations in fig.~\ref{fig:mu_Dz}. In other words, we model the parameter space across the range of values of $1\leq \mu \leq 1.2$ and $1.26 \leq D \leq 1.45$. With the variation of the amplitude of the concentration-mass relation, we are able to achieve $k_{\rm fail}> 1.5\,h^{-1}{\rm Mpc}$ for an accuracy threshold of $1\%$ and $k_{\rm fail}> 3\,h\,{\rm Mpc}^{-1}$ for an accuracy threshold of $3\%$.

We wish to create a fit for the concentration modification as a function of bin midpoint, and do this as a function of $\mu$ and $D$, thus covering the full phenomenological parameter space we are considering. The general functional form for the fit is given by
\begin{equation}\label{eq;MainFit}
    A/A_{\Lambda \rm CDM} = \begin{cases}
    f_1(\bar{z}) & \text{if } \bar{z}\geq 1.2\\
    f_2(\bar{z}),& \text{if } \bar{z}\leq 1.2  
\end{cases}
\end{equation}
where $\bar{z} = z_{\rm mp} - z_{\rm pk}$ and $z_{\rm pk}$ is the redshift at which the power spectrum is computed. We divide the redshift range into two regimes. In the first regime, i.e., $1.2\leq \bar{z} \leq 7$,  we use an exponential function 
\begin{eqnarray}
    f_1(\bar{z}) = C_1e^{-\bar{z}} + 1\, ,
\end{eqnarray}
 where $C_1<0$ when $\mu>1$ (and vice-versa for $\mu<1$). This form ties back to the qualitative prediction that we made from hierarchical structure formation, i.e., modifying $\mu$ at later times leads to larger halos and therefore smaller concentrations relative $\Lambda$CDM. Note that for calculating the power spectrum at $z_{\rm pk}$ when $\mu \neq 1$ in a particular bin, $A/A_{\rm \Lambda CDM}$ has a single constant value that is a function of $z_{\rm pk}$ and the bin parameters $\{z_{mp}, \mu, D\}$ (see appendix \ref{app:Fitting_Function} for details). In particular, $A/A_{\rm \Lambda CDM}$ is not taken as a function of redshift during the integration in redshift that occurs in the halo model calculation within ReACT. Our general fit for $z_{\rm pk}$ could be used to make $A/A_{\rm \Lambda CDM}$ a function of the redshift in the halo model calculation; we leave an investigation of whether this further improves the accuracy to future work.

We find that this trend is broken at very low redshift, i.e., at $z_{\rm mp}<1$. We fit this upturn in the behaviour of $A/A_{\rm \Lambda CDM}$ using a cubic polynomial 
\begin{equation}
    f_2(z) = (C_2 + C_3\bar{z} + C_4\bar{z}^2 + C_5\bar{z}^3) \, .
\end{equation}
In the transition region between these two cases ($0.7\leq z_{\rm mp} \leq 1.2$) we use a cubic spline to interpolate between $f_1$ and $f_2$. Thus, we have a list of parameters $C_i = \{C_1, C_2, C_3, C_4, C_5\}$ that vary as a function of $\{\mu, D\}$.

We find that the $C_i$ scale linearly with $D$ at fixed values of $\mu$, and the dependence on $\mu$ is a second order polynomial at fixed values of $D$ (see appendix \ref{app:Fitting_Function} for more details). We consider $\mu$ first and  derive the fitting parameters $\tilde{C}_i$ and $\bar{C}_i$, as a function of arbitrary $\mu$ at the extreme values of $D = \{1.26, 1.35\}$, respectively. We then linearly interpolate in $D$ to obtain the general form $\hat{C}_i$ in the two-dimensional $(\mu, D)$ space. This allows us to explicitly construct the final fitting function for $A$ at $z_{\rm pk} = 0$ to be 
\begin{eqnarray}\label{eq:Cfit}
    x & = & 1 - (1.45-D)/(0.19) \, , \\
    \hat{C}_i & = & \tilde{C}_i (1-x) + \bar{C}_i x \,, \\ 
    f_1(\bar{z}) & = & \hat{C}_1 e^{-\bar{z}} + 1  \, , \\ 
    f_2(\bar{z}) & = &  \hat{C}_2 + \hat{C}_3 \bar{z} + \hat{C}_4 \bar{z}^2 + \hat{C}_5 \bar{z}^3 \,,  
    \end{eqnarray}
where we present the explicit functional forms for $\tilde{C}_i$ and $\bar{C}_i$ in appendix \ref{app:Fitting_Function}. Note that equation  gives an accurate calculation of the non-linear matter power spectrum  at z=0 for arbitrary $\mu$, $D(z)$ in the range $0 \leq |\mu -1| \leq 0.2$ and $1.26 \leq D(z) \leq  1.45 $. See appendix \ref{app:Fitting_Function} for how this is generalised to power spectra at non-zero z and for $\mu<1$. This general form is implemented in our extended ReACT code.

\begin{figure}
    \centering
    \includegraphics[width = 0.45\textwidth]{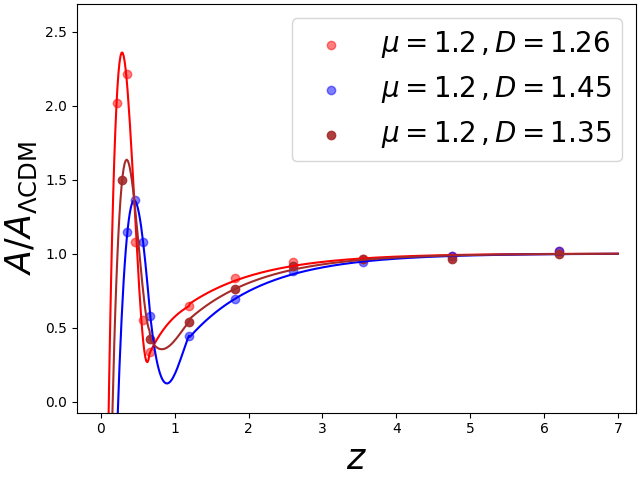}
    \includegraphics[width = 0.45\textwidth]{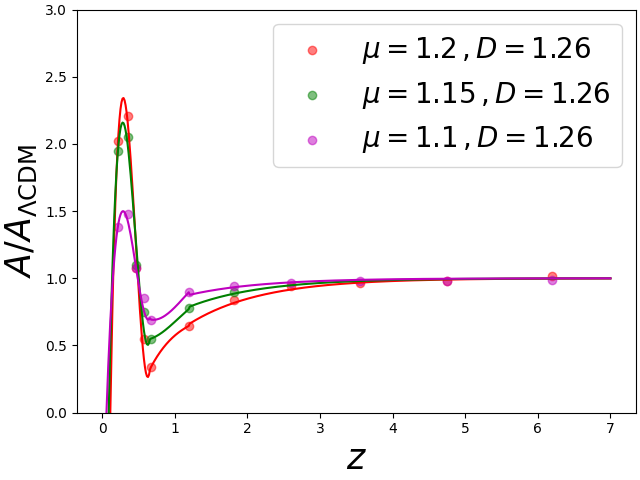}
    \caption{The fitting function for the ratio of the amplitude of the concentration-mass parameter relative to $\Lambda$CDM as a function of $z_{\rm mp}$. The data points are computed from our simulations at the midpoint of the redshift bin at which $\mu$ is switched on in the simulation. In the left panel, we present our fit for varying $D(z)$ and in the right panel, for varying $\mu$. We see that the shape of the fit remains consistent across parameter space, which ties back to our discussion regarding the two regimes (see sec.~\ref{sec:pipeline}). The exponential functional form at $z>1$ follows from the hierarchical structure formation (see appendix \ref{app:spherical_collapse}) while the strong upturn in the concentration at low redshifts is likely due to the freezing in of structure formation as dark energy domination occurs. }
    \label{fig:fit}
\end{figure}

We show the performance of the fitting function described in this section at $z=0$ as a function of $\mu$ and $D(z)$ in fig.\ref{fig:fit}. The two redshift regimes are clearly visible in both the panels. Each data point is obtained from the simulations, and the $k_{\rm fail}$ values associated with this `extended' implementation can be seen in fig.~\ref{fig:kfails}. The robustness of this fitting function to simulation and cosmological parameters is discussed in appendix \ref{app:kfail_analysis}. We also present a qualitative justification of the general observation of the modification at low redshift being more substantial than when $\mu$ is modified at high redshift in appendix \ref{app:spherical_collapse}. 

 Interestingly, the sharp upturn in $A$ coincides with the redshift at which $\Omega_{\Lambda}$ starts becoming comparable to the critical density of the Universe. Once dark energy domination is firmly established, structure formation is frozen in and each modification of gravity affects structure formation uniformly across the range of scales of convergence in our simulations. This numerical recipe is to our knowledge the only validated one in the literature for model-independent modified gravity. 

We have already presented the 1\% $k_{\rm fail}$ for all of our simulations in fig.~\ref{fig:kfails}. In particular, we present the $k_{\rm fail}$ for varying accuracy thresholds in appendix \ref{app:Fitting_Function}. We find that across all of our simulations, we achieve a 1\% $k_{\rm fail} \geq 2\,h\,{\rm Mpc}^{-1}$. Note that this is considerably improved for the bins with $z_{\rm mp}>1$, where $k_{\rm fail} > 3\,h\,{\rm Mpc}^{-1}$ can be achieved. Most notably, we see that a factor $\gtrsim 2$ improvement in $k_{\rm fail}$ with our concentration modification compared to the vanilla \texttt{ReACT}. Finally, $k_{\rm fail}$ values of $\gtrsim 3\,h\,{\rm Mpc}^{-1}$ can be achieved to within 3\% consistently across all our simulations.  

\subsection{Quantifying the performance of the fit for lensing observations}\label{sec:lensing}

\begin{figure}
    \centering
    \includegraphics[width=0.45\textwidth]{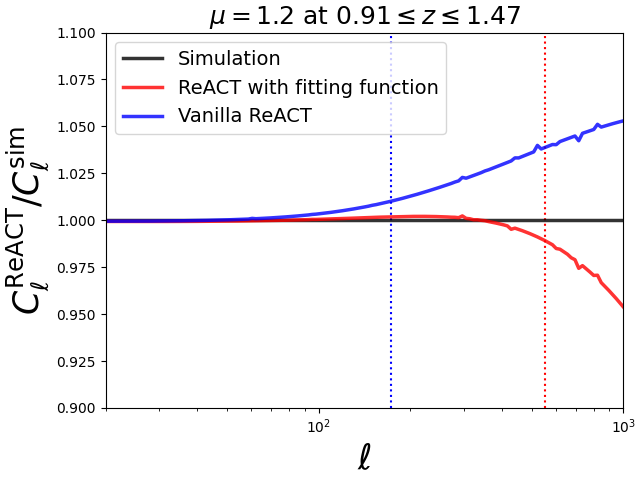}
    \includegraphics[width=0.45\textwidth]{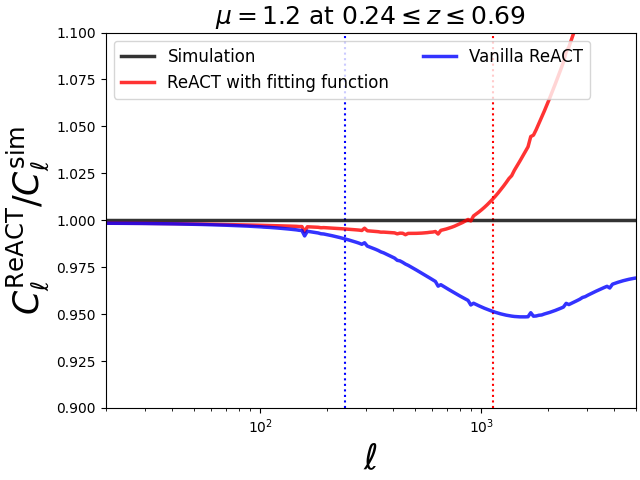}
    \includegraphics[width=0.45\textwidth]{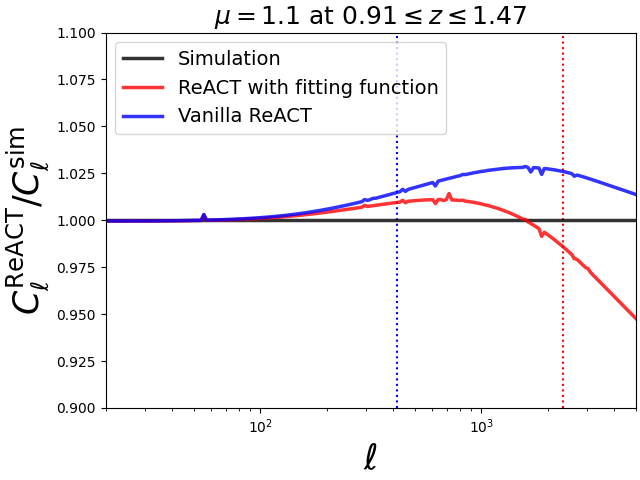}
    \includegraphics[width=0.45\textwidth]{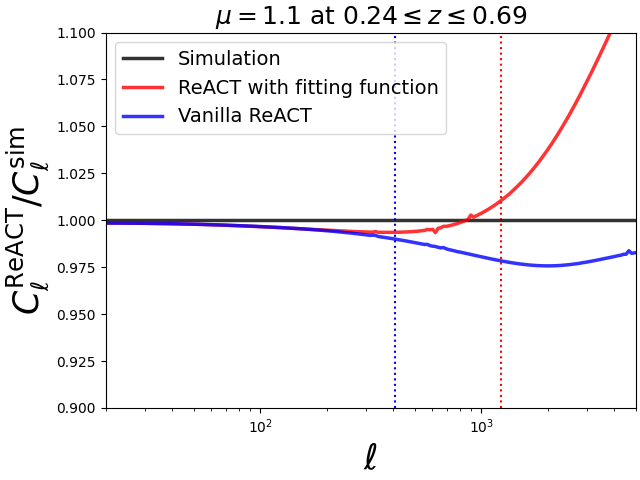}
    \caption{We show the ratio of the weak-lensing convergence spectrum computed using the power spectra directly measured from our simulations to that computed from \texttt{ReACT}. The vertical lines are the cut-off in $\ell$ where the disagreement exceeds 1\% for each case. In the left panel, we present the most pessimistic scenario, i.e., the $\mu(z)$ corresponding to the lowest value of $k_{\rm fail}$. In the right panel, we present a more `realistic' scenario with a $k_{\rm fail}$ is around the typical value in our simulations. For larger values of $\mu$ the improvement is from $\ell\sim 200$ to $\ell \sim 500-1000$. For smaller $\mu$ the improvement is from $\ell \sim 400$ to $\ell>1000$. }
    \label{fig:lfail}
\end{figure}

\begin{figure}
    \centering
    \includegraphics[width = 0.7\textwidth]{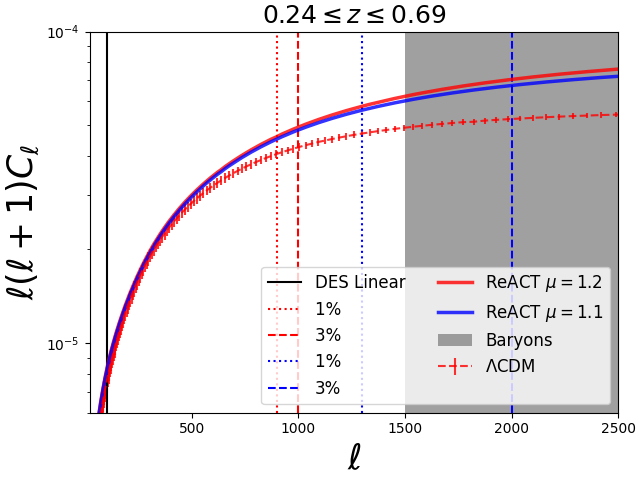}
    \caption{The weak-lensing convergence power spectrum cut-off computed for the bin with the smallest $k_{\rm fail}$ compared to the $\Lambda$CDM spectrum with \textit{Euclid}-like error bars. As presented in table \ref{tab:lfail_table}, we show the cutoff that DES employed in their beyond-$\Lambda$CDM analysis which was restricted to linear scales, 1\% and 3\% values of $\ell_{\rm fail}$ for $\mu = 1.2$ (red) and $\mu=1.1$ (blue). We also indicate the regions where baryons are most likely to play a significant role \cite{mancini2023degeneracies}. Here, we see explicitly here that the amount of information scales as $\ell^2$, and therefore cutting off the $C_{\ell}$ at low value of $\ell$ leads to information loss and therefore smaller constraining power. Note also that the error bars become smaller as one goes from small $\ell$ to large $\ell$, i.e., from large scales to small scales.  }
    \label{fig:kcut_lensing}
\end{figure}

We now quantify the accuracy and range of validity of our fit for future lensing surveys and deduce $\ell_{\rm fail}$, analogous to $k_{\rm fail}$ in angular space\footnote{Note that the angular projection means that there is not a one-to-one correspondence between $k_{\rm fail}$ and $\ell_{\rm fail}$.} We compute the weak-lensing convergence power spectrum from the matter power spectrum and the modified gravity parameters
\begin{equation}\label{eq:convergenceSpec}
    C_{\ell} = \frac{9H_0^4\Omega_{\rm m}^2}{4c^4}\int_0^{\chi_{\rm max}}\mathrm{d}\chi\frac{1}{a^2(\chi)} g^2(z)\frac{\mu^2(1 + \eta)^2}{4}P_{\delta}(\ell/\chi)\, ,
\end{equation}
where $P_{\delta}$ is the matter power spectrum, $\eta$ is the second modified gravity parameter that affects the photon geodesics and $\chi$ is the comoving angular diameter distance to the source along the line of sight. We do this for the two conservative and realistic scenarios from our simulations. 

The function $g(z)$ is a filter function that depends on the redshift distribution of the background galaxies and is typically written as
\begin{equation}
 g_i(z) = \int_z^{\infty}\mathrm{d}z^{\prime}\left(1- \frac{\chi(z)}{\chi(z^{\prime})}\right)n_i(z^{\prime})\,.
\end{equation}
We use the standard expression for the source galaxy redshift distribution given by \cite{ref:Smail1994}
\begin{equation}\label{eq:smail}
 n(z) \propto \left(\frac{z}{z_0}\right)^{\alpha}\exp\left[-\left(\frac{z}{z_0}\right)^{\beta}\right]\,.
\end{equation}
We adopt a {\it Euclid}-like binning of the source number density into 10 equi-populated bins according to eq.~(\ref{eq:smail}) with $z_0 = 0.9/\sqrt{2}$, $\alpha = 2$ and $\beta = 3/2$, where we have assumed an average source density $\bar{n}_{\rm g} = 30\,{\rm arcmin}^{-2}$ \cite{ref:Hu1999, ref:Castro2005, ref:Casas2017, ref:SpurioMancini2018, ref:EuclidWL}. The tomographic bins are such that our fitting function may be used across the entire weak-lensing data-set. The projected errorbars are obtained by computing the following \cite{ref:EuclidWL}
\begin{equation}
 \delta C_\ell^{ij} = \sqrt{\frac{2}{f_{\rm sky}(2\ell + 1)}}\left(C_l^{ij} + \frac{\sigma_{\epsilon}^2}{\bar{n}_i} \delta^{ij}\right) \, , 
\end{equation}
where $f_{\rm sky} = 0.7$ is the fractional sky coverage, $\sigma_{\epsilon} = 0.21$ is the variance of the observed ellipticities and $\bar{n}_i$ is the surface galaxy density of each bin.

We compute the $C_{\ell}$ from the matter power spectra using both our simulation snapshots as well as from \texttt{ReACT}. We compare the prediction from vanilla \texttt{ReACT} and with our concentration fitting function implemented in the code with that obtained from the simulations to obtain an $\ell_{\rm fail}$ value up to which we formally validate \texttt{ReACT} for weak-lensing. 

While we cannot associate a one-to-one mapping between $k_{\rm fail}$ and $\ell_{\rm fail}$, we would expect naively that simulations with small values of $k_{\rm fail}$ will be correlated to the ones with low $\ell_{\rm fail}$ values. We report the 1\% $\ell_{\rm fail}$ value for this case as our most conservative estimate of the scale up to which one can reliably compute the convergence power spectrum $C_{\ell}$. However, we also quote the 3\% $\ell_{\rm fail}$ as a more `realistic' estimate of the physical scales that can be accurately modelled with \texttt{ReACT}. Note that our 1\% accuracy results may be viewed to be conservative because the expected level of accuracy that can be achieved across the parameter space validated in our simulations accounting for baryonic feedback \cite{Bose_2021} in \texttt{ReACT} is 3\%.  We show an example of these two cases in fig.~\ref{fig:lfail}. Note that applying our extended implementation of \texttt{ReACT} always leads to a significant improvement in the range of scales we can accurately model. In addition, For smaller values of $\mu$, even our vanilla implementation of \texttt{ReACT} is able to achieve sub-3\% accuracy.   

Table \ref{tab:lfail_table} shows that with our vanilla implementation, we are able to consistently model the weak-lensing convergence power spectrum only up to $\ell\sim 500$. In comparison, with our extended implementation, we are able to reliably model up to $\ell \sim 1300$. To put this in context, the DES analysis that reported 3x2pt constraints \cite{DES_Y1_2018} on extensions to $\Lambda$CDM  employed stringent scale-cuts that only data on linear scales survived. This corresponds to an $\ell_{\rm fail} \sim 100$ in our analysis. We note that this leads to a significant loss in constraining power. In a recent study \cite{mancini2023degeneracies}, it was found that the modelling of baryons limited the constraining power that can be gained on scales corresponding to $\ell > 1500$. We condense all these results into fig.~\ref{fig:kcut_lensing}, where it is clear the that using our extended formalism allows one to achieve values of $\ell_{\rm fail}$ close to the limit quoted above which baryons will likely dominate the constraining power. We note that we are able to achieve $\ell_{\rm fail}$ within a factor of $\sim 2$ of this limit even in the most conservative cases, and indeed achieve $\ell_{\rm fail} >1500$ for the vast majority of our simulations. Without our concentration fitting function, we consistently achieve $\ell_{\rm fail}<500$.

\begin{table}
\centering
\begin{tabular}{|c|c|c|c|c|c|c|c|c|}
\hline
\multirow{2}{*}{$|\mu - 1|$} & \multicolumn{2}{c|}{Vanilla $\ell_{\rm fail}$}  & %
    \multicolumn{2}{c|}{Extended $\ell_{\rm fail}$}  & \multicolumn{2}{c|}{Vanilla $k_{\rm fail}$} & %
    \multicolumn{2}{c|}{Extended $k_{\rm fail}$}  \\
\cline{2-9}
 & 1\% & 3\%  & 1\% & 3\% & 1\% & 3\% & 1\% & 3\%\\
\hline
 0.2 & 400 & 900 & 900 & 1200 & 0.4 & 0.8 & 1.2 & 3.0\\
\hline
 0.1 & 600 & 2100 & 2400 & 3500 & 0.7 & 1.0 & 1.4 & 4.3\\
\hline
\end{tabular}
\caption{We report the value of $\ell_{\rm fail}$ that we find across the range of values of $\mu$ in our simulations. We find that the variation in $\ell_{\rm fail}$ is much more significant with $\mu$ compared to the linear growth factor $D(z)$. Therefore, we report a conservative 1\% $\ell_{\rm fail}$ corresponding to the simulation with the smallest 1\% $k_{\rm fail}$ and a `realistic' 3\% $\ell_{\rm fail}$ that corresponds to a more typical value of $k_{\rm fail}$ that can be achieved to within 3\% (see appendix \ref{app:Fitting_Function} for more details on $k_{\rm fail}$ as a function of $\mu$ and accuracy threshold). We pick these numbers to correspond to the bin corresponding to the smallest values of $k_{\rm fail}$ as the limiting value for given value of $\mu$, which is $0.91 \leq z \leq 1.47$.}
\label{tab:lfail_table}
\end{table}

\section{Conclusion}\label{sec:conclusion}
The wealth of data provided by upcoming surveys such as the \textit{Euclid} satellite, NRT, VRO and SKA creates an unprecedented opportunity to test modified gravity in a model-independent way with cosmology, which could provide crucial evidence for the nature of the dark sector. The majority of this survey data is on non-linear scales, so robust theoretical predictions for this regime are crucial for maximising the gain from these surveys.

In this work we extend the preliminary investigation in \cite{Srinivasan2021} to perform a thorough validation of our \texttt{ReACT} approach for predicting the non-linear matter power spectrum in phenomenological modified gravity. We cover the modified gravity parameter space by treating $\mu$ as a piece-wise constant function, and we create a substantial suite of N-body simulations varying the value in each piece, the width of each piece, and the location (in redshift) of each piece.

We analyse the performance of our ``vanilla'' implementation of \texttt{ReACT} across this simulation suite, finding that we are able to model the weak lensing convergence power spectrum measured in our simulations across parameter space reliably up to $\ell \sim 500$. There is a significant loss in constraining power associated to this cutoff, as seen in fig.~\ref{fig:kcut_lensing}. Therefore, we provide a fitting function to improve this. 

We find that the concentration-mass relation is one of the key ingredients in \texttt{ReACT} and have derived a fit for the ratio of its amplitude relative to $\Lambda$CDM across the MG parameter space. The key results are in equations \eqref{eq;MainFit}, \eqref{eq:Cfit}, generalised to arbitrary redshift in appendix \ref{app:Fitting_Function}, which we implement as a modification to the $\Lambda$CDM $c(M)$ relation in our extended version of \texttt{ReACT}. See table \ref{tab:lfail_table} for the range of validity of this expression for the matter power spectrum and weak-lensing calculations, given different choices of accuracy level and the region of phenomenological MG parameter space. We are able to achieve 1\% accuracy in predicting the matter power spectrum up to $k= 1\,h\,{\rm Mpc}^{-1}$ for the extreme case of $\mu=1.2$ in each of our bins and 3\% accuracy up to  $k= 3\,h\,{\rm Mpc}^{-1}$. We achieve larger values of $k_{\rm fail}$ for $\mu=1.1$ and this can be seen in fig.~\ref{fig:kfails}. We show for a Euclid-like survey, how this translates to a 1\% (3\%) cut-off of $\ell = 900 \, (1200)$ for bins with $\mu=1.2$ and $\ell = 1300\, (2000)$ for $\mu=1.1$.  



Most importantly, the tools developed here allow the data from existing and future surveys to be used to constrain phenomenological modified gravity. Performing this analysis robustly will be facilitated by the existing work demonstrating how baryonic effects and massive neutrions can be included in the \texttt{ReACT} approach  \cite{Carrilho_2022}. 

Our simulations present a useful dataset from which one can recover the peculiar velocity field and therefore compute the small scale redshift-space-distortions of galaxy clustering \cite{Ruan:2021rqv}. Cross-correlating the galaxy-clustering and lensing data can prove to be a powerful tool in the future to probe gravity on non-linear scales. Such an analysis would provide valuable insight into the nature of the pixelised $\mu$ approach, which will inform the full time and scale dependent analysis. In other words, we provide a way to produce 3x2pt forecasts and constraints in a model-independent way. 


The time-dependent case studied here and in \cite{Srinivasan2021} exhibits rich phenomenology on non-linear scales, and some theories (such as those encompassed by the Parameterised-Post-Newtonian Cosmology approach \cite{Thomas_2023, Clifton_2019, Sanghai_2015}), have no scale dependence in $\mu$ on non-linear scales. As such a detection or tight constraints on this purely time-dependent case would be a significant step forward in terms of testing gravity on cosmological scales. Nonetheless, now that the time-dependent analysis has been understood and fully validated, this approach can be extended to a wider analysis that includes the extra complexity from additionally allowing scale dependent behaviour of $\mu$.

\section*{Acknowledgments}
This work was performed partly using the DiRAC Data Intensive service at Leicester, operated by the University of Leicester IT Services, which forms part of the STFC DiRAC HPC Facility (www.dirac.ac.uk). The equipment was funded by BEIS capital funding via STFC capital grants ST/K000373/1 and ST/R002363/1 and STFC DiRAC Operations grant ST/R001014/1. DiRAC is part of the National e-Infrastructure. We would like to acknowledge the assistance given by Research IT, and the use of The HPC Pool funded by the Research Lifecycle Programme at The University of Manchester. DBT acknowledges support from the Science and Technology Facilities Council (STFC, grant numbers ST/P000592/1 and ST/X006344/1). SS was supported formerly by a George Rigg Scholarship and more recently by the UK Science and Technology Facilities Council (STFC).  
\newpage
\clearpage

\bibliography{References.bib}

\begin{thebibliography}{42}%
\makeatletter
\providecommand \@ifxundefined [1]{%
 \@ifx{#1\undefined}
}%
\providecommand \@ifnum [1]{%
 \ifnum #1\expandafter \@firstoftwo
 \else \expandafter \@secondoftwo
 \fi
}%
\providecommand \@ifx [1]{%
 \ifx #1\expandafter \@firstoftwo
 \else \expandafter \@secondoftwo
 \fi
}%
\providecommand \natexlab [1]{#1}%
\providecommand \enquote  [1]{``#1''}%
\providecommand \bibnamefont  [1]{#1}%
\providecommand \bibfnamefont [1]{#1}%
\providecommand \citenamefont [1]{#1}%
\providecommand \href@noop [0]{\@secondoftwo}%
\providecommand \href [0]{\begingroup \@sanitize@url \@href}%
\providecommand \@href[1]{\@@startlink{#1}\@@href}%
\providecommand \@@href[1]{\endgroup#1\@@endlink}%
\providecommand \@sanitize@url [0]{\catcode `\\12\catcode `\$12\catcode
  `\&12\catcode `\#12\catcode `\^12\catcode `\_12\catcode `\%12\relax}%
\providecommand \@@startlink[1]{}%
\providecommand \@@endlink[0]{}%
\providecommand \url  [0]{\begingroup\@sanitize@url \@url }%
\providecommand \@url [1]{\endgroup\@href {#1}{\urlprefix }}%
\providecommand \urlprefix  [0]{URL }%
\providecommand \Eprint [0]{\href }%
\providecommand \doibase [0]{http://dx.doi.org/}%
\providecommand \selectlanguage [0]{\@gobble}%
\providecommand \bibinfo  [0]{\@secondoftwo}%
\providecommand \bibfield  [0]{\@secondoftwo}%
\providecommand \translation [1]{[#1]}%
\providecommand \BibitemOpen [0]{}%
\providecommand \bibitemStop [0]{}%
\providecommand \bibitemNoStop [0]{.\EOS\space}%
\providecommand \EOS [0]{\spacefactor3000\relax}%
\providecommand \BibitemShut  [1]{\csname bibitem#1\endcsname}%
\let\auto@bib@innerbib\@empty
\bibitem [{\citenamefont {{Clifton}}\ \emph {et~al.}(2012)\citenamefont
  {{Clifton}}, \citenamefont {{Ferreira}}, \citenamefont {{Padilla}},\ and\
  \citenamefont {{Skordis}}}]{ref:CliftonReview}%
  \BibitemOpen
  \bibfield  {author} {\bibinfo {author} {\bibfnamefont {T.}~\bibnamefont
  {{Clifton}}}, \bibinfo {author} {\bibfnamefont {P.~G.}\ \bibnamefont
  {{Ferreira}}}, \bibinfo {author} {\bibfnamefont {A.}~\bibnamefont
  {{Padilla}}}, \ and\ \bibinfo {author} {\bibfnamefont {C.}~\bibnamefont
  {{Skordis}}},\ }\href {\doibase 10.1016/j.physrep.2012.01.001} {\bibfield
  {journal} {\bibinfo  {journal} {\physrep}\ }\textbf {\bibinfo {volume}
  {513}},\ \bibinfo {pages} {1} (\bibinfo {year} {2012})},\ \Eprint
  {http://arxiv.org/abs/1106.2476} {arXiv:1106.2476 [astro-ph.CO]} \BibitemShut
  {NoStop}%
\bibitem [{\citenamefont {Nojiri}\ \emph {et~al.}(2017)\citenamefont {Nojiri},
  \citenamefont {Odintsov},\ and\ \citenamefont {Oikonomou}}]{Nojiri_2017}%
  \BibitemOpen
  \bibfield  {author} {\bibinfo {author} {\bibfnamefont {S.}~\bibnamefont
  {Nojiri}}, \bibinfo {author} {\bibfnamefont {S.}~\bibnamefont {Odintsov}}, \
  and\ \bibinfo {author} {\bibfnamefont {V.}~\bibnamefont {Oikonomou}},\ }\href
  {\doibase 10.1016/j.physrep.2017.06.001} {\bibfield  {journal} {\bibinfo
  {journal} {Physics Reports}\ }\textbf {\bibinfo {volume} {692}},\ \bibinfo
  {pages} {1} (\bibinfo {year} {2017})}\BibitemShut {NoStop}%
\bibitem [{\citenamefont {{Laureijs}}\ \emph {et~al.}(2011)\citenamefont
  {{Laureijs}}, \citenamefont {{Amiaux}}, \citenamefont {{Arduini}},
  \citenamefont {{Augu{\`e}res}}, \citenamefont {{Brinchmann}}, \citenamefont
  {{Cole}}, \citenamefont {{Cropper}}, \citenamefont {{Dabin}}, \citenamefont
  {{Duvet}}, \citenamefont {{Ealet}}, \citenamefont {{Garilli}}, \citenamefont
  {{Gondoin}}, \citenamefont {{Guzzo}}, \citenamefont {{Hoar}}, \citenamefont
  {{Hoekstra}}, \citenamefont {{Holmes}}, \citenamefont {{Kitching}},
  \citenamefont {{Maciaszek}}, \citenamefont {{Mellier}}, \citenamefont
  {{Pasian}}, \citenamefont {{Percival}}, \citenamefont {{Rhodes}},
  \citenamefont {{Saavedra Criado}}, \citenamefont {{Sauvage}}, \citenamefont
  {{Scaramella}}, \citenamefont {{Valenziano}}, \citenamefont {{Warren}},
  \citenamefont {{Bender}}, \citenamefont {{Castander}}, \citenamefont
  {{Cimatti}}, \citenamefont {{Le F{\`e}vre}}, \citenamefont {{Kurki-Suonio}},
  \citenamefont {{Levi}}, \citenamefont {{Lilje}}, \citenamefont {{Meylan}},
  \citenamefont {{Nichol}}, \citenamefont {{Pedersen}}, \citenamefont {{Popa}},
  \citenamefont {{Rebolo Lopez}}, \citenamefont {{Rix}}, \citenamefont
  {{Rottgering}}, \citenamefont {{Zeilinger}}, \citenamefont {{Grupp}},
  \citenamefont {{Hudelot}}, \citenamefont {{Massey}}, \citenamefont
  {{Meneghetti}}, \citenamefont {{Miller}}, \citenamefont {{Paltani}},
  \citenamefont {{Paulin-Henriksson}}, \citenamefont {{Pires}}, \citenamefont
  {{Saxton}}, \citenamefont {{Schrabback}}, \citenamefont {{Seidel}},
  \citenamefont {{Walsh}}, \citenamefont {{Aghanim}}, \citenamefont
  {{Amendola}}, \citenamefont {{Bartlett}}, \citenamefont {{Baccigalupi}},
  \citenamefont {{Beaulieu}}, \citenamefont {{Benabed}}, \citenamefont
  {{Cuby}}, \citenamefont {{Elbaz}}, \citenamefont {{Fosalba}}, \citenamefont
  {{Gavazzi}}, \citenamefont {{Helmi}}, \citenamefont {{Hook}}, \citenamefont
  {{Irwin}}, \citenamefont {{Kneib}}, \citenamefont {{Kunz}}, \citenamefont
  {{Mannucci}}, \citenamefont {{Moscardini}}, \citenamefont {{Tao}},
  \citenamefont {{Teyssier}}, \citenamefont {{Weller}}, \citenamefont
  {{Zamorani}}, \citenamefont {{Zapatero Osorio}}, \citenamefont {{Boulade}},
  \citenamefont {{Foumond}}, \citenamefont {{Di Giorgio}}, \citenamefont
  {{Guttridge}}, \citenamefont {{James}}, \citenamefont {{Kemp}}, \citenamefont
  {{Martignac}}, \citenamefont {{Spencer}}, \citenamefont {{Walton}},
  \citenamefont {{Bl{\"u}mchen}}, \citenamefont {{Bonoli}}, \citenamefont
  {{Bortoletto}}, \citenamefont {{Cerna}}, \citenamefont {{Corcione}},
  \citenamefont {{Fabron}}, \citenamefont {{Jahnke}}, \citenamefont {{Ligori}},
  \citenamefont {{Madrid}}, \citenamefont {{Martin}}, \citenamefont
  {{Morgante}}, \citenamefont {{Pamplona}}, \citenamefont {{Prieto}},
  \citenamefont {{Riva}}, \citenamefont {{Toledo}}, \citenamefont
  {{Trifoglio}}, \citenamefont {{Zerbi}}, \citenamefont {{Abdalla}},
  \citenamefont {{Douspis}}, \citenamefont {{Grenet}}, \citenamefont
  {{Borgani}}, \citenamefont {{Bouwens}}, \citenamefont {{Courbin}},
  \citenamefont {{Delouis}}, \citenamefont {{Dubath}}, \citenamefont
  {{Fontana}}, \citenamefont {{Frailis}}, \citenamefont {{Grazian}},
  \citenamefont {{Koppenh{\"o}fer}}, \citenamefont {{Mansutti}}, \citenamefont
  {{Melchior}}, \citenamefont {{Mignoli}}, \citenamefont {{Mohr}},
  \citenamefont {{Neissner}}, \citenamefont {{Noddle}}, \citenamefont
  {{Poncet}}, \citenamefont {{Scodeggio}}, \citenamefont {{Serrano}},
  \citenamefont {{Shane}}, \citenamefont {{Starck}}, \citenamefont {{Surace}},
  \citenamefont {{Taylor}}, \citenamefont {{Verdoes-Kleijn}}, \citenamefont
  {{Vuerli}}, \citenamefont {{Williams}}, \citenamefont {{Zacchei}},
  \citenamefont {{Altieri}}, \citenamefont {{Escudero Sanz}}, \citenamefont
  {{Kohley}}, \citenamefont {{Oosterbroek}}, \citenamefont {{Astier}},
  \citenamefont {{Bacon}}, \citenamefont {{Bardelli}}, \citenamefont {{Baugh}},
  \citenamefont {{Bellagamba}}, \citenamefont {{Benoist}}, \citenamefont
  {{Bianchi}}, \citenamefont {{Biviano}}, \citenamefont {{Branchini}},
  \citenamefont {{Carbone}}, \citenamefont {{Cardone}}, \citenamefont
  {{Clements}}, \citenamefont {{Colombi}}, \citenamefont {{Conselice}},
  \citenamefont {{Cresci}}, \citenamefont {{Deacon}}, \citenamefont {{Dunlop}},
  \citenamefont {{Fedeli}}, \citenamefont {{Fontanot}}, \citenamefont
  {{Franzetti}}, \citenamefont {{Giocoli}}, \citenamefont {{Garcia-Bellido}},
  \citenamefont {{Gow}}, \citenamefont {{Heavens}}, \citenamefont {{Hewett}},
  \citenamefont {{Heymans}}, \citenamefont {{Holland}}, \citenamefont
  {{Huang}}, \citenamefont {{Ilbert}}, \citenamefont {{Joachimi}},
  \citenamefont {{Jennins}}, \citenamefont {{Kerins}}, \citenamefont
  {{Kiessling}}, \citenamefont {{Kirk}}, \citenamefont {{Kotak}}, \citenamefont
  {{Krause}}, \citenamefont {{Lahav}}, \citenamefont {{van Leeuwen}},
  \citenamefont {{Lesgourgues}}, \citenamefont {{Lombardi}}, \citenamefont
  {{Magliocchetti}}, \citenamefont {{Maguire}}, \citenamefont {{Majerotto}},
  \citenamefont {{Maoli}}, \citenamefont {{Marulli}}, \citenamefont
  {{Maurogordato}}, \citenamefont {{McCracken}}, \citenamefont {{McLure}},
  \citenamefont {{Melchiorri}}, \citenamefont {{Merson}}, \citenamefont
  {{Moresco}}, \citenamefont {{Nonino}}, \citenamefont {{Norberg}},
  \citenamefont {{Peacock}}, \citenamefont {{Pello}}, \citenamefont {{Penny}},
  \citenamefont {{Pettorino}}, \citenamefont {{Di Porto}}, \citenamefont
  {{Pozzetti}}, \citenamefont {{Quercellini}}, \citenamefont {{Radovich}},
  \citenamefont {{Rassat}}, \citenamefont {{Roche}}, \citenamefont
  {{Ronayette}}, \citenamefont {{Rossetti}}, \citenamefont {{Sartoris}},
  \citenamefont {{Schneider}}, \citenamefont {{Semboloni}}, \citenamefont
  {{Serjeant}}, \citenamefont {{Simpson}}, \citenamefont {{Skordis}},
  \citenamefont {{Smadja}}, \citenamefont {{Smartt}}, \citenamefont {{Spano}},
  \citenamefont {{Spiro}}, \citenamefont {{Sullivan}}, \citenamefont
  {{Tilquin}}, \citenamefont {{Trotta}}, \citenamefont {{Verde}}, \citenamefont
  {{Wang}}, \citenamefont {{Williger}}, \citenamefont {{Zhao}}, \citenamefont
  {{Zoubian}},\ and\ \citenamefont {{Zucca}}}]{ref:Euclid}%
  \BibitemOpen
  \bibfield  {author} {\bibinfo {author} {\bibfnamefont {R.}~\bibnamefont
  {{Laureijs}}}, \bibinfo {author} {\bibfnamefont {J.}~\bibnamefont
  {{Amiaux}}}, \bibinfo {author} {\bibfnamefont {S.}~\bibnamefont {{Arduini}}},
  \bibinfo {author} {\bibfnamefont {J.~L.}\ \bibnamefont {{Augu{\`e}res}}},
  \bibinfo {author} {\bibfnamefont {J.}~\bibnamefont {{Brinchmann}}}, \bibinfo
  {author} {\bibfnamefont {R.}~\bibnamefont {{Cole}}}, \bibinfo {author}
  {\bibfnamefont {M.}~\bibnamefont {{Cropper}}}, \bibinfo {author}
  {\bibfnamefont {C.}~\bibnamefont {{Dabin}}}, \bibinfo {author} {\bibfnamefont
  {L.}~\bibnamefont {{Duvet}}}, \bibinfo {author} {\bibfnamefont
  {A.}~\bibnamefont {{Ealet}}}, \bibinfo {author} {\bibfnamefont
  {B.}~\bibnamefont {{Garilli}}}, \bibinfo {author} {\bibfnamefont
  {P.}~\bibnamefont {{Gondoin}}}, \bibinfo {author} {\bibfnamefont
  {L.}~\bibnamefont {{Guzzo}}}, \bibinfo {author} {\bibfnamefont
  {J.}~\bibnamefont {{Hoar}}}, \bibinfo {author} {\bibfnamefont
  {H.}~\bibnamefont {{Hoekstra}}}, \bibinfo {author} {\bibfnamefont
  {R.}~\bibnamefont {{Holmes}}}, \bibinfo {author} {\bibfnamefont
  {T.}~\bibnamefont {{Kitching}}}, \bibinfo {author} {\bibfnamefont
  {T.}~\bibnamefont {{Maciaszek}}}, \bibinfo {author} {\bibfnamefont
  {Y.}~\bibnamefont {{Mellier}}}, \bibinfo {author} {\bibfnamefont
  {F.}~\bibnamefont {{Pasian}}}, \bibinfo {author} {\bibfnamefont
  {W.}~\bibnamefont {{Percival}}}, \bibinfo {author} {\bibfnamefont
  {J.}~\bibnamefont {{Rhodes}}}, \bibinfo {author} {\bibfnamefont
  {G.}~\bibnamefont {{Saavedra Criado}}}, \bibinfo {author} {\bibfnamefont
  {M.}~\bibnamefont {{Sauvage}}}, \bibinfo {author} {\bibfnamefont
  {R.}~\bibnamefont {{Scaramella}}}, \bibinfo {author} {\bibfnamefont
  {L.}~\bibnamefont {{Valenziano}}}, \bibinfo {author} {\bibfnamefont
  {S.}~\bibnamefont {{Warren}}}, \bibinfo {author} {\bibfnamefont
  {R.}~\bibnamefont {{Bender}}}, \bibinfo {author} {\bibfnamefont
  {F.}~\bibnamefont {{Castander}}}, \bibinfo {author} {\bibfnamefont
  {A.}~\bibnamefont {{Cimatti}}}, \bibinfo {author} {\bibfnamefont
  {O.}~\bibnamefont {{Le F{\`e}vre}}}, \bibinfo {author} {\bibfnamefont
  {H.}~\bibnamefont {{Kurki-Suonio}}}, \bibinfo {author} {\bibfnamefont
  {M.}~\bibnamefont {{Levi}}}, \bibinfo {author} {\bibfnamefont
  {P.}~\bibnamefont {{Lilje}}}, \bibinfo {author} {\bibfnamefont
  {G.}~\bibnamefont {{Meylan}}}, \bibinfo {author} {\bibfnamefont
  {R.}~\bibnamefont {{Nichol}}}, \bibinfo {author} {\bibfnamefont
  {K.}~\bibnamefont {{Pedersen}}}, \bibinfo {author} {\bibfnamefont
  {V.}~\bibnamefont {{Popa}}}, \bibinfo {author} {\bibfnamefont
  {R.}~\bibnamefont {{Rebolo Lopez}}}, \bibinfo {author} {\bibfnamefont
  {H.~W.}\ \bibnamefont {{Rix}}}, \bibinfo {author} {\bibfnamefont
  {H.}~\bibnamefont {{Rottgering}}}, \bibinfo {author} {\bibfnamefont
  {W.}~\bibnamefont {{Zeilinger}}}, \bibinfo {author} {\bibfnamefont
  {F.}~\bibnamefont {{Grupp}}}, \bibinfo {author} {\bibfnamefont
  {P.}~\bibnamefont {{Hudelot}}}, \bibinfo {author} {\bibfnamefont
  {R.}~\bibnamefont {{Massey}}}, \bibinfo {author} {\bibfnamefont
  {M.}~\bibnamefont {{Meneghetti}}}, \bibinfo {author} {\bibfnamefont
  {L.}~\bibnamefont {{Miller}}}, \bibinfo {author} {\bibfnamefont
  {S.}~\bibnamefont {{Paltani}}}, \bibinfo {author} {\bibfnamefont
  {S.}~\bibnamefont {{Paulin-Henriksson}}}, \bibinfo {author} {\bibfnamefont
  {S.}~\bibnamefont {{Pires}}}, \bibinfo {author} {\bibfnamefont
  {C.}~\bibnamefont {{Saxton}}}, \bibinfo {author} {\bibfnamefont
  {T.}~\bibnamefont {{Schrabback}}}, \bibinfo {author} {\bibfnamefont
  {G.}~\bibnamefont {{Seidel}}}, \bibinfo {author} {\bibfnamefont
  {J.}~\bibnamefont {{Walsh}}}, \bibinfo {author} {\bibfnamefont
  {N.}~\bibnamefont {{Aghanim}}}, \bibinfo {author} {\bibfnamefont
  {L.}~\bibnamefont {{Amendola}}}, \bibinfo {author} {\bibfnamefont
  {J.}~\bibnamefont {{Bartlett}}}, \bibinfo {author} {\bibfnamefont
  {C.}~\bibnamefont {{Baccigalupi}}}, \bibinfo {author} {\bibfnamefont {J.~P.}\
  \bibnamefont {{Beaulieu}}}, \bibinfo {author} {\bibfnamefont
  {K.}~\bibnamefont {{Benabed}}}, \bibinfo {author} {\bibfnamefont {J.~G.}\
  \bibnamefont {{Cuby}}}, \bibinfo {author} {\bibfnamefont {D.}~\bibnamefont
  {{Elbaz}}}, \bibinfo {author} {\bibfnamefont {P.}~\bibnamefont {{Fosalba}}},
  \bibinfo {author} {\bibfnamefont {G.}~\bibnamefont {{Gavazzi}}}, \bibinfo
  {author} {\bibfnamefont {A.}~\bibnamefont {{Helmi}}}, \bibinfo {author}
  {\bibfnamefont {I.}~\bibnamefont {{Hook}}}, \bibinfo {author} {\bibfnamefont
  {M.}~\bibnamefont {{Irwin}}}, \bibinfo {author} {\bibfnamefont {J.~P.}\
  \bibnamefont {{Kneib}}}, \bibinfo {author} {\bibfnamefont {M.}~\bibnamefont
  {{Kunz}}}, \bibinfo {author} {\bibfnamefont {F.}~\bibnamefont {{Mannucci}}},
  \bibinfo {author} {\bibfnamefont {L.}~\bibnamefont {{Moscardini}}}, \bibinfo
  {author} {\bibfnamefont {C.}~\bibnamefont {{Tao}}}, \bibinfo {author}
  {\bibfnamefont {R.}~\bibnamefont {{Teyssier}}}, \bibinfo {author}
  {\bibfnamefont {J.}~\bibnamefont {{Weller}}}, \bibinfo {author}
  {\bibfnamefont {G.}~\bibnamefont {{Zamorani}}}, \bibinfo {author}
  {\bibfnamefont {M.~R.}\ \bibnamefont {{Zapatero Osorio}}}, \bibinfo {author}
  {\bibfnamefont {O.}~\bibnamefont {{Boulade}}}, \bibinfo {author}
  {\bibfnamefont {J.~J.}\ \bibnamefont {{Foumond}}}, \bibinfo {author}
  {\bibfnamefont {A.}~\bibnamefont {{Di Giorgio}}}, \bibinfo {author}
  {\bibfnamefont {P.}~\bibnamefont {{Guttridge}}}, \bibinfo {author}
  {\bibfnamefont {A.}~\bibnamefont {{James}}}, \bibinfo {author} {\bibfnamefont
  {M.}~\bibnamefont {{Kemp}}}, \bibinfo {author} {\bibfnamefont
  {J.}~\bibnamefont {{Martignac}}}, \bibinfo {author} {\bibfnamefont
  {A.}~\bibnamefont {{Spencer}}}, \bibinfo {author} {\bibfnamefont
  {D.}~\bibnamefont {{Walton}}}, \bibinfo {author} {\bibfnamefont
  {T.}~\bibnamefont {{Bl{\"u}mchen}}}, \bibinfo {author} {\bibfnamefont
  {C.}~\bibnamefont {{Bonoli}}}, \bibinfo {author} {\bibfnamefont
  {F.}~\bibnamefont {{Bortoletto}}}, \bibinfo {author} {\bibfnamefont
  {C.}~\bibnamefont {{Cerna}}}, \bibinfo {author} {\bibfnamefont
  {L.}~\bibnamefont {{Corcione}}}, \bibinfo {author} {\bibfnamefont
  {C.}~\bibnamefont {{Fabron}}}, \bibinfo {author} {\bibfnamefont
  {K.}~\bibnamefont {{Jahnke}}}, \bibinfo {author} {\bibfnamefont
  {S.}~\bibnamefont {{Ligori}}}, \bibinfo {author} {\bibfnamefont
  {F.}~\bibnamefont {{Madrid}}}, \bibinfo {author} {\bibfnamefont
  {L.}~\bibnamefont {{Martin}}}, \bibinfo {author} {\bibfnamefont
  {G.}~\bibnamefont {{Morgante}}}, \bibinfo {author} {\bibfnamefont
  {T.}~\bibnamefont {{Pamplona}}}, \bibinfo {author} {\bibfnamefont
  {E.}~\bibnamefont {{Prieto}}}, \bibinfo {author} {\bibfnamefont
  {M.}~\bibnamefont {{Riva}}}, \bibinfo {author} {\bibfnamefont
  {R.}~\bibnamefont {{Toledo}}}, \bibinfo {author} {\bibfnamefont
  {M.}~\bibnamefont {{Trifoglio}}}, \bibinfo {author} {\bibfnamefont
  {F.}~\bibnamefont {{Zerbi}}}, \bibinfo {author} {\bibfnamefont
  {F.}~\bibnamefont {{Abdalla}}}, \bibinfo {author} {\bibfnamefont
  {M.}~\bibnamefont {{Douspis}}}, \bibinfo {author} {\bibfnamefont
  {C.}~\bibnamefont {{Grenet}}}, \bibinfo {author} {\bibfnamefont
  {S.}~\bibnamefont {{Borgani}}}, \bibinfo {author} {\bibfnamefont
  {R.}~\bibnamefont {{Bouwens}}}, \bibinfo {author} {\bibfnamefont
  {F.}~\bibnamefont {{Courbin}}}, \bibinfo {author} {\bibfnamefont {J.~M.}\
  \bibnamefont {{Delouis}}}, \bibinfo {author} {\bibfnamefont {P.}~\bibnamefont
  {{Dubath}}}, \bibinfo {author} {\bibfnamefont {A.}~\bibnamefont {{Fontana}}},
  \bibinfo {author} {\bibfnamefont {M.}~\bibnamefont {{Frailis}}}, \bibinfo
  {author} {\bibfnamefont {A.}~\bibnamefont {{Grazian}}}, \bibinfo {author}
  {\bibfnamefont {J.}~\bibnamefont {{Koppenh{\"o}fer}}}, \bibinfo {author}
  {\bibfnamefont {O.}~\bibnamefont {{Mansutti}}}, \bibinfo {author}
  {\bibfnamefont {M.}~\bibnamefont {{Melchior}}}, \bibinfo {author}
  {\bibfnamefont {M.}~\bibnamefont {{Mignoli}}}, \bibinfo {author}
  {\bibfnamefont {J.}~\bibnamefont {{Mohr}}}, \bibinfo {author} {\bibfnamefont
  {C.}~\bibnamefont {{Neissner}}}, \bibinfo {author} {\bibfnamefont
  {K.}~\bibnamefont {{Noddle}}}, \bibinfo {author} {\bibfnamefont
  {M.}~\bibnamefont {{Poncet}}}, \bibinfo {author} {\bibfnamefont
  {M.}~\bibnamefont {{Scodeggio}}}, \bibinfo {author} {\bibfnamefont
  {S.}~\bibnamefont {{Serrano}}}, \bibinfo {author} {\bibfnamefont
  {N.}~\bibnamefont {{Shane}}}, \bibinfo {author} {\bibfnamefont {J.~L.}\
  \bibnamefont {{Starck}}}, \bibinfo {author} {\bibfnamefont {C.}~\bibnamefont
  {{Surace}}}, \bibinfo {author} {\bibfnamefont {A.}~\bibnamefont {{Taylor}}},
  \bibinfo {author} {\bibfnamefont {G.}~\bibnamefont {{Verdoes-Kleijn}}},
  \bibinfo {author} {\bibfnamefont {C.}~\bibnamefont {{Vuerli}}}, \bibinfo
  {author} {\bibfnamefont {O.~R.}\ \bibnamefont {{Williams}}}, \bibinfo
  {author} {\bibfnamefont {A.}~\bibnamefont {{Zacchei}}}, \bibinfo {author}
  {\bibfnamefont {B.}~\bibnamefont {{Altieri}}}, \bibinfo {author}
  {\bibfnamefont {I.}~\bibnamefont {{Escudero Sanz}}}, \bibinfo {author}
  {\bibfnamefont {R.}~\bibnamefont {{Kohley}}}, \bibinfo {author}
  {\bibfnamefont {T.}~\bibnamefont {{Oosterbroek}}}, \bibinfo {author}
  {\bibfnamefont {P.}~\bibnamefont {{Astier}}}, \bibinfo {author}
  {\bibfnamefont {D.}~\bibnamefont {{Bacon}}}, \bibinfo {author} {\bibfnamefont
  {S.}~\bibnamefont {{Bardelli}}}, \bibinfo {author} {\bibfnamefont
  {C.}~\bibnamefont {{Baugh}}}, \bibinfo {author} {\bibfnamefont
  {F.}~\bibnamefont {{Bellagamba}}}, \bibinfo {author} {\bibfnamefont
  {C.}~\bibnamefont {{Benoist}}}, \bibinfo {author} {\bibfnamefont
  {D.}~\bibnamefont {{Bianchi}}}, \bibinfo {author} {\bibfnamefont
  {A.}~\bibnamefont {{Biviano}}}, \bibinfo {author} {\bibfnamefont
  {E.}~\bibnamefont {{Branchini}}}, \bibinfo {author} {\bibfnamefont
  {C.}~\bibnamefont {{Carbone}}}, \bibinfo {author} {\bibfnamefont
  {V.}~\bibnamefont {{Cardone}}}, \bibinfo {author} {\bibfnamefont
  {D.}~\bibnamefont {{Clements}}}, \bibinfo {author} {\bibfnamefont
  {S.}~\bibnamefont {{Colombi}}}, \bibinfo {author} {\bibfnamefont
  {C.}~\bibnamefont {{Conselice}}}, \bibinfo {author} {\bibfnamefont
  {G.}~\bibnamefont {{Cresci}}}, \bibinfo {author} {\bibfnamefont
  {N.}~\bibnamefont {{Deacon}}}, \bibinfo {author} {\bibfnamefont
  {J.}~\bibnamefont {{Dunlop}}}, \bibinfo {author} {\bibfnamefont
  {C.}~\bibnamefont {{Fedeli}}}, \bibinfo {author} {\bibfnamefont
  {F.}~\bibnamefont {{Fontanot}}}, \bibinfo {author} {\bibfnamefont
  {P.}~\bibnamefont {{Franzetti}}}, \bibinfo {author} {\bibfnamefont
  {C.}~\bibnamefont {{Giocoli}}}, \bibinfo {author} {\bibfnamefont
  {J.}~\bibnamefont {{Garcia-Bellido}}}, \bibinfo {author} {\bibfnamefont
  {J.}~\bibnamefont {{Gow}}}, \bibinfo {author} {\bibfnamefont
  {A.}~\bibnamefont {{Heavens}}}, \bibinfo {author} {\bibfnamefont
  {P.}~\bibnamefont {{Hewett}}}, \bibinfo {author} {\bibfnamefont
  {C.}~\bibnamefont {{Heymans}}}, \bibinfo {author} {\bibfnamefont
  {A.}~\bibnamefont {{Holland}}}, \bibinfo {author} {\bibfnamefont
  {Z.}~\bibnamefont {{Huang}}}, \bibinfo {author} {\bibfnamefont
  {O.}~\bibnamefont {{Ilbert}}}, \bibinfo {author} {\bibfnamefont
  {B.}~\bibnamefont {{Joachimi}}}, \bibinfo {author} {\bibfnamefont
  {E.}~\bibnamefont {{Jennins}}}, \bibinfo {author} {\bibfnamefont
  {E.}~\bibnamefont {{Kerins}}}, \bibinfo {author} {\bibfnamefont
  {A.}~\bibnamefont {{Kiessling}}}, \bibinfo {author} {\bibfnamefont
  {D.}~\bibnamefont {{Kirk}}}, \bibinfo {author} {\bibfnamefont
  {R.}~\bibnamefont {{Kotak}}}, \bibinfo {author} {\bibfnamefont
  {O.}~\bibnamefont {{Krause}}}, \bibinfo {author} {\bibfnamefont
  {O.}~\bibnamefont {{Lahav}}}, \bibinfo {author} {\bibfnamefont
  {F.}~\bibnamefont {{van Leeuwen}}}, \bibinfo {author} {\bibfnamefont
  {J.}~\bibnamefont {{Lesgourgues}}}, \bibinfo {author} {\bibfnamefont
  {M.}~\bibnamefont {{Lombardi}}}, \bibinfo {author} {\bibfnamefont
  {M.}~\bibnamefont {{Magliocchetti}}}, \bibinfo {author} {\bibfnamefont
  {K.}~\bibnamefont {{Maguire}}}, \bibinfo {author} {\bibfnamefont
  {E.}~\bibnamefont {{Majerotto}}}, \bibinfo {author} {\bibfnamefont
  {R.}~\bibnamefont {{Maoli}}}, \bibinfo {author} {\bibfnamefont
  {F.}~\bibnamefont {{Marulli}}}, \bibinfo {author} {\bibfnamefont
  {S.}~\bibnamefont {{Maurogordato}}}, \bibinfo {author} {\bibfnamefont
  {H.}~\bibnamefont {{McCracken}}}, \bibinfo {author} {\bibfnamefont
  {R.}~\bibnamefont {{McLure}}}, \bibinfo {author} {\bibfnamefont
  {A.}~\bibnamefont {{Melchiorri}}}, \bibinfo {author} {\bibfnamefont
  {A.}~\bibnamefont {{Merson}}}, \bibinfo {author} {\bibfnamefont
  {M.}~\bibnamefont {{Moresco}}}, \bibinfo {author} {\bibfnamefont
  {M.}~\bibnamefont {{Nonino}}}, \bibinfo {author} {\bibfnamefont
  {P.}~\bibnamefont {{Norberg}}}, \bibinfo {author} {\bibfnamefont
  {J.}~\bibnamefont {{Peacock}}}, \bibinfo {author} {\bibfnamefont
  {R.}~\bibnamefont {{Pello}}}, \bibinfo {author} {\bibfnamefont
  {M.}~\bibnamefont {{Penny}}}, \bibinfo {author} {\bibfnamefont
  {V.}~\bibnamefont {{Pettorino}}}, \bibinfo {author} {\bibfnamefont
  {C.}~\bibnamefont {{Di Porto}}}, \bibinfo {author} {\bibfnamefont
  {L.}~\bibnamefont {{Pozzetti}}}, \bibinfo {author} {\bibfnamefont
  {C.}~\bibnamefont {{Quercellini}}}, \bibinfo {author} {\bibfnamefont
  {M.}~\bibnamefont {{Radovich}}}, \bibinfo {author} {\bibfnamefont
  {A.}~\bibnamefont {{Rassat}}}, \bibinfo {author} {\bibfnamefont
  {N.}~\bibnamefont {{Roche}}}, \bibinfo {author} {\bibfnamefont
  {S.}~\bibnamefont {{Ronayette}}}, \bibinfo {author} {\bibfnamefont
  {E.}~\bibnamefont {{Rossetti}}}, \bibinfo {author} {\bibfnamefont
  {B.}~\bibnamefont {{Sartoris}}}, \bibinfo {author} {\bibfnamefont
  {P.}~\bibnamefont {{Schneider}}}, \bibinfo {author} {\bibfnamefont
  {E.}~\bibnamefont {{Semboloni}}}, \bibinfo {author} {\bibfnamefont
  {S.}~\bibnamefont {{Serjeant}}}, \bibinfo {author} {\bibfnamefont
  {F.}~\bibnamefont {{Simpson}}}, \bibinfo {author} {\bibfnamefont
  {C.}~\bibnamefont {{Skordis}}}, \bibinfo {author} {\bibfnamefont
  {G.}~\bibnamefont {{Smadja}}}, \bibinfo {author} {\bibfnamefont
  {S.}~\bibnamefont {{Smartt}}}, \bibinfo {author} {\bibfnamefont
  {P.}~\bibnamefont {{Spano}}}, \bibinfo {author} {\bibfnamefont
  {S.}~\bibnamefont {{Spiro}}}, \bibinfo {author} {\bibfnamefont
  {M.}~\bibnamefont {{Sullivan}}}, \bibinfo {author} {\bibfnamefont
  {A.}~\bibnamefont {{Tilquin}}}, \bibinfo {author} {\bibfnamefont
  {R.}~\bibnamefont {{Trotta}}}, \bibinfo {author} {\bibfnamefont
  {L.}~\bibnamefont {{Verde}}}, \bibinfo {author} {\bibfnamefont
  {Y.}~\bibnamefont {{Wang}}}, \bibinfo {author} {\bibfnamefont
  {G.}~\bibnamefont {{Williger}}}, \bibinfo {author} {\bibfnamefont
  {G.}~\bibnamefont {{Zhao}}}, \bibinfo {author} {\bibfnamefont
  {J.}~\bibnamefont {{Zoubian}}}, \ and\ \bibinfo {author} {\bibfnamefont
  {E.}~\bibnamefont {{Zucca}}},\ }\href@noop {} {\bibfield  {journal} {\bibinfo
   {journal} {arXiv e-prints}\ ,\ \bibinfo {eid} {arXiv:1110.3193}} (\bibinfo
  {year} {2011})},\ \Eprint {http://arxiv.org/abs/1110.3193} {arXiv:1110.3193
  [astro-ph.CO]} \BibitemShut {NoStop}%
\bibitem [{\citenamefont {{Eifler}}\ \emph {et~al.}(2021)\citenamefont
  {{Eifler}}, \citenamefont {{Miyatake}}, \citenamefont {{Krause}},
  \citenamefont {{Heinrich}}, \citenamefont {{Miranda}}, \citenamefont
  {{Hirata}}, \citenamefont {{Xu}}, \citenamefont {{Hemmati}}, \citenamefont
  {{Simet}}, \citenamefont {{Capak}}, \citenamefont {{Choi}}, \citenamefont
  {{Dor{\'e}}}, \citenamefont {{Doux}}, \citenamefont {{Fang}}, \citenamefont
  {{Hounsell}}, \citenamefont {{Huff}}, \citenamefont {{Huang}}, \citenamefont
  {{Jarvis}}, \citenamefont {{Kruk}}, \citenamefont {{Masters}}, \citenamefont
  {{Rozo}}, \citenamefont {{Scolnic}}, \citenamefont {{Spergel}}, \citenamefont
  {{Troxel}}, \citenamefont {{von der Linden}}, \citenamefont {{Wang}},
  \citenamefont {{Weinberg}}, \citenamefont {{Wenzl}},\ and\ \citenamefont
  {{Wu}}}]{NancyRoman}%
  \BibitemOpen
  \bibfield  {author} {\bibinfo {author} {\bibfnamefont {T.}~\bibnamefont
  {{Eifler}}}, \bibinfo {author} {\bibfnamefont {H.}~\bibnamefont
  {{Miyatake}}}, \bibinfo {author} {\bibfnamefont {E.}~\bibnamefont
  {{Krause}}}, \bibinfo {author} {\bibfnamefont {C.}~\bibnamefont
  {{Heinrich}}}, \bibinfo {author} {\bibfnamefont {V.}~\bibnamefont
  {{Miranda}}}, \bibinfo {author} {\bibfnamefont {C.}~\bibnamefont {{Hirata}}},
  \bibinfo {author} {\bibfnamefont {J.}~\bibnamefont {{Xu}}}, \bibinfo {author}
  {\bibfnamefont {S.}~\bibnamefont {{Hemmati}}}, \bibinfo {author}
  {\bibfnamefont {M.}~\bibnamefont {{Simet}}}, \bibinfo {author} {\bibfnamefont
  {P.}~\bibnamefont {{Capak}}}, \bibinfo {author} {\bibfnamefont
  {A.}~\bibnamefont {{Choi}}}, \bibinfo {author} {\bibfnamefont
  {O.}~\bibnamefont {{Dor{\'e}}}}, \bibinfo {author} {\bibfnamefont
  {C.}~\bibnamefont {{Doux}}}, \bibinfo {author} {\bibfnamefont
  {X.}~\bibnamefont {{Fang}}}, \bibinfo {author} {\bibfnamefont
  {R.}~\bibnamefont {{Hounsell}}}, \bibinfo {author} {\bibfnamefont
  {E.}~\bibnamefont {{Huff}}}, \bibinfo {author} {\bibfnamefont {H.-J.}\
  \bibnamefont {{Huang}}}, \bibinfo {author} {\bibfnamefont {M.}~\bibnamefont
  {{Jarvis}}}, \bibinfo {author} {\bibfnamefont {J.}~\bibnamefont {{Kruk}}},
  \bibinfo {author} {\bibfnamefont {D.}~\bibnamefont {{Masters}}}, \bibinfo
  {author} {\bibfnamefont {E.}~\bibnamefont {{Rozo}}}, \bibinfo {author}
  {\bibfnamefont {D.}~\bibnamefont {{Scolnic}}}, \bibinfo {author}
  {\bibfnamefont {D.~N.}\ \bibnamefont {{Spergel}}}, \bibinfo {author}
  {\bibfnamefont {M.}~\bibnamefont {{Troxel}}}, \bibinfo {author}
  {\bibfnamefont {A.}~\bibnamefont {{von der Linden}}}, \bibinfo {author}
  {\bibfnamefont {Y.}~\bibnamefont {{Wang}}}, \bibinfo {author} {\bibfnamefont
  {D.~H.}\ \bibnamefont {{Weinberg}}}, \bibinfo {author} {\bibfnamefont
  {L.}~\bibnamefont {{Wenzl}}}, \ and\ \bibinfo {author} {\bibfnamefont
  {H.-Y.}\ \bibnamefont {{Wu}}},\ }\href {\doibase 10.1093/mnras/stab1762}
  {\bibfield  {journal} {\bibinfo  {journal} {\mnras}\ }\textbf {\bibinfo
  {volume} {507}},\ \bibinfo {pages} {1746} (\bibinfo {year} {2021})},\ \Eprint
  {http://arxiv.org/abs/2004.05271} {arXiv:2004.05271 [astro-ph.CO]}
  \BibitemShut {NoStop}%
\bibitem [{\citenamefont {{LSST Dark Energy Science
  Collaboration}}(2012)}]{LSST}%
  \BibitemOpen
  \bibfield  {author} {\bibinfo {author} {\bibnamefont {{LSST Dark Energy
  Science Collaboration}}},\ }\href {\doibase 10.48550/arXiv.1211.0310}
  {\bibfield  {journal} {\bibinfo  {journal} {arXiv e-prints}\ ,\ \bibinfo
  {eid} {arXiv:1211.0310}} (\bibinfo {year} {2012})},\ \Eprint
  {http://arxiv.org/abs/1211.0310} {arXiv:1211.0310 [astro-ph.CO]} \BibitemShut
  {NoStop}%
\bibitem [{\citenamefont {{Square Kilometre Array Cosmology Science Working
  Group}}\ \emph {et~al.}(2018)\citenamefont {{Square Kilometre Array Cosmology
  Science Working Group}}, \citenamefont {{Bacon}}, \citenamefont {{Battye}},
  \citenamefont {{Bull}}, \citenamefont {{Camera}}, \citenamefont {{Ferreira}},
  \citenamefont {{Harrison}}, \citenamefont {{Parkinson}}, \citenamefont
  {{Pourtsidou}}, \citenamefont {{Santos}}, \citenamefont {{Wolz}},
  \citenamefont {{Abdalla}}, \citenamefont {{Akrami}}, \citenamefont
  {{Alonso}}, \citenamefont {{Andrianomena}}, \citenamefont {{Ballardini}},
  \citenamefont {{Bernal}}, \citenamefont {{Bertacca}}, \citenamefont
  {{Bengaly}}, \citenamefont {{Bonaldi}}, \citenamefont {{Bonvin}},
  \citenamefont {{Brown}}, \citenamefont {{Chapman}}, \citenamefont {{Chen}},
  \citenamefont {{Chen}}, \citenamefont {{Cunnington}}, \citenamefont
  {{Davis}}, \citenamefont {{Dickinson}}, \citenamefont {{Fonseca}},
  \citenamefont {{Grainge}}, \citenamefont {{Harper}}, \citenamefont
  {{Jarvis}}, \citenamefont {{Maartens}}, \citenamefont {{Maddox}},
  \citenamefont {{Padmanabhan}}, \citenamefont {{Pritchard}}, \citenamefont
  {{Raccanelli}}, \citenamefont {{Rivi}}, \citenamefont {{Roychowdhury}},
  \citenamefont {{Sahlen}}, \citenamefont {{Schwarz}}, \citenamefont
  {{Siewert}}, \citenamefont {{Viel}}, \citenamefont {{Villaescusa-Navarro}},
  \citenamefont {{Xu}}, \citenamefont {{Yamauchi}},\ and\ \citenamefont
  {{Zuntz}}}]{ref:SKACosmo}%
  \BibitemOpen
  \bibfield  {author} {\bibinfo {author} {\bibnamefont {{Square Kilometre Array
  Cosmology Science Working Group}}}, \bibinfo {author} {\bibfnamefont {D.~J.}\
  \bibnamefont {{Bacon}}}, \bibinfo {author} {\bibfnamefont {R.~A.}\
  \bibnamefont {{Battye}}}, \bibinfo {author} {\bibfnamefont {P.}~\bibnamefont
  {{Bull}}}, \bibinfo {author} {\bibfnamefont {S.}~\bibnamefont {{Camera}}},
  \bibinfo {author} {\bibfnamefont {P.~G.}\ \bibnamefont {{Ferreira}}},
  \bibinfo {author} {\bibfnamefont {I.}~\bibnamefont {{Harrison}}}, \bibinfo
  {author} {\bibfnamefont {D.}~\bibnamefont {{Parkinson}}}, \bibinfo {author}
  {\bibfnamefont {A.}~\bibnamefont {{Pourtsidou}}}, \bibinfo {author}
  {\bibfnamefont {M.~G.}\ \bibnamefont {{Santos}}}, \bibinfo {author}
  {\bibfnamefont {L.}~\bibnamefont {{Wolz}}}, \bibinfo {author} {\bibfnamefont
  {F.}~\bibnamefont {{Abdalla}}}, \bibinfo {author} {\bibfnamefont
  {Y.}~\bibnamefont {{Akrami}}}, \bibinfo {author} {\bibfnamefont
  {D.}~\bibnamefont {{Alonso}}}, \bibinfo {author} {\bibfnamefont
  {S.}~\bibnamefont {{Andrianomena}}}, \bibinfo {author} {\bibfnamefont
  {M.}~\bibnamefont {{Ballardini}}}, \bibinfo {author} {\bibfnamefont {J.~L.}\
  \bibnamefont {{Bernal}}}, \bibinfo {author} {\bibfnamefont {D.}~\bibnamefont
  {{Bertacca}}}, \bibinfo {author} {\bibfnamefont {C.~A.~P.}\ \bibnamefont
  {{Bengaly}}}, \bibinfo {author} {\bibfnamefont {A.}~\bibnamefont
  {{Bonaldi}}}, \bibinfo {author} {\bibfnamefont {C.}~\bibnamefont {{Bonvin}}},
  \bibinfo {author} {\bibfnamefont {M.~L.}\ \bibnamefont {{Brown}}}, \bibinfo
  {author} {\bibfnamefont {E.}~\bibnamefont {{Chapman}}}, \bibinfo {author}
  {\bibfnamefont {S.}~\bibnamefont {{Chen}}}, \bibinfo {author} {\bibfnamefont
  {X.}~\bibnamefont {{Chen}}}, \bibinfo {author} {\bibfnamefont
  {S.}~\bibnamefont {{Cunnington}}}, \bibinfo {author} {\bibfnamefont {T.~M.}\
  \bibnamefont {{Davis}}}, \bibinfo {author} {\bibfnamefont {C.}~\bibnamefont
  {{Dickinson}}}, \bibinfo {author} {\bibfnamefont {J.}~\bibnamefont
  {{Fonseca}}}, \bibinfo {author} {\bibfnamefont {K.}~\bibnamefont
  {{Grainge}}}, \bibinfo {author} {\bibfnamefont {S.}~\bibnamefont {{Harper}}},
  \bibinfo {author} {\bibfnamefont {M.~J.}\ \bibnamefont {{Jarvis}}}, \bibinfo
  {author} {\bibfnamefont {R.}~\bibnamefont {{Maartens}}}, \bibinfo {author}
  {\bibfnamefont {N.}~\bibnamefont {{Maddox}}}, \bibinfo {author}
  {\bibfnamefont {H.}~\bibnamefont {{Padmanabhan}}}, \bibinfo {author}
  {\bibfnamefont {J.~R.}\ \bibnamefont {{Pritchard}}}, \bibinfo {author}
  {\bibfnamefont {A.}~\bibnamefont {{Raccanelli}}}, \bibinfo {author}
  {\bibfnamefont {M.}~\bibnamefont {{Rivi}}}, \bibinfo {author} {\bibfnamefont
  {S.}~\bibnamefont {{Roychowdhury}}}, \bibinfo {author} {\bibfnamefont
  {M.}~\bibnamefont {{Sahlen}}}, \bibinfo {author} {\bibfnamefont {D.~J.}\
  \bibnamefont {{Schwarz}}}, \bibinfo {author} {\bibfnamefont {T.~M.}\
  \bibnamefont {{Siewert}}}, \bibinfo {author} {\bibfnamefont {M.}~\bibnamefont
  {{Viel}}}, \bibinfo {author} {\bibfnamefont {F.}~\bibnamefont
  {{Villaescusa-Navarro}}}, \bibinfo {author} {\bibfnamefont {Y.}~\bibnamefont
  {{Xu}}}, \bibinfo {author} {\bibfnamefont {D.}~\bibnamefont {{Yamauchi}}}, \
  and\ \bibinfo {author} {\bibfnamefont {J.}~\bibnamefont {{Zuntz}}},\
  }\href@noop {} {\bibfield  {journal} {\bibinfo  {journal} {arXiv e-prints}\ }
  (\bibinfo {year} {2018})},\ \Eprint {http://arxiv.org/abs/1811.02743}
  {arXiv:1811.02743} \BibitemShut {NoStop}%
\bibitem [{\citenamefont {{Battye}}\ and\ \citenamefont
  {{Pearson}}(2012)}]{ref:BattyePearson}%
  \BibitemOpen
  \bibfield  {author} {\bibinfo {author} {\bibfnamefont {R.~A.}\ \bibnamefont
  {{Battye}}}\ and\ \bibinfo {author} {\bibfnamefont {J.~A.}\ \bibnamefont
  {{Pearson}}},\ }\href {\doibase 10.1088/1475-7516/2012/07/019} {\bibfield
  {journal} {\bibinfo  {journal} {Journal of Cosmology and Astro-Particle
  Physics}\ }\textbf {\bibinfo {volume} {2012}},\ \bibinfo {eid} {019}
  (\bibinfo {year} {2012})},\ \Eprint {http://arxiv.org/abs/1203.0398}
  {arXiv:1203.0398 [hep-th]} \BibitemShut {NoStop}%
\bibitem [{\citenamefont {{Gleyzes}}\ \emph {et~al.}(2014)\citenamefont
  {{Gleyzes}}, \citenamefont {{Langlois}},\ and\ \citenamefont
  {{Vernizzi}}}]{ref:Gleyzes}%
  \BibitemOpen
  \bibfield  {author} {\bibinfo {author} {\bibfnamefont {J.}~\bibnamefont
  {{Gleyzes}}}, \bibinfo {author} {\bibfnamefont {D.}~\bibnamefont
  {{Langlois}}}, \ and\ \bibinfo {author} {\bibfnamefont {F.}~\bibnamefont
  {{Vernizzi}}},\ }\href {\doibase 10.1142/S021827181443010X} {\bibfield
  {journal} {\bibinfo  {journal} {International Journal of Modern Physics D}\
  }\textbf {\bibinfo {volume} {23}},\ \bibinfo {eid} {1443010} (\bibinfo {year}
  {2014})},\ \Eprint {http://arxiv.org/abs/1411.3712} {arXiv:1411.3712
  [hep-th]} \BibitemShut {NoStop}%
\bibitem [{\citenamefont {{DES Collaboration}}(2018)}]{DES_Y1_2018}%
  \BibitemOpen
  \bibfield  {author} {\bibinfo {author} {\bibnamefont {{DES Collaboration}}},\
  }\href {\doibase 10.1103/PhysRevD.98.043526} {\bibfield  {journal} {\bibinfo
  {journal} {\prd}\ }\textbf {\bibinfo {volume} {98}},\ \bibinfo {eid} {043526}
  (\bibinfo {year} {2018})},\ \Eprint {http://arxiv.org/abs/1708.01530}
  {arXiv:1708.01530} \BibitemShut {NoStop}%
\bibitem [{\citenamefont {{Arnold}}\ \emph {et~al.}(2019)\citenamefont
  {{Arnold}}, \citenamefont {{Leo}},\ and\ \citenamefont {{Li}}}]{Arnold2019}%
  \BibitemOpen
  \bibfield  {author} {\bibinfo {author} {\bibfnamefont {C.}~\bibnamefont
  {{Arnold}}}, \bibinfo {author} {\bibfnamefont {M.}~\bibnamefont {{Leo}}}, \
  and\ \bibinfo {author} {\bibfnamefont {B.}~\bibnamefont {{Li}}},\ }\href
  {\doibase 10.1038/s41550-019-0823-y} {\bibfield  {journal} {\bibinfo
  {journal} {Nature Astronomy}\ }\textbf {\bibinfo {volume} {3}},\ \bibinfo
  {pages} {945} (\bibinfo {year} {2019})},\ \Eprint
  {http://arxiv.org/abs/1907.02977} {arXiv:1907.02977 [astro-ph.CO]}
  \BibitemShut {NoStop}%
\bibitem [{\citenamefont {{Arnold}}\ \emph {et~al.}(2022)\citenamefont
  {{Arnold}}, \citenamefont {{Li}}, \citenamefont {{Giblin}}, \citenamefont
  {{Harnois-D{\'e}raps}},\ and\ \citenamefont {{Cai}}}]{Arnold2022}%
  \BibitemOpen
  \bibfield  {author} {\bibinfo {author} {\bibfnamefont {C.}~\bibnamefont
  {{Arnold}}}, \bibinfo {author} {\bibfnamefont {B.}~\bibnamefont {{Li}}},
  \bibinfo {author} {\bibfnamefont {B.}~\bibnamefont {{Giblin}}}, \bibinfo
  {author} {\bibfnamefont {J.}~\bibnamefont {{Harnois-D{\'e}raps}}}, \ and\
  \bibinfo {author} {\bibfnamefont {Y.-C.}\ \bibnamefont {{Cai}}},\ }\href
  {\doibase 10.1093/mnras/stac1091} {\bibfield  {journal} {\bibinfo  {journal}
  {\mnras}\ }\textbf {\bibinfo {volume} {515}},\ \bibinfo {pages} {4161}
  (\bibinfo {year} {2022})},\ \Eprint {http://arxiv.org/abs/2109.04984}
  {arXiv:2109.04984 [astro-ph.CO]} \BibitemShut {NoStop}%
\bibitem [{\citenamefont {{Hassani}}\ and\ \citenamefont
  {{Lombriser}}(2020)}]{ref:HassaniNBodyMG}%
  \BibitemOpen
  \bibfield  {author} {\bibinfo {author} {\bibfnamefont {F.}~\bibnamefont
  {{Hassani}}}\ and\ \bibinfo {author} {\bibfnamefont {L.}~\bibnamefont
  {{Lombriser}}},\ }\href@noop {} {\bibfield  {journal} {\bibinfo  {journal}
  {arXiv e-prints}\ ,\ \bibinfo {eid} {arXiv:2003.05927}} (\bibinfo {year}
  {2020})},\ \Eprint {http://arxiv.org/abs/2003.05927} {arXiv:2003.05927
  [astro-ph.CO]} \BibitemShut {NoStop}%
\bibitem [{\citenamefont {Thomas}(2020)}]{ref:DanPF}%
  \BibitemOpen
  \bibfield  {author} {\bibinfo {author} {\bibfnamefont {D.~B.}\ \bibnamefont
  {Thomas}},\ }\href {\doibase 10.1103/physrevd.101.123517} {\bibfield
  {journal} {\bibinfo  {journal} {Physical Review D}\ }\textbf {\bibinfo
  {volume} {101}} (\bibinfo {year} {2020}),\
  10.1103/physrevd.101.123517}\BibitemShut {NoStop}%
\bibitem [{\citenamefont {{Srinivasan}}\ \emph {et~al.}(2021)\citenamefont
  {{Srinivasan}}, \citenamefont {{Thomas}}, \citenamefont {{Pace}},\ and\
  \citenamefont {{Battye}}}]{Srinivasan2021}%
  \BibitemOpen
  \bibfield  {author} {\bibinfo {author} {\bibfnamefont {S.}~\bibnamefont
  {{Srinivasan}}}, \bibinfo {author} {\bibfnamefont {D.~B.}\ \bibnamefont
  {{Thomas}}}, \bibinfo {author} {\bibfnamefont {F.}~\bibnamefont {{Pace}}}, \
  and\ \bibinfo {author} {\bibfnamefont {R.}~\bibnamefont {{Battye}}},\ }\href
  {\doibase 10.1088/1475-7516/2021/06/016} {\bibfield  {journal} {\bibinfo
  {journal} {\jcap}\ }\textbf {\bibinfo {volume} {2021}},\ \bibinfo {eid} {016}
  (\bibinfo {year} {2021})},\ \Eprint {http://arxiv.org/abs/2103.05051}
  {arXiv:2103.05051 [astro-ph.CO]} \BibitemShut {NoStop}%
\bibitem [{\citenamefont {{Cui}}\ \emph {et~al.}(2010)\citenamefont {{Cui}},
  \citenamefont {{Zhang}},\ and\ \citenamefont {{Yang}}}]{ref:Cui1}%
  \BibitemOpen
  \bibfield  {author} {\bibinfo {author} {\bibfnamefont {W.}~\bibnamefont
  {{Cui}}}, \bibinfo {author} {\bibfnamefont {P.}~\bibnamefont {{Zhang}}}, \
  and\ \bibinfo {author} {\bibfnamefont {X.}~\bibnamefont {{Yang}}},\ }\href
  {\doibase 10.1103/PhysRevD.81.103528} {\bibfield  {journal} {\bibinfo
  {journal} {\prd}\ }\textbf {\bibinfo {volume} {81}},\ \bibinfo {eid} {103528}
  (\bibinfo {year} {2010})},\ \Eprint {http://arxiv.org/abs/1001.5184}
  {arXiv:1001.5184 [astro-ph.CO]} \BibitemShut {NoStop}%
\bibitem [{\citenamefont {{Zhang}}\ \emph {et~al.}(2013)\citenamefont
  {{Zhang}}, \citenamefont {{Zhang}}, \citenamefont {{Yang}},\ and\
  \citenamefont {{Cui}}}]{ref:Cui2}%
  \BibitemOpen
  \bibfield  {author} {\bibinfo {author} {\bibfnamefont {Y.}~\bibnamefont
  {{Zhang}}}, \bibinfo {author} {\bibfnamefont {P.}~\bibnamefont {{Zhang}}},
  \bibinfo {author} {\bibfnamefont {X.}~\bibnamefont {{Yang}}}, \ and\ \bibinfo
  {author} {\bibfnamefont {W.}~\bibnamefont {{Cui}}},\ }\href {\doibase
  10.1103/PhysRevD.87.023521} {\bibfield  {journal} {\bibinfo  {journal}
  {\prd}\ }\textbf {\bibinfo {volume} {87}},\ \bibinfo {eid} {023521} (\bibinfo
  {year} {2013})},\ \Eprint {http://arxiv.org/abs/1301.3255} {arXiv:1301.3255
  [astro-ph.CO]} \BibitemShut {NoStop}%
\bibitem [{\citenamefont {{Bose}}\ \emph {et~al.}(2021)\citenamefont {{Bose}},
  \citenamefont {{Wright}}, \citenamefont {{Cataneo}}, \citenamefont
  {{Pourtsidou}}, \citenamefont {{Giocoli}}, \citenamefont {{Lombriser}},
  \citenamefont {{McCarthy}}, \citenamefont {{Baldi}}, \citenamefont
  {{Pfeifer}},\ and\ \citenamefont {{Xia.}}}]{Bose_2021}%
  \BibitemOpen
  \bibfield  {author} {\bibinfo {author} {\bibfnamefont {B.}~\bibnamefont
  {{Bose}}}, \bibinfo {author} {\bibfnamefont {B.~S.}\ \bibnamefont
  {{Wright}}}, \bibinfo {author} {\bibfnamefont {M.}~\bibnamefont {{Cataneo}}},
  \bibinfo {author} {\bibfnamefont {A.}~\bibnamefont {{Pourtsidou}}}, \bibinfo
  {author} {\bibfnamefont {C.}~\bibnamefont {{Giocoli}}}, \bibinfo {author}
  {\bibfnamefont {L.}~\bibnamefont {{Lombriser}}}, \bibinfo {author}
  {\bibfnamefont {I.~G.}\ \bibnamefont {{McCarthy}}}, \bibinfo {author}
  {\bibfnamefont {M.}~\bibnamefont {{Baldi}}}, \bibinfo {author} {\bibfnamefont
  {S.}~\bibnamefont {{Pfeifer}}}, \ and\ \bibinfo {author} {\bibfnamefont
  {Q.}~\bibnamefont {{Xia.}}},\ }\href {\doibase 10.1093/mnras/stab2731}
  {\bibfield  {journal} {\bibinfo  {journal} {\mnras}\ }\textbf {\bibinfo
  {volume} {508}},\ \bibinfo {pages} {2479} (\bibinfo {year} {2021})},\ \Eprint
  {http://arxiv.org/abs/2105.12114} {arXiv:2105.12114 [astro-ph.CO]}
  \BibitemShut {NoStop}%
\bibitem [{\citenamefont {{Bose}}\ \emph
  {et~al.}(2020{\natexlab{a}})\citenamefont {{Bose}}, \citenamefont
  {{Cataneo}}, \citenamefont {{Tr{\"o}ster}}, \citenamefont {{Xia}},
  \citenamefont {{Heymans}},\ and\ \citenamefont {{Lombriser}}}]{BoseReACT}%
  \BibitemOpen
  \bibfield  {author} {\bibinfo {author} {\bibfnamefont {B.}~\bibnamefont
  {{Bose}}}, \bibinfo {author} {\bibfnamefont {M.}~\bibnamefont {{Cataneo}}},
  \bibinfo {author} {\bibfnamefont {T.}~\bibnamefont {{Tr{\"o}ster}}}, \bibinfo
  {author} {\bibfnamefont {Q.}~\bibnamefont {{Xia}}}, \bibinfo {author}
  {\bibfnamefont {C.}~\bibnamefont {{Heymans}}}, \ and\ \bibinfo {author}
  {\bibfnamefont {L.}~\bibnamefont {{Lombriser}}},\ }\href {\doibase
  10.1093/mnras/staa2696} {\bibfield  {journal} {\bibinfo  {journal} {\mnras}\
  }\textbf {\bibinfo {volume} {498}},\ \bibinfo {pages} {4650} (\bibinfo {year}
  {2020}{\natexlab{a}})},\ \Eprint {http://arxiv.org/abs/2005.12184}
  {arXiv:2005.12184 [astro-ph.CO]} \BibitemShut {NoStop}%
\bibitem [{\citenamefont {{Bose}}\ \emph
  {et~al.}(2020{\natexlab{b}})\citenamefont {{Bose}}, \citenamefont
  {{Cataneo}}, \citenamefont {{Tr{\"o}ster}}, \citenamefont {{Xia}},
  \citenamefont {{Heymans}},\ and\ \citenamefont
  {{Lombriser}}}]{ref:reactionCataneo}%
  \BibitemOpen
  \bibfield  {author} {\bibinfo {author} {\bibfnamefont {B.}~\bibnamefont
  {{Bose}}}, \bibinfo {author} {\bibfnamefont {M.}~\bibnamefont {{Cataneo}}},
  \bibinfo {author} {\bibfnamefont {T.}~\bibnamefont {{Tr{\"o}ster}}}, \bibinfo
  {author} {\bibfnamefont {Q.}~\bibnamefont {{Xia}}}, \bibinfo {author}
  {\bibfnamefont {C.}~\bibnamefont {{Heymans}}}, \ and\ \bibinfo {author}
  {\bibfnamefont {L.}~\bibnamefont {{Lombriser}}},\ }\href@noop {} {\bibfield
  {journal} {\bibinfo  {journal} {arXiv e-prints}\ ,\ \bibinfo {eid}
  {arXiv:2005.12184}} (\bibinfo {year} {2020}{\natexlab{b}})},\ \Eprint
  {http://arxiv.org/abs/2005.12184} {arXiv:2005.12184 [astro-ph.CO]}
  \BibitemShut {NoStop}%
\bibitem [{\citenamefont {{Mead}}(2017)}]{Mead2017}%
  \BibitemOpen
  \bibfield  {author} {\bibinfo {author} {\bibfnamefont {A.~J.}\ \bibnamefont
  {{Mead}}},\ }\href {\doibase 10.1093/mnras/stw2312} {\bibfield  {journal}
  {\bibinfo  {journal} {\mnras}\ }\textbf {\bibinfo {volume} {464}},\ \bibinfo
  {pages} {1282} (\bibinfo {year} {2017})},\ \Eprint
  {http://arxiv.org/abs/1606.05345} {arXiv:1606.05345 [astro-ph.CO]}
  \BibitemShut {NoStop}%
\bibitem [{\citenamefont {{Cataneo}}\ \emph {et~al.}(2019)\citenamefont
  {{Cataneo}}, \citenamefont {{Lombriser}}, \citenamefont {{Heymans}},
  \citenamefont {{Mead}}, \citenamefont {{Barreira}}, \citenamefont {{Bose}},\
  and\ \citenamefont {{Li}}}]{ref:ReactTheory}%
  \BibitemOpen
  \bibfield  {author} {\bibinfo {author} {\bibfnamefont {M.}~\bibnamefont
  {{Cataneo}}}, \bibinfo {author} {\bibfnamefont {L.}~\bibnamefont
  {{Lombriser}}}, \bibinfo {author} {\bibfnamefont {C.}~\bibnamefont
  {{Heymans}}}, \bibinfo {author} {\bibfnamefont {A.~J.}\ \bibnamefont
  {{Mead}}}, \bibinfo {author} {\bibfnamefont {A.}~\bibnamefont {{Barreira}}},
  \bibinfo {author} {\bibfnamefont {S.}~\bibnamefont {{Bose}}}, \ and\ \bibinfo
  {author} {\bibfnamefont {B.}~\bibnamefont {{Li}}},\ }\href {\doibase
  10.1093/mnras/stz1836} {\bibfield  {journal} {\bibinfo  {journal} {\mnras}\
  }\textbf {\bibinfo {volume} {488}},\ \bibinfo {pages} {2121} (\bibinfo {year}
  {2019})},\ \Eprint {http://arxiv.org/abs/1812.05594} {arXiv:1812.05594
  [astro-ph.CO]} \BibitemShut {NoStop}%
\bibitem [{\citenamefont {{Carrilho}}\ \emph {et~al.}(2022)\citenamefont
  {{Carrilho}}, \citenamefont {{Carrion}}, \citenamefont {{Bose}},
  \citenamefont {{Pourtsidou}}, \citenamefont {{Hidalgo}}, \citenamefont
  {{Lombriser}},\ and\ \citenamefont {{Baldi}}}]{Carrilho_2022}%
  \BibitemOpen
  \bibfield  {author} {\bibinfo {author} {\bibfnamefont {P.}~\bibnamefont
  {{Carrilho}}}, \bibinfo {author} {\bibfnamefont {K.}~\bibnamefont
  {{Carrion}}}, \bibinfo {author} {\bibfnamefont {B.}~\bibnamefont {{Bose}}},
  \bibinfo {author} {\bibfnamefont {A.}~\bibnamefont {{Pourtsidou}}}, \bibinfo
  {author} {\bibfnamefont {J.~C.}\ \bibnamefont {{Hidalgo}}}, \bibinfo {author}
  {\bibfnamefont {L.}~\bibnamefont {{Lombriser}}}, \ and\ \bibinfo {author}
  {\bibfnamefont {M.}~\bibnamefont {{Baldi}}},\ }\href {\doibase
  10.1093/mnras/stac641} {\bibfield  {journal} {\bibinfo  {journal} {\mnras}\
  }\textbf {\bibinfo {volume} {512}},\ \bibinfo {pages} {3691} (\bibinfo {year}
  {2022})},\ \Eprint {http://arxiv.org/abs/2111.13598} {arXiv:2111.13598
  [astro-ph.CO]} \BibitemShut {NoStop}%
\bibitem [{\citenamefont {{Navarro}}\ \emph {et~al.}(1996)\citenamefont
  {{Navarro}}, \citenamefont {{Frenk}},\ and\ \citenamefont
  {{White}}}]{ref:NFW}%
  \BibitemOpen
  \bibfield  {author} {\bibinfo {author} {\bibfnamefont {J.~F.}\ \bibnamefont
  {{Navarro}}}, \bibinfo {author} {\bibfnamefont {C.~S.}\ \bibnamefont
  {{Frenk}}}, \ and\ \bibinfo {author} {\bibfnamefont {S.~D.~M.}\ \bibnamefont
  {{White}}},\ }\href {\doibase 10.1086/177173} {\bibfield  {journal} {\bibinfo
   {journal} {\apj}\ }\textbf {\bibinfo {volume} {462}},\ \bibinfo {pages}
  {563} (\bibinfo {year} {1996})},\ \Eprint
  {http://arxiv.org/abs/astro-ph/9508025} {astro-ph/9508025} \BibitemShut
  {NoStop}%
\bibitem [{\citenamefont {Klypin}\ \emph {et~al.}(2016)\citenamefont {Klypin},
  \citenamefont {Yepes}, \citenamefont {Gottlöber}, \citenamefont {Prada},\
  and\ \citenamefont {Heß}}]{ref:MultiDark}%
  \BibitemOpen
  \bibfield  {author} {\bibinfo {author} {\bibfnamefont {A.}~\bibnamefont
  {Klypin}}, \bibinfo {author} {\bibfnamefont {G.}~\bibnamefont {Yepes}},
  \bibinfo {author} {\bibfnamefont {S.}~\bibnamefont {Gottlöber}}, \bibinfo
  {author} {\bibfnamefont {F.}~\bibnamefont {Prada}}, \ and\ \bibinfo {author}
  {\bibfnamefont {S.}~\bibnamefont {Heß}},\ }\href {\doibase
  10.1093/mnras/stw248} {\bibfield  {journal} {\bibinfo  {journal} {Monthly
  Notices of the Royal Astronomical Society}\ }\textbf {\bibinfo {volume}
  {457}},\ \bibinfo {pages} {4340} (\bibinfo {year} {2016})},\ \Eprint
  {http://arxiv.org/abs/http://oup.prod.sis.lan/mnras/article-pdf/457/4/4340/18515365/stw248.pdf}
  {http://oup.prod.sis.lan/mnras/article-pdf/457/4/4340/18515365/stw248.pdf}
  \BibitemShut {NoStop}%
\bibitem [{\citenamefont {{Planck Collaboration}}\ \emph
  {et~al.}(2018)\citenamefont {{Planck Collaboration}}, \citenamefont
  {{Aghanim}}, \citenamefont {{Akrami}}, \citenamefont {{Ashdown}},
  \citenamefont {{Aumont}}, \citenamefont {{Baccigalupi}}, \citenamefont
  {{Ballardini}}, \citenamefont {{Banday}}, \citenamefont {{Barreiro}},
  \citenamefont {{Bartolo}}, \citenamefont {{Basak}}, \citenamefont {{Battye}},
  \citenamefont {{Benabed}}, \citenamefont {{Bernard}}, \citenamefont
  {{Bersanelli}}, \citenamefont {{Bielewicz}}, \citenamefont {{Bock}},
  \citenamefont {{Bond}}, \citenamefont {{Borrill}}, \citenamefont {{Bouchet}},
  \citenamefont {{Boulanger}}, \citenamefont {{Bucher}}, \citenamefont
  {{Burigana}}, \citenamefont {{Butler}}, \citenamefont {{Calabrese}},
  \citenamefont {{Cardoso}}, \citenamefont {{Carron}}, \citenamefont
  {{Challinor}}, \citenamefont {{Chiang}}, \citenamefont {{Chluba}},
  \citenamefont {{Colombo}}, \citenamefont {{Combet}}, \citenamefont
  {{Contreras}}, \citenamefont {{Crill}}, \citenamefont {{Cuttaia}},
  \citenamefont {{de Bernardis}}, \citenamefont {{de Zotti}}, \citenamefont
  {{Delabrouille}}, \citenamefont {{Delouis}}, \citenamefont {{Di Valentino}},
  \citenamefont {{Diego}}, \citenamefont {{Dor{\'e}}}, \citenamefont
  {{Douspis}}, \citenamefont {{Ducout}}, \citenamefont {{Dupac}}, \citenamefont
  {{Dusini}}, \citenamefont {{Efstathiou}}, \citenamefont {{Elsner}},
  \citenamefont {{En{\ss}lin}}, \citenamefont {{Eriksen}}, \citenamefont
  {{Fantaye}}, \citenamefont {{Farhang}}, \citenamefont {{Fergusson}},
  \citenamefont {{Fernandez-Cobos}}, \citenamefont {{Finelli}}, \citenamefont
  {{Forastieri}}, \citenamefont {{Frailis}}, \citenamefont {{Franceschi}},
  \citenamefont {{Frolov}}, \citenamefont {{Galeotta}}, \citenamefont
  {{Galli}}, \citenamefont {{Ganga}}, \citenamefont {{G{\'e}nova-Santos}},
  \citenamefont {{Gerbino}}, \citenamefont {{Ghosh}}, \citenamefont
  {{Gonz{\'a}lez-Nuevo}}, \citenamefont {{G{\'o}rski}}, \citenamefont
  {{Gratton}}, \citenamefont {{Gruppuso}}, \citenamefont {{Gudmundsson}},
  \citenamefont {{Hamann}}, \citenamefont {{Handley}}, \citenamefont
  {{Herranz}}, \citenamefont {{Hivon}}, \citenamefont {{Huang}}, \citenamefont
  {{Jaffe}}, \citenamefont {{Jones}}, \citenamefont {{Karakci}}, \citenamefont
  {{Keih{\"a}nen}}, \citenamefont {{Keskitalo}}, \citenamefont {{Kiiveri}},
  \citenamefont {{Kim}}, \citenamefont {{Kisner}}, \citenamefont {{Knox}},
  \citenamefont {{Krachmalnicoff}}, \citenamefont {{Kunz}}, \citenamefont
  {{Kurki-Suonio}}, \citenamefont {{Lagache}}, \citenamefont {{Lamarre}},
  \citenamefont {{Lasenby}}, \citenamefont {{Lattanzi}}, \citenamefont
  {{Lawrence}}, \citenamefont {{Le Jeune}}, \citenamefont {{Lemos}},
  \citenamefont {{Lesgourgues}}, \citenamefont {{Levrier}}, \citenamefont
  {{Lewis}}, \citenamefont {{Liguori}}, \citenamefont {{Lilje}}, \citenamefont
  {{Lilley}}, \citenamefont {{Lindholm}}, \citenamefont {{L{\'o}pez-Caniego}},
  \citenamefont {{Lubin}}, \citenamefont {{Ma}}, \citenamefont
  {{Mac{\'{\i}}as-P{\'e}rez}}, \citenamefont {{Maggio}}, \citenamefont
  {{Maino}}, \citenamefont {{Mandolesi}}, \citenamefont {{Mangilli}},
  \citenamefont {{Marcos-Caballero}}, \citenamefont {{Maris}}, \citenamefont
  {{Martin}}, \citenamefont {{Martinelli}}, \citenamefont
  {{Mart{\'{\i}}nez-Gonz{\'a}lez}}, \citenamefont {{Matarrese}}, \citenamefont
  {{Mauri}}, \citenamefont {{McEwen}}, \citenamefont {{Meinhold}},
  \citenamefont {{Melchiorri}}, \citenamefont {{Mennella}}, \citenamefont
  {{Migliaccio}}, \citenamefont {{Millea}}, \citenamefont {{Mitra}},
  \citenamefont {{Miville-Desch{\^e}nes}}, \citenamefont {{Molinari}},
  \citenamefont {{Montier}}, \citenamefont {{Morgante}}, \citenamefont
  {{Moss}}, \citenamefont {{Natoli}}, \citenamefont {{N{\o}rgaard-Nielsen}},
  \citenamefont {{Pagano}}, \citenamefont {{Paoletti}}, \citenamefont
  {{Partridge}}, \citenamefont {{Patanchon}}, \citenamefont {{Peiris}},
  \citenamefont {{Perrotta}}, \citenamefont {{Pettorino}}, \citenamefont
  {{Piacentini}}, \citenamefont {{Polastri}}, \citenamefont {{Polenta}},
  \citenamefont {{Puget}}, \citenamefont {{Rachen}}, \citenamefont
  {{Reinecke}}, \citenamefont {{Remazeilles}}, \citenamefont {{Renzi}},
  \citenamefont {{Rocha}}, \citenamefont {{Rosset}}, \citenamefont {{Roudier}},
  \citenamefont {{Rubi{\~n}o-Mart{\'{\i}}n}}, \citenamefont {{Ruiz-Granados}},
  \citenamefont {{Salvati}}, \citenamefont {{Sandri}}, \citenamefont
  {{Savelainen}}, \citenamefont {{Scott}}, \citenamefont {{Shellard}},
  \citenamefont {{Sirignano}}, \citenamefont {{Sirri}}, \citenamefont
  {{Spencer}}, \citenamefont {{Sunyaev}}, \citenamefont {{Suur-Uski}},
  \citenamefont {{Tauber}}, \citenamefont {{Tavagnacco}}, \citenamefont
  {{Tenti}}, \citenamefont {{Toffolatti}}, \citenamefont {{Tomasi}},
  \citenamefont {{Trombetti}}, \citenamefont {{Valenziano}}, \citenamefont
  {{Valiviita}}, \citenamefont {{Van Tent}}, \citenamefont {{Vibert}},
  \citenamefont {{Vielva}}, \citenamefont {{Villa}}, \citenamefont
  {{Vittorio}}, \citenamefont {{Wandelt}}, \citenamefont {{Wehus}},
  \citenamefont {{White}}, \citenamefont {{White}}, \citenamefont {{Zacchei}},\
  and\ \citenamefont {{Zonca}}}]{ref:Planck2018}%
  \BibitemOpen
  \bibfield  {author} {\bibinfo {author} {\bibnamefont {{Planck
  Collaboration}}}, \bibinfo {author} {\bibfnamefont {N.}~\bibnamefont
  {{Aghanim}}}, \bibinfo {author} {\bibfnamefont {Y.}~\bibnamefont {{Akrami}}},
  \bibinfo {author} {\bibfnamefont {M.}~\bibnamefont {{Ashdown}}}, \bibinfo
  {author} {\bibfnamefont {J.}~\bibnamefont {{Aumont}}}, \bibinfo {author}
  {\bibfnamefont {C.}~\bibnamefont {{Baccigalupi}}}, \bibinfo {author}
  {\bibfnamefont {M.}~\bibnamefont {{Ballardini}}}, \bibinfo {author}
  {\bibfnamefont {A.~J.}\ \bibnamefont {{Banday}}}, \bibinfo {author}
  {\bibfnamefont {R.~B.}\ \bibnamefont {{Barreiro}}}, \bibinfo {author}
  {\bibfnamefont {N.}~\bibnamefont {{Bartolo}}}, \bibinfo {author}
  {\bibfnamefont {S.}~\bibnamefont {{Basak}}}, \bibinfo {author} {\bibfnamefont
  {R.}~\bibnamefont {{Battye}}}, \bibinfo {author} {\bibfnamefont
  {K.}~\bibnamefont {{Benabed}}}, \bibinfo {author} {\bibfnamefont {J.-P.}\
  \bibnamefont {{Bernard}}}, \bibinfo {author} {\bibfnamefont {M.}~\bibnamefont
  {{Bersanelli}}}, \bibinfo {author} {\bibfnamefont {P.}~\bibnamefont
  {{Bielewicz}}}, \bibinfo {author} {\bibfnamefont {J.~J.}\ \bibnamefont
  {{Bock}}}, \bibinfo {author} {\bibfnamefont {J.~R.}\ \bibnamefont {{Bond}}},
  \bibinfo {author} {\bibfnamefont {J.}~\bibnamefont {{Borrill}}}, \bibinfo
  {author} {\bibfnamefont {F.~R.}\ \bibnamefont {{Bouchet}}}, \bibinfo {author}
  {\bibfnamefont {F.}~\bibnamefont {{Boulanger}}}, \bibinfo {author}
  {\bibfnamefont {M.}~\bibnamefont {{Bucher}}}, \bibinfo {author}
  {\bibfnamefont {C.}~\bibnamefont {{Burigana}}}, \bibinfo {author}
  {\bibfnamefont {R.~C.}\ \bibnamefont {{Butler}}}, \bibinfo {author}
  {\bibfnamefont {E.}~\bibnamefont {{Calabrese}}}, \bibinfo {author}
  {\bibfnamefont {J.-F.}\ \bibnamefont {{Cardoso}}}, \bibinfo {author}
  {\bibfnamefont {J.}~\bibnamefont {{Carron}}}, \bibinfo {author}
  {\bibfnamefont {A.}~\bibnamefont {{Challinor}}}, \bibinfo {author}
  {\bibfnamefont {H.~C.}\ \bibnamefont {{Chiang}}}, \bibinfo {author}
  {\bibfnamefont {J.}~\bibnamefont {{Chluba}}}, \bibinfo {author}
  {\bibfnamefont {L.~P.~L.}\ \bibnamefont {{Colombo}}}, \bibinfo {author}
  {\bibfnamefont {C.}~\bibnamefont {{Combet}}}, \bibinfo {author}
  {\bibfnamefont {D.}~\bibnamefont {{Contreras}}}, \bibinfo {author}
  {\bibfnamefont {B.~P.}\ \bibnamefont {{Crill}}}, \bibinfo {author}
  {\bibfnamefont {F.}~\bibnamefont {{Cuttaia}}}, \bibinfo {author}
  {\bibfnamefont {P.}~\bibnamefont {{de Bernardis}}}, \bibinfo {author}
  {\bibfnamefont {G.}~\bibnamefont {{de Zotti}}}, \bibinfo {author}
  {\bibfnamefont {J.}~\bibnamefont {{Delabrouille}}}, \bibinfo {author}
  {\bibfnamefont {J.-M.}\ \bibnamefont {{Delouis}}}, \bibinfo {author}
  {\bibfnamefont {E.}~\bibnamefont {{Di Valentino}}}, \bibinfo {author}
  {\bibfnamefont {J.~M.}\ \bibnamefont {{Diego}}}, \bibinfo {author}
  {\bibfnamefont {O.}~\bibnamefont {{Dor{\'e}}}}, \bibinfo {author}
  {\bibfnamefont {M.}~\bibnamefont {{Douspis}}}, \bibinfo {author}
  {\bibfnamefont {A.}~\bibnamefont {{Ducout}}}, \bibinfo {author}
  {\bibfnamefont {X.}~\bibnamefont {{Dupac}}}, \bibinfo {author} {\bibfnamefont
  {S.}~\bibnamefont {{Dusini}}}, \bibinfo {author} {\bibfnamefont
  {G.}~\bibnamefont {{Efstathiou}}}, \bibinfo {author} {\bibfnamefont
  {F.}~\bibnamefont {{Elsner}}}, \bibinfo {author} {\bibfnamefont {T.~A.}\
  \bibnamefont {{En{\ss}lin}}}, \bibinfo {author} {\bibfnamefont {H.~K.}\
  \bibnamefont {{Eriksen}}}, \bibinfo {author} {\bibfnamefont {Y.}~\bibnamefont
  {{Fantaye}}}, \bibinfo {author} {\bibfnamefont {M.}~\bibnamefont
  {{Farhang}}}, \bibinfo {author} {\bibfnamefont {J.}~\bibnamefont
  {{Fergusson}}}, \bibinfo {author} {\bibfnamefont {R.}~\bibnamefont
  {{Fernandez-Cobos}}}, \bibinfo {author} {\bibfnamefont {F.}~\bibnamefont
  {{Finelli}}}, \bibinfo {author} {\bibfnamefont {F.}~\bibnamefont
  {{Forastieri}}}, \bibinfo {author} {\bibfnamefont {M.}~\bibnamefont
  {{Frailis}}}, \bibinfo {author} {\bibfnamefont {E.}~\bibnamefont
  {{Franceschi}}}, \bibinfo {author} {\bibfnamefont {A.}~\bibnamefont
  {{Frolov}}}, \bibinfo {author} {\bibfnamefont {S.}~\bibnamefont
  {{Galeotta}}}, \bibinfo {author} {\bibfnamefont {S.}~\bibnamefont {{Galli}}},
  \bibinfo {author} {\bibfnamefont {K.}~\bibnamefont {{Ganga}}}, \bibinfo
  {author} {\bibfnamefont {R.~T.}\ \bibnamefont {{G{\'e}nova-Santos}}},
  \bibinfo {author} {\bibfnamefont {M.}~\bibnamefont {{Gerbino}}}, \bibinfo
  {author} {\bibfnamefont {T.}~\bibnamefont {{Ghosh}}}, \bibinfo {author}
  {\bibfnamefont {J.}~\bibnamefont {{Gonz{\'a}lez-Nuevo}}}, \bibinfo {author}
  {\bibfnamefont {K.~M.}\ \bibnamefont {{G{\'o}rski}}}, \bibinfo {author}
  {\bibfnamefont {S.}~\bibnamefont {{Gratton}}}, \bibinfo {author}
  {\bibfnamefont {A.}~\bibnamefont {{Gruppuso}}}, \bibinfo {author}
  {\bibfnamefont {J.~E.}\ \bibnamefont {{Gudmundsson}}}, \bibinfo {author}
  {\bibfnamefont {J.}~\bibnamefont {{Hamann}}}, \bibinfo {author}
  {\bibfnamefont {W.}~\bibnamefont {{Handley}}}, \bibinfo {author}
  {\bibfnamefont {D.}~\bibnamefont {{Herranz}}}, \bibinfo {author}
  {\bibfnamefont {E.}~\bibnamefont {{Hivon}}}, \bibinfo {author} {\bibfnamefont
  {Z.}~\bibnamefont {{Huang}}}, \bibinfo {author} {\bibfnamefont {A.~H.}\
  \bibnamefont {{Jaffe}}}, \bibinfo {author} {\bibfnamefont {W.~C.}\
  \bibnamefont {{Jones}}}, \bibinfo {author} {\bibfnamefont {A.}~\bibnamefont
  {{Karakci}}}, \bibinfo {author} {\bibfnamefont {E.}~\bibnamefont
  {{Keih{\"a}nen}}}, \bibinfo {author} {\bibfnamefont {R.}~\bibnamefont
  {{Keskitalo}}}, \bibinfo {author} {\bibfnamefont {K.}~\bibnamefont
  {{Kiiveri}}}, \bibinfo {author} {\bibfnamefont {J.}~\bibnamefont {{Kim}}},
  \bibinfo {author} {\bibfnamefont {T.~S.}\ \bibnamefont {{Kisner}}}, \bibinfo
  {author} {\bibfnamefont {L.}~\bibnamefont {{Knox}}}, \bibinfo {author}
  {\bibfnamefont {N.}~\bibnamefont {{Krachmalnicoff}}}, \bibinfo {author}
  {\bibfnamefont {M.}~\bibnamefont {{Kunz}}}, \bibinfo {author} {\bibfnamefont
  {H.}~\bibnamefont {{Kurki-Suonio}}}, \bibinfo {author} {\bibfnamefont
  {G.}~\bibnamefont {{Lagache}}}, \bibinfo {author} {\bibfnamefont {J.-M.}\
  \bibnamefont {{Lamarre}}}, \bibinfo {author} {\bibfnamefont {A.}~\bibnamefont
  {{Lasenby}}}, \bibinfo {author} {\bibfnamefont {M.}~\bibnamefont
  {{Lattanzi}}}, \bibinfo {author} {\bibfnamefont {C.~R.}\ \bibnamefont
  {{Lawrence}}}, \bibinfo {author} {\bibfnamefont {M.}~\bibnamefont {{Le
  Jeune}}}, \bibinfo {author} {\bibfnamefont {P.}~\bibnamefont {{Lemos}}},
  \bibinfo {author} {\bibfnamefont {J.}~\bibnamefont {{Lesgourgues}}}, \bibinfo
  {author} {\bibfnamefont {F.}~\bibnamefont {{Levrier}}}, \bibinfo {author}
  {\bibfnamefont {A.}~\bibnamefont {{Lewis}}}, \bibinfo {author} {\bibfnamefont
  {M.}~\bibnamefont {{Liguori}}}, \bibinfo {author} {\bibfnamefont {P.~B.}\
  \bibnamefont {{Lilje}}}, \bibinfo {author} {\bibfnamefont {M.}~\bibnamefont
  {{Lilley}}}, \bibinfo {author} {\bibfnamefont {V.}~\bibnamefont
  {{Lindholm}}}, \bibinfo {author} {\bibfnamefont {M.}~\bibnamefont
  {{L{\'o}pez-Caniego}}}, \bibinfo {author} {\bibfnamefont {P.~M.}\
  \bibnamefont {{Lubin}}}, \bibinfo {author} {\bibfnamefont {Y.-Z.}\
  \bibnamefont {{Ma}}}, \bibinfo {author} {\bibfnamefont {J.~F.}\ \bibnamefont
  {{Mac{\'{\i}}as-P{\'e}rez}}}, \bibinfo {author} {\bibfnamefont
  {G.}~\bibnamefont {{Maggio}}}, \bibinfo {author} {\bibfnamefont
  {D.}~\bibnamefont {{Maino}}}, \bibinfo {author} {\bibfnamefont
  {N.}~\bibnamefont {{Mandolesi}}}, \bibinfo {author} {\bibfnamefont
  {A.}~\bibnamefont {{Mangilli}}}, \bibinfo {author} {\bibfnamefont
  {A.}~\bibnamefont {{Marcos-Caballero}}}, \bibinfo {author} {\bibfnamefont
  {M.}~\bibnamefont {{Maris}}}, \bibinfo {author} {\bibfnamefont {P.~G.}\
  \bibnamefont {{Martin}}}, \bibinfo {author} {\bibfnamefont {M.}~\bibnamefont
  {{Martinelli}}}, \bibinfo {author} {\bibfnamefont {E.}~\bibnamefont
  {{Mart{\'{\i}}nez-Gonz{\'a}lez}}}, \bibinfo {author} {\bibfnamefont
  {S.}~\bibnamefont {{Matarrese}}}, \bibinfo {author} {\bibfnamefont
  {N.}~\bibnamefont {{Mauri}}}, \bibinfo {author} {\bibfnamefont {J.~D.}\
  \bibnamefont {{McEwen}}}, \bibinfo {author} {\bibfnamefont {P.~R.}\
  \bibnamefont {{Meinhold}}}, \bibinfo {author} {\bibfnamefont
  {A.}~\bibnamefont {{Melchiorri}}}, \bibinfo {author} {\bibfnamefont
  {A.}~\bibnamefont {{Mennella}}}, \bibinfo {author} {\bibfnamefont
  {M.}~\bibnamefont {{Migliaccio}}}, \bibinfo {author} {\bibfnamefont
  {M.}~\bibnamefont {{Millea}}}, \bibinfo {author} {\bibfnamefont
  {S.}~\bibnamefont {{Mitra}}}, \bibinfo {author} {\bibfnamefont {M.-A.}\
  \bibnamefont {{Miville-Desch{\^e}nes}}}, \bibinfo {author} {\bibfnamefont
  {D.}~\bibnamefont {{Molinari}}}, \bibinfo {author} {\bibfnamefont
  {L.}~\bibnamefont {{Montier}}}, \bibinfo {author} {\bibfnamefont
  {G.}~\bibnamefont {{Morgante}}}, \bibinfo {author} {\bibfnamefont
  {A.}~\bibnamefont {{Moss}}}, \bibinfo {author} {\bibfnamefont
  {P.}~\bibnamefont {{Natoli}}}, \bibinfo {author} {\bibfnamefont {H.~U.}\
  \bibnamefont {{N{\o}rgaard-Nielsen}}}, \bibinfo {author} {\bibfnamefont
  {L.}~\bibnamefont {{Pagano}}}, \bibinfo {author} {\bibfnamefont
  {D.}~\bibnamefont {{Paoletti}}}, \bibinfo {author} {\bibfnamefont
  {B.}~\bibnamefont {{Partridge}}}, \bibinfo {author} {\bibfnamefont
  {G.}~\bibnamefont {{Patanchon}}}, \bibinfo {author} {\bibfnamefont {H.~V.}\
  \bibnamefont {{Peiris}}}, \bibinfo {author} {\bibfnamefont {F.}~\bibnamefont
  {{Perrotta}}}, \bibinfo {author} {\bibfnamefont {V.}~\bibnamefont
  {{Pettorino}}}, \bibinfo {author} {\bibfnamefont {F.}~\bibnamefont
  {{Piacentini}}}, \bibinfo {author} {\bibfnamefont {L.}~\bibnamefont
  {{Polastri}}}, \bibinfo {author} {\bibfnamefont {G.}~\bibnamefont
  {{Polenta}}}, \bibinfo {author} {\bibfnamefont {J.-L.}\ \bibnamefont
  {{Puget}}}, \bibinfo {author} {\bibfnamefont {J.~P.}\ \bibnamefont
  {{Rachen}}}, \bibinfo {author} {\bibfnamefont {M.}~\bibnamefont
  {{Reinecke}}}, \bibinfo {author} {\bibfnamefont {M.}~\bibnamefont
  {{Remazeilles}}}, \bibinfo {author} {\bibfnamefont {A.}~\bibnamefont
  {{Renzi}}}, \bibinfo {author} {\bibfnamefont {G.}~\bibnamefont {{Rocha}}},
  \bibinfo {author} {\bibfnamefont {C.}~\bibnamefont {{Rosset}}}, \bibinfo
  {author} {\bibfnamefont {G.}~\bibnamefont {{Roudier}}}, \bibinfo {author}
  {\bibfnamefont {J.~A.}\ \bibnamefont {{Rubi{\~n}o-Mart{\'{\i}}n}}}, \bibinfo
  {author} {\bibfnamefont {B.}~\bibnamefont {{Ruiz-Granados}}}, \bibinfo
  {author} {\bibfnamefont {L.}~\bibnamefont {{Salvati}}}, \bibinfo {author}
  {\bibfnamefont {M.}~\bibnamefont {{Sandri}}}, \bibinfo {author}
  {\bibfnamefont {M.}~\bibnamefont {{Savelainen}}}, \bibinfo {author}
  {\bibfnamefont {D.}~\bibnamefont {{Scott}}}, \bibinfo {author} {\bibfnamefont
  {E.~P.~S.}\ \bibnamefont {{Shellard}}}, \bibinfo {author} {\bibfnamefont
  {C.}~\bibnamefont {{Sirignano}}}, \bibinfo {author} {\bibfnamefont
  {G.}~\bibnamefont {{Sirri}}}, \bibinfo {author} {\bibfnamefont {L.~D.}\
  \bibnamefont {{Spencer}}}, \bibinfo {author} {\bibfnamefont {R.}~\bibnamefont
  {{Sunyaev}}}, \bibinfo {author} {\bibfnamefont {A.-S.}\ \bibnamefont
  {{Suur-Uski}}}, \bibinfo {author} {\bibfnamefont {J.~A.}\ \bibnamefont
  {{Tauber}}}, \bibinfo {author} {\bibfnamefont {D.}~\bibnamefont
  {{Tavagnacco}}}, \bibinfo {author} {\bibfnamefont {M.}~\bibnamefont
  {{Tenti}}}, \bibinfo {author} {\bibfnamefont {L.}~\bibnamefont
  {{Toffolatti}}}, \bibinfo {author} {\bibfnamefont {M.}~\bibnamefont
  {{Tomasi}}}, \bibinfo {author} {\bibfnamefont {T.}~\bibnamefont
  {{Trombetti}}}, \bibinfo {author} {\bibfnamefont {L.}~\bibnamefont
  {{Valenziano}}}, \bibinfo {author} {\bibfnamefont {J.}~\bibnamefont
  {{Valiviita}}}, \bibinfo {author} {\bibfnamefont {B.}~\bibnamefont {{Van
  Tent}}}, \bibinfo {author} {\bibfnamefont {L.}~\bibnamefont {{Vibert}}},
  \bibinfo {author} {\bibfnamefont {P.}~\bibnamefont {{Vielva}}}, \bibinfo
  {author} {\bibfnamefont {F.}~\bibnamefont {{Villa}}}, \bibinfo {author}
  {\bibfnamefont {N.}~\bibnamefont {{Vittorio}}}, \bibinfo {author}
  {\bibfnamefont {B.~D.}\ \bibnamefont {{Wandelt}}}, \bibinfo {author}
  {\bibfnamefont {I.~K.}\ \bibnamefont {{Wehus}}}, \bibinfo {author}
  {\bibfnamefont {M.}~\bibnamefont {{White}}}, \bibinfo {author} {\bibfnamefont
  {S.~D.~M.}\ \bibnamefont {{White}}}, \bibinfo {author} {\bibfnamefont
  {A.}~\bibnamefont {{Zacchei}}}, \ and\ \bibinfo {author} {\bibfnamefont
  {A.}~\bibnamefont {{Zonca}}},\ }\href@noop {} {\bibfield  {journal} {\bibinfo
   {journal} {ArXiv e-prints}\ } (\bibinfo {year} {2018})},\ \Eprint
  {http://arxiv.org/abs/1807.06209} {arXiv:1807.06209} \BibitemShut {NoStop}%
\bibitem [{\citenamefont {{Dolag}}\ \emph {et~al.}(2004)\citenamefont
  {{Dolag}}, \citenamefont {{Bartelmann}}, \citenamefont {{Perrotta}},
  \citenamefont {{Baccigalupi}}, \citenamefont {{Moscardini}}, \citenamefont
  {{Meneghetti}},\ and\ \citenamefont {{Tormen}}}]{Dolag2004}%
  \BibitemOpen
  \bibfield  {author} {\bibinfo {author} {\bibfnamefont {K.}~\bibnamefont
  {{Dolag}}}, \bibinfo {author} {\bibfnamefont {M.}~\bibnamefont
  {{Bartelmann}}}, \bibinfo {author} {\bibfnamefont {F.}~\bibnamefont
  {{Perrotta}}}, \bibinfo {author} {\bibfnamefont {C.}~\bibnamefont
  {{Baccigalupi}}}, \bibinfo {author} {\bibfnamefont {L.}~\bibnamefont
  {{Moscardini}}}, \bibinfo {author} {\bibfnamefont {M.}~\bibnamefont
  {{Meneghetti}}}, \ and\ \bibinfo {author} {\bibfnamefont {G.}~\bibnamefont
  {{Tormen}}},\ }\href {\doibase 10.1051/0004-6361:20031757} {\bibfield
  {journal} {\bibinfo  {journal} {\aap}\ }\textbf {\bibinfo {volume} {416}},\
  \bibinfo {pages} {853} (\bibinfo {year} {2004})},\ \Eprint
  {http://arxiv.org/abs/astro-ph/0309771} {arXiv:astro-ph/0309771 [astro-ph]}
  \BibitemShut {NoStop}%
\bibitem [{\citenamefont {{Lombriser}}\ \emph {et~al.}(2014)\citenamefont
  {{Lombriser}}, \citenamefont {{Koyama}},\ and\ \citenamefont
  {{Li}}}]{Lombriser2014}%
  \BibitemOpen
  \bibfield  {author} {\bibinfo {author} {\bibfnamefont {L.}~\bibnamefont
  {{Lombriser}}}, \bibinfo {author} {\bibfnamefont {K.}~\bibnamefont
  {{Koyama}}}, \ and\ \bibinfo {author} {\bibfnamefont {B.}~\bibnamefont
  {{Li}}},\ }\href {\doibase 10.1088/1475-7516/2014/03/021} {\bibfield
  {journal} {\bibinfo  {journal} {\jcap}\ }\textbf {\bibinfo {volume} {2014}},\
  \bibinfo {eid} {021} (\bibinfo {year} {2014})},\ \Eprint
  {http://arxiv.org/abs/1312.1292} {arXiv:1312.1292 [astro-ph.CO]} \BibitemShut
  {NoStop}%
\bibitem [{\citenamefont {{Shi}}\ \emph {et~al.}(2015)\citenamefont {{Shi}},
  \citenamefont {{Li}}, \citenamefont {{Han}}, \citenamefont {{Gao}},\ and\
  \citenamefont {{Hellwing}}}]{Shi2015}%
  \BibitemOpen
  \bibfield  {author} {\bibinfo {author} {\bibfnamefont {D.}~\bibnamefont
  {{Shi}}}, \bibinfo {author} {\bibfnamefont {B.}~\bibnamefont {{Li}}},
  \bibinfo {author} {\bibfnamefont {J.}~\bibnamefont {{Han}}}, \bibinfo
  {author} {\bibfnamefont {L.}~\bibnamefont {{Gao}}}, \ and\ \bibinfo {author}
  {\bibfnamefont {W.~A.}\ \bibnamefont {{Hellwing}}},\ }\href {\doibase
  10.1093/mnras/stv1549} {\bibfield  {journal} {\bibinfo  {journal} {\mnras}\
  }\textbf {\bibinfo {volume} {452}},\ \bibinfo {pages} {3179} (\bibinfo {year}
  {2015})},\ \Eprint {http://arxiv.org/abs/1503.01109} {arXiv:1503.01109
  [astro-ph.CO]} \BibitemShut {NoStop}%
\bibitem [{\citenamefont {{Ruan}}\ \emph {et~al.}(2023)\citenamefont {{Ruan}},
  \citenamefont {{Cuesta-Lazaro}}, \citenamefont {{Eggemeier}}, \citenamefont
  {{Li}}, \citenamefont {{Baugh}}, \citenamefont {{Arnold}}, \citenamefont
  {{Bose}}, \citenamefont {{Hern{\'a}ndez-Aguayo}}, \citenamefont {{Zarrouk}},\
  and\ \citenamefont {{Davies}}}]{Ruan2023}%
  \BibitemOpen
  \bibfield  {author} {\bibinfo {author} {\bibfnamefont {C.-Z.}\ \bibnamefont
  {{Ruan}}}, \bibinfo {author} {\bibfnamefont {C.}~\bibnamefont
  {{Cuesta-Lazaro}}}, \bibinfo {author} {\bibfnamefont {A.}~\bibnamefont
  {{Eggemeier}}}, \bibinfo {author} {\bibfnamefont {B.}~\bibnamefont {{Li}}},
  \bibinfo {author} {\bibfnamefont {C.~M.}\ \bibnamefont {{Baugh}}}, \bibinfo
  {author} {\bibfnamefont {C.}~\bibnamefont {{Arnold}}}, \bibinfo {author}
  {\bibfnamefont {S.}~\bibnamefont {{Bose}}}, \bibinfo {author} {\bibfnamefont
  {C.}~\bibnamefont {{Hern{\'a}ndez-Aguayo}}}, \bibinfo {author} {\bibfnamefont
  {P.}~\bibnamefont {{Zarrouk}}}, \ and\ \bibinfo {author} {\bibfnamefont
  {C.~T.}\ \bibnamefont {{Davies}}},\ }\href {\doibase
  10.48550/arXiv.2301.02970} {\bibfield  {journal} {\bibinfo  {journal} {arXiv
  e-prints}\ ,\ \bibinfo {eid} {arXiv:2301.02970}} (\bibinfo {year} {2023})},\
  \Eprint {http://arxiv.org/abs/2301.02970} {arXiv:2301.02970 [astro-ph.CO]}
  \BibitemShut {NoStop}%
\bibitem [{\citenamefont {{Casas}}\ \emph {et~al.}(2017)\citenamefont
  {{Casas}}, \citenamefont {{Kunz}}, \citenamefont {{Martinelli}},\ and\
  \citenamefont {{Pettorino}}}]{ref:Casas2017}%
  \BibitemOpen
  \bibfield  {author} {\bibinfo {author} {\bibfnamefont {S.}~\bibnamefont
  {{Casas}}}, \bibinfo {author} {\bibfnamefont {M.}~\bibnamefont {{Kunz}}},
  \bibinfo {author} {\bibfnamefont {M.}~\bibnamefont {{Martinelli}}}, \ and\
  \bibinfo {author} {\bibfnamefont {V.}~\bibnamefont {{Pettorino}}},\ }\href
  {\doibase 10.1016/j.dark.2017.09.009} {\bibfield  {journal} {\bibinfo
  {journal} {Physics of the Dark Universe}\ }\textbf {\bibinfo {volume} {18}},\
  \bibinfo {pages} {73} (\bibinfo {year} {2017})},\ \Eprint
  {http://arxiv.org/abs/1703.01271} {arXiv:1703.01271 [astro-ph.CO]}
  \BibitemShut {NoStop}%
\bibitem [{\citenamefont {{McDonald}}\ \emph {et~al.}(2006)\citenamefont
  {{McDonald}}, \citenamefont {{Trac}},\ and\ \citenamefont
  {{Contaldi}}}]{ref:McDonaldRatio2006}%
  \BibitemOpen
  \bibfield  {author} {\bibinfo {author} {\bibfnamefont {P.}~\bibnamefont
  {{McDonald}}}, \bibinfo {author} {\bibfnamefont {H.}~\bibnamefont {{Trac}}},
  \ and\ \bibinfo {author} {\bibfnamefont {C.}~\bibnamefont {{Contaldi}}},\
  }\href {\doibase 10.1111/j.1365-2966.2005.09881.x} {\bibfield  {journal}
  {\bibinfo  {journal} {\mnras}\ }\textbf {\bibinfo {volume} {366}},\ \bibinfo
  {pages} {547} (\bibinfo {year} {2006})},\ \Eprint
  {http://arxiv.org/abs/astro-ph/0505565} {arXiv:astro-ph/0505565 [astro-ph]}
  \BibitemShut {NoStop}%
\bibitem [{\citenamefont {Mancini}\ and\ \citenamefont
  {Bose}(2023)}]{mancini2023degeneracies}%
  \BibitemOpen
  \bibfield  {author} {\bibinfo {author} {\bibfnamefont {A.~S.}\ \bibnamefont
  {Mancini}}\ and\ \bibinfo {author} {\bibfnamefont {B.}~\bibnamefont {Bose}},\
  }\href@noop {} {\enquote {\bibinfo {title} {On the degeneracies between
  baryons, massive neutrinos and f(r) gravity in stage iv cosmic shear
  analyses},}\ } (\bibinfo {year} {2023}),\ \Eprint
  {http://arxiv.org/abs/2305.06350} {arXiv:2305.06350 [astro-ph.CO]}
  \BibitemShut {NoStop}%
\bibitem [{\citenamefont {{Smail}}\ \emph {et~al.}(1994)\citenamefont
  {{Smail}}, \citenamefont {{Ellis}},\ and\ \citenamefont
  {{Fitchett}}}]{ref:Smail1994}%
  \BibitemOpen
  \bibfield  {author} {\bibinfo {author} {\bibfnamefont {I.}~\bibnamefont
  {{Smail}}}, \bibinfo {author} {\bibfnamefont {R.~S.}\ \bibnamefont
  {{Ellis}}}, \ and\ \bibinfo {author} {\bibfnamefont {M.~J.}\ \bibnamefont
  {{Fitchett}}},\ }\href {\doibase 10.1093/mnras/270.2.245} {\bibfield
  {journal} {\bibinfo  {journal} {\mnras}\ }\textbf {\bibinfo {volume} {270}},\
  \bibinfo {pages} {245} (\bibinfo {year} {1994})},\ \Eprint
  {http://arxiv.org/abs/astro-ph/9402048} {arXiv:astro-ph/9402048 [astro-ph]}
  \BibitemShut {NoStop}%
\bibitem [{\citenamefont {{Hu}}(1999)}]{ref:Hu1999}%
  \BibitemOpen
  \bibfield  {author} {\bibinfo {author} {\bibfnamefont {W.}~\bibnamefont
  {{Hu}}},\ }\href {\doibase 10.1086/312210} {\bibfield  {journal} {\bibinfo
  {journal} {\apjl}\ }\textbf {\bibinfo {volume} {522}},\ \bibinfo {pages}
  {L21} (\bibinfo {year} {1999})},\ \Eprint
  {http://arxiv.org/abs/astro-ph/9904153} {arXiv:astro-ph/9904153 [astro-ph]}
  \BibitemShut {NoStop}%
\bibitem [{\citenamefont {{Castro}}\ \emph {et~al.}(2005)\citenamefont
  {{Castro}}, \citenamefont {{Heavens}},\ and\ \citenamefont
  {{Kitching}}}]{ref:Castro2005}%
  \BibitemOpen
  \bibfield  {author} {\bibinfo {author} {\bibfnamefont {P.~G.}\ \bibnamefont
  {{Castro}}}, \bibinfo {author} {\bibfnamefont {A.~F.}\ \bibnamefont
  {{Heavens}}}, \ and\ \bibinfo {author} {\bibfnamefont {T.~D.}\ \bibnamefont
  {{Kitching}}},\ }\href {\doibase 10.1103/PhysRevD.72.023516} {\bibfield
  {journal} {\bibinfo  {journal} {\prd}\ }\textbf {\bibinfo {volume} {72}},\
  \bibinfo {eid} {023516} (\bibinfo {year} {2005})},\ \Eprint
  {http://arxiv.org/abs/astro-ph/0503479} {arXiv:astro-ph/0503479 [astro-ph]}
  \BibitemShut {NoStop}%
\bibitem [{\citenamefont {{Spurio Mancini}}\ \emph {et~al.}(2018)\citenamefont
  {{Spurio Mancini}}, \citenamefont {{Reischke}}, \citenamefont {{Pettorino}},
  \citenamefont {{Sch{\"a}fer}},\ and\ \citenamefont
  {{Zumalac{\'a}rregui}}}]{ref:SpurioMancini2018}%
  \BibitemOpen
  \bibfield  {author} {\bibinfo {author} {\bibfnamefont {A.}~\bibnamefont
  {{Spurio Mancini}}}, \bibinfo {author} {\bibfnamefont {R.}~\bibnamefont
  {{Reischke}}}, \bibinfo {author} {\bibfnamefont {V.}~\bibnamefont
  {{Pettorino}}}, \bibinfo {author} {\bibfnamefont {B.~M.}\ \bibnamefont
  {{Sch{\"a}fer}}}, \ and\ \bibinfo {author} {\bibfnamefont {M.}~\bibnamefont
  {{Zumalac{\'a}rregui}}},\ }\href {\doibase 10.1093/mnras/sty2092} {\bibfield
  {journal} {\bibinfo  {journal} {\mnras}\ }\textbf {\bibinfo {volume} {480}},\
  \bibinfo {pages} {3725} (\bibinfo {year} {2018})},\ \Eprint
  {http://arxiv.org/abs/1801.04251} {arXiv:1801.04251 [astro-ph.CO]}
  \BibitemShut {NoStop}%
\bibitem [{\citenamefont {{Euclid Collaboration}}\ \emph
  {et~al.}(2020)\citenamefont {{Euclid Collaboration}}, \citenamefont
  {{Blanchard}}, \citenamefont {{Camera}}, \citenamefont {{Carbone}},
  \citenamefont {{Cardone}}, \citenamefont {{Casas}}, \citenamefont {{Clesse}},
  \citenamefont {{Ili{\'c}}}, \citenamefont {{Kilbinger}}, \citenamefont
  {{Kitching}}, \citenamefont {{Kunz}}, \citenamefont {{Lacasa}}, \citenamefont
  {{Linder}}, \citenamefont {{Majerotto}}, \citenamefont {{Markovi{\v{c}}}},
  \citenamefont {{Martinelli}}, \citenamefont {{Pettorino}}, \citenamefont
  {{Pourtsidou}}, \citenamefont {{Sakr}}, \citenamefont {{S{\'a}nchez}},
  \citenamefont {{Sapone}}, \citenamefont {{Tutusaus}}, \citenamefont
  {{Yahia-Cherif}}, \citenamefont {{Yankelevich}}, \citenamefont {{Andreon}},
  \citenamefont {{Aussel}}, \citenamefont {{Balaguera-Antol{\'\i}nez}},
  \citenamefont {{Baldi}}, \citenamefont {{Bardelli}}, \citenamefont
  {{Bender}}, \citenamefont {{Biviano}}, \citenamefont {{Bonino}},
  \citenamefont {{Boucaud}}, \citenamefont {{Bozzo}}, \citenamefont
  {{Branchini}}, \citenamefont {{Brau-Nogue}}, \citenamefont {{Brescia}},
  \citenamefont {{Brinchmann}}, \citenamefont {{Burigana}}, \citenamefont
  {{Cabanac}}, \citenamefont {{Capobianco}}, \citenamefont {{Cappi}},
  \citenamefont {{Carretero}}, \citenamefont {{Carvalho}}, \citenamefont
  {{Casas}}, \citenamefont {{Castander}}, \citenamefont {{Castellano}},
  \citenamefont {{Cavuoti}}, \citenamefont {{Cimatti}}, \citenamefont
  {{Cledassou}}, \citenamefont {{Colodro-Conde}}, \citenamefont {{Congedo}},
  \citenamefont {{Conselice}}, \citenamefont {{Conversi}}, \citenamefont
  {{Copin}}, \citenamefont {{Corcione}}, \citenamefont {{Coupon}},
  \citenamefont {{Courtois}}, \citenamefont {{Cropper}}, \citenamefont {{Da
  Silva}}, \citenamefont {{de la Torre}}, \citenamefont {{Di Ferdinando}},
  \citenamefont {{Dubath}}, \citenamefont {{Ducret}}, \citenamefont {{Duncan}},
  \citenamefont {{Dupac}}, \citenamefont {{Dusini}}, \citenamefont {{Fabbian}},
  \citenamefont {{Fabricius}}, \citenamefont {{Farrens}}, \citenamefont
  {{Fosalba}}, \citenamefont {{Fotopoulou}}, \citenamefont {{Fourmanoit}},
  \citenamefont {{Frailis}}, \citenamefont {{Franceschi}}, \citenamefont
  {{Franzetti}}, \citenamefont {{Fumana}}, \citenamefont {{Galeotta}},
  \citenamefont {{Gillard}}, \citenamefont {{Gillis}}, \citenamefont
  {{Giocoli}}, \citenamefont {{G{\'o}mez-Alvarez}}, \citenamefont
  {{Graci{\'a}-Carpio}}, \citenamefont {{Grupp}}, \citenamefont {{Guzzo}},
  \citenamefont {{Hoekstra}}, \citenamefont {{Hormuth}}, \citenamefont
  {{Israel}}, \citenamefont {{Jahnke}}, \citenamefont {{Keihanen}},
  \citenamefont {{Kermiche}}, \citenamefont {{Kirkpatrick}}, \citenamefont
  {{Kohley}}, \citenamefont {{Kubik}}, \citenamefont {{Kurki-Suonio}},
  \citenamefont {{Ligori}}, \citenamefont {{Lilje}}, \citenamefont {{Lloro}},
  \citenamefont {{Maino}}, \citenamefont {{Maiorano}}, \citenamefont
  {{Marggraf}}, \citenamefont {{Martinet}}, \citenamefont {{Marulli}},
  \citenamefont {{Massey}}, \citenamefont {{Medinaceli}}, \citenamefont
  {{Mei}}, \citenamefont {{Mellier}}, \citenamefont {{Metcalf}}, \citenamefont
  {{Metge}}, \citenamefont {{Meylan}}, \citenamefont {{Moresco}}, \citenamefont
  {{Moscardini}}, \citenamefont {{Munari}}, \citenamefont {{Nichol}},
  \citenamefont {{Niemi}}, \citenamefont {{Nucita}}, \citenamefont {{Padilla}},
  \citenamefont {{Paltani}}, \citenamefont {{Pasian}}, \citenamefont
  {{Percival}}, \citenamefont {{Pires}}, \citenamefont {{Polenta}},
  \citenamefont {{Poncet}}, \citenamefont {{Pozzetti}}, \citenamefont
  {{Racca}}, \citenamefont {{Raison}}, \citenamefont {{Renzi}}, \citenamefont
  {{Rhodes}}, \citenamefont {{Romelli}}, \citenamefont {{Roncarelli}},
  \citenamefont {{Rossetti}}, \citenamefont {{Saglia}}, \citenamefont
  {{Schneider}}, \citenamefont {{Scottez}}, \citenamefont {{Secroun}},
  \citenamefont {{Sirri}}, \citenamefont {{Stanco}}, \citenamefont {{Starck}},
  \citenamefont {{Sureau}}, \citenamefont {{Tallada-Cresp{\'\i}}},
  \citenamefont {{Tavagnacco}}, \citenamefont {{Taylor}}, \citenamefont
  {{Tenti}}, \citenamefont {{Tereno}}, \citenamefont {{Toledo-Moreo}},
  \citenamefont {{Torradeflot}}, \citenamefont {{Valenziano}}, \citenamefont
  {{Vassallo}}, \citenamefont {{Verdoes Kleijn}}, \citenamefont {{Viel}},
  \citenamefont {{Wang}}, \citenamefont {{Zacchei}}, \citenamefont
  {{Zoubian}},\ and\ \citenamefont {{Zucca}}}]{ref:EuclidWL}%
  \BibitemOpen
  \bibfield  {author} {\bibinfo {author} {\bibnamefont {{Euclid
  Collaboration}}}, \bibinfo {author} {\bibfnamefont {A.}~\bibnamefont
  {{Blanchard}}}, \bibinfo {author} {\bibfnamefont {S.}~\bibnamefont
  {{Camera}}}, \bibinfo {author} {\bibfnamefont {C.}~\bibnamefont {{Carbone}}},
  \bibinfo {author} {\bibfnamefont {V.~F.}\ \bibnamefont {{Cardone}}}, \bibinfo
  {author} {\bibfnamefont {S.}~\bibnamefont {{Casas}}}, \bibinfo {author}
  {\bibfnamefont {S.}~\bibnamefont {{Clesse}}}, \bibinfo {author}
  {\bibfnamefont {S.}~\bibnamefont {{Ili{\'c}}}}, \bibinfo {author}
  {\bibfnamefont {M.}~\bibnamefont {{Kilbinger}}}, \bibinfo {author}
  {\bibfnamefont {T.}~\bibnamefont {{Kitching}}}, \bibinfo {author}
  {\bibfnamefont {M.}~\bibnamefont {{Kunz}}}, \bibinfo {author} {\bibfnamefont
  {F.}~\bibnamefont {{Lacasa}}}, \bibinfo {author} {\bibfnamefont
  {E.}~\bibnamefont {{Linder}}}, \bibinfo {author} {\bibfnamefont
  {E.}~\bibnamefont {{Majerotto}}}, \bibinfo {author} {\bibfnamefont
  {K.}~\bibnamefont {{Markovi{\v{c}}}}}, \bibinfo {author} {\bibfnamefont
  {M.}~\bibnamefont {{Martinelli}}}, \bibinfo {author} {\bibfnamefont
  {V.}~\bibnamefont {{Pettorino}}}, \bibinfo {author} {\bibfnamefont
  {A.}~\bibnamefont {{Pourtsidou}}}, \bibinfo {author} {\bibfnamefont
  {Z.}~\bibnamefont {{Sakr}}}, \bibinfo {author} {\bibfnamefont {A.~G.}\
  \bibnamefont {{S{\'a}nchez}}}, \bibinfo {author} {\bibfnamefont
  {D.}~\bibnamefont {{Sapone}}}, \bibinfo {author} {\bibfnamefont
  {I.}~\bibnamefont {{Tutusaus}}}, \bibinfo {author} {\bibfnamefont
  {S.}~\bibnamefont {{Yahia-Cherif}}}, \bibinfo {author} {\bibfnamefont
  {V.}~\bibnamefont {{Yankelevich}}}, \bibinfo {author} {\bibfnamefont
  {S.}~\bibnamefont {{Andreon}}}, \bibinfo {author} {\bibfnamefont
  {H.}~\bibnamefont {{Aussel}}}, \bibinfo {author} {\bibfnamefont
  {A.}~\bibnamefont {{Balaguera-Antol{\'\i}nez}}}, \bibinfo {author}
  {\bibfnamefont {M.}~\bibnamefont {{Baldi}}}, \bibinfo {author} {\bibfnamefont
  {S.}~\bibnamefont {{Bardelli}}}, \bibinfo {author} {\bibfnamefont
  {R.}~\bibnamefont {{Bender}}}, \bibinfo {author} {\bibfnamefont
  {A.}~\bibnamefont {{Biviano}}}, \bibinfo {author} {\bibfnamefont
  {D.}~\bibnamefont {{Bonino}}}, \bibinfo {author} {\bibfnamefont
  {A.}~\bibnamefont {{Boucaud}}}, \bibinfo {author} {\bibfnamefont
  {E.}~\bibnamefont {{Bozzo}}}, \bibinfo {author} {\bibfnamefont
  {E.}~\bibnamefont {{Branchini}}}, \bibinfo {author} {\bibfnamefont
  {S.}~\bibnamefont {{Brau-Nogue}}}, \bibinfo {author} {\bibfnamefont
  {M.}~\bibnamefont {{Brescia}}}, \bibinfo {author} {\bibfnamefont
  {J.}~\bibnamefont {{Brinchmann}}}, \bibinfo {author} {\bibfnamefont
  {C.}~\bibnamefont {{Burigana}}}, \bibinfo {author} {\bibfnamefont
  {R.}~\bibnamefont {{Cabanac}}}, \bibinfo {author} {\bibfnamefont
  {V.}~\bibnamefont {{Capobianco}}}, \bibinfo {author} {\bibfnamefont
  {A.}~\bibnamefont {{Cappi}}}, \bibinfo {author} {\bibfnamefont
  {J.}~\bibnamefont {{Carretero}}}, \bibinfo {author} {\bibfnamefont {C.~S.}\
  \bibnamefont {{Carvalho}}}, \bibinfo {author} {\bibfnamefont
  {R.}~\bibnamefont {{Casas}}}, \bibinfo {author} {\bibfnamefont {F.~J.}\
  \bibnamefont {{Castander}}}, \bibinfo {author} {\bibfnamefont
  {M.}~\bibnamefont {{Castellano}}}, \bibinfo {author} {\bibfnamefont
  {S.}~\bibnamefont {{Cavuoti}}}, \bibinfo {author} {\bibfnamefont
  {A.}~\bibnamefont {{Cimatti}}}, \bibinfo {author} {\bibfnamefont
  {R.}~\bibnamefont {{Cledassou}}}, \bibinfo {author} {\bibfnamefont
  {C.}~\bibnamefont {{Colodro-Conde}}}, \bibinfo {author} {\bibfnamefont
  {G.}~\bibnamefont {{Congedo}}}, \bibinfo {author} {\bibfnamefont {C.~J.}\
  \bibnamefont {{Conselice}}}, \bibinfo {author} {\bibfnamefont
  {L.}~\bibnamefont {{Conversi}}}, \bibinfo {author} {\bibfnamefont
  {Y.}~\bibnamefont {{Copin}}}, \bibinfo {author} {\bibfnamefont
  {L.}~\bibnamefont {{Corcione}}}, \bibinfo {author} {\bibfnamefont
  {J.}~\bibnamefont {{Coupon}}}, \bibinfo {author} {\bibfnamefont {H.~M.}\
  \bibnamefont {{Courtois}}}, \bibinfo {author} {\bibfnamefont
  {M.}~\bibnamefont {{Cropper}}}, \bibinfo {author} {\bibfnamefont
  {A.}~\bibnamefont {{Da Silva}}}, \bibinfo {author} {\bibfnamefont
  {S.}~\bibnamefont {{de la Torre}}}, \bibinfo {author} {\bibfnamefont
  {D.}~\bibnamefont {{Di Ferdinando}}}, \bibinfo {author} {\bibfnamefont
  {F.}~\bibnamefont {{Dubath}}}, \bibinfo {author} {\bibfnamefont
  {F.}~\bibnamefont {{Ducret}}}, \bibinfo {author} {\bibfnamefont {C.~A.~J.}\
  \bibnamefont {{Duncan}}}, \bibinfo {author} {\bibfnamefont {X.}~\bibnamefont
  {{Dupac}}}, \bibinfo {author} {\bibfnamefont {S.}~\bibnamefont {{Dusini}}},
  \bibinfo {author} {\bibfnamefont {G.}~\bibnamefont {{Fabbian}}}, \bibinfo
  {author} {\bibfnamefont {M.}~\bibnamefont {{Fabricius}}}, \bibinfo {author}
  {\bibfnamefont {S.}~\bibnamefont {{Farrens}}}, \bibinfo {author}
  {\bibfnamefont {P.}~\bibnamefont {{Fosalba}}}, \bibinfo {author}
  {\bibfnamefont {S.}~\bibnamefont {{Fotopoulou}}}, \bibinfo {author}
  {\bibfnamefont {N.}~\bibnamefont {{Fourmanoit}}}, \bibinfo {author}
  {\bibfnamefont {M.}~\bibnamefont {{Frailis}}}, \bibinfo {author}
  {\bibfnamefont {E.}~\bibnamefont {{Franceschi}}}, \bibinfo {author}
  {\bibfnamefont {P.}~\bibnamefont {{Franzetti}}}, \bibinfo {author}
  {\bibfnamefont {M.}~\bibnamefont {{Fumana}}}, \bibinfo {author}
  {\bibfnamefont {S.}~\bibnamefont {{Galeotta}}}, \bibinfo {author}
  {\bibfnamefont {W.}~\bibnamefont {{Gillard}}}, \bibinfo {author}
  {\bibfnamefont {B.}~\bibnamefont {{Gillis}}}, \bibinfo {author}
  {\bibfnamefont {C.}~\bibnamefont {{Giocoli}}}, \bibinfo {author}
  {\bibfnamefont {P.}~\bibnamefont {{G{\'o}mez-Alvarez}}}, \bibinfo {author}
  {\bibfnamefont {J.}~\bibnamefont {{Graci{\'a}-Carpio}}}, \bibinfo {author}
  {\bibfnamefont {F.}~\bibnamefont {{Grupp}}}, \bibinfo {author} {\bibfnamefont
  {L.}~\bibnamefont {{Guzzo}}}, \bibinfo {author} {\bibfnamefont
  {H.}~\bibnamefont {{Hoekstra}}}, \bibinfo {author} {\bibfnamefont
  {F.}~\bibnamefont {{Hormuth}}}, \bibinfo {author} {\bibfnamefont
  {H.}~\bibnamefont {{Israel}}}, \bibinfo {author} {\bibfnamefont
  {K.}~\bibnamefont {{Jahnke}}}, \bibinfo {author} {\bibfnamefont
  {E.}~\bibnamefont {{Keihanen}}}, \bibinfo {author} {\bibfnamefont
  {S.}~\bibnamefont {{Kermiche}}}, \bibinfo {author} {\bibfnamefont {C.~C.}\
  \bibnamefont {{Kirkpatrick}}}, \bibinfo {author} {\bibfnamefont
  {R.}~\bibnamefont {{Kohley}}}, \bibinfo {author} {\bibfnamefont
  {B.}~\bibnamefont {{Kubik}}}, \bibinfo {author} {\bibfnamefont
  {H.}~\bibnamefont {{Kurki-Suonio}}}, \bibinfo {author} {\bibfnamefont
  {S.}~\bibnamefont {{Ligori}}}, \bibinfo {author} {\bibfnamefont {P.~B.}\
  \bibnamefont {{Lilje}}}, \bibinfo {author} {\bibfnamefont {I.}~\bibnamefont
  {{Lloro}}}, \bibinfo {author} {\bibfnamefont {D.}~\bibnamefont {{Maino}}},
  \bibinfo {author} {\bibfnamefont {E.}~\bibnamefont {{Maiorano}}}, \bibinfo
  {author} {\bibfnamefont {O.}~\bibnamefont {{Marggraf}}}, \bibinfo {author}
  {\bibfnamefont {N.}~\bibnamefont {{Martinet}}}, \bibinfo {author}
  {\bibfnamefont {F.}~\bibnamefont {{Marulli}}}, \bibinfo {author}
  {\bibfnamefont {R.}~\bibnamefont {{Massey}}}, \bibinfo {author}
  {\bibfnamefont {E.}~\bibnamefont {{Medinaceli}}}, \bibinfo {author}
  {\bibfnamefont {S.}~\bibnamefont {{Mei}}}, \bibinfo {author} {\bibfnamefont
  {Y.}~\bibnamefont {{Mellier}}}, \bibinfo {author} {\bibfnamefont
  {B.}~\bibnamefont {{Metcalf}}}, \bibinfo {author} {\bibfnamefont {J.~J.}\
  \bibnamefont {{Metge}}}, \bibinfo {author} {\bibfnamefont {G.}~\bibnamefont
  {{Meylan}}}, \bibinfo {author} {\bibfnamefont {M.}~\bibnamefont {{Moresco}}},
  \bibinfo {author} {\bibfnamefont {L.}~\bibnamefont {{Moscardini}}}, \bibinfo
  {author} {\bibfnamefont {E.}~\bibnamefont {{Munari}}}, \bibinfo {author}
  {\bibfnamefont {R.~C.}\ \bibnamefont {{Nichol}}}, \bibinfo {author}
  {\bibfnamefont {S.}~\bibnamefont {{Niemi}}}, \bibinfo {author} {\bibfnamefont
  {A.~A.}\ \bibnamefont {{Nucita}}}, \bibinfo {author} {\bibfnamefont
  {C.}~\bibnamefont {{Padilla}}}, \bibinfo {author} {\bibfnamefont
  {S.}~\bibnamefont {{Paltani}}}, \bibinfo {author} {\bibfnamefont
  {F.}~\bibnamefont {{Pasian}}}, \bibinfo {author} {\bibfnamefont {W.~J.}\
  \bibnamefont {{Percival}}}, \bibinfo {author} {\bibfnamefont
  {S.}~\bibnamefont {{Pires}}}, \bibinfo {author} {\bibfnamefont
  {G.}~\bibnamefont {{Polenta}}}, \bibinfo {author} {\bibfnamefont
  {M.}~\bibnamefont {{Poncet}}}, \bibinfo {author} {\bibfnamefont
  {L.}~\bibnamefont {{Pozzetti}}}, \bibinfo {author} {\bibfnamefont {G.~D.}\
  \bibnamefont {{Racca}}}, \bibinfo {author} {\bibfnamefont {F.}~\bibnamefont
  {{Raison}}}, \bibinfo {author} {\bibfnamefont {A.}~\bibnamefont {{Renzi}}},
  \bibinfo {author} {\bibfnamefont {J.}~\bibnamefont {{Rhodes}}}, \bibinfo
  {author} {\bibfnamefont {E.}~\bibnamefont {{Romelli}}}, \bibinfo {author}
  {\bibfnamefont {M.}~\bibnamefont {{Roncarelli}}}, \bibinfo {author}
  {\bibfnamefont {E.}~\bibnamefont {{Rossetti}}}, \bibinfo {author}
  {\bibfnamefont {R.}~\bibnamefont {{Saglia}}}, \bibinfo {author}
  {\bibfnamefont {P.}~\bibnamefont {{Schneider}}}, \bibinfo {author}
  {\bibfnamefont {V.}~\bibnamefont {{Scottez}}}, \bibinfo {author}
  {\bibfnamefont {A.}~\bibnamefont {{Secroun}}}, \bibinfo {author}
  {\bibfnamefont {G.}~\bibnamefont {{Sirri}}}, \bibinfo {author} {\bibfnamefont
  {L.}~\bibnamefont {{Stanco}}}, \bibinfo {author} {\bibfnamefont {J.~L.}\
  \bibnamefont {{Starck}}}, \bibinfo {author} {\bibfnamefont {F.}~\bibnamefont
  {{Sureau}}}, \bibinfo {author} {\bibfnamefont {P.}~\bibnamefont
  {{Tallada-Cresp{\'\i}}}}, \bibinfo {author} {\bibfnamefont {D.}~\bibnamefont
  {{Tavagnacco}}}, \bibinfo {author} {\bibfnamefont {A.~N.}\ \bibnamefont
  {{Taylor}}}, \bibinfo {author} {\bibfnamefont {M.}~\bibnamefont {{Tenti}}},
  \bibinfo {author} {\bibfnamefont {I.}~\bibnamefont {{Tereno}}}, \bibinfo
  {author} {\bibfnamefont {R.}~\bibnamefont {{Toledo-Moreo}}}, \bibinfo
  {author} {\bibfnamefont {F.}~\bibnamefont {{Torradeflot}}}, \bibinfo {author}
  {\bibfnamefont {L.}~\bibnamefont {{Valenziano}}}, \bibinfo {author}
  {\bibfnamefont {T.}~\bibnamefont {{Vassallo}}}, \bibinfo {author}
  {\bibfnamefont {G.~A.}\ \bibnamefont {{Verdoes Kleijn}}}, \bibinfo {author}
  {\bibfnamefont {M.}~\bibnamefont {{Viel}}}, \bibinfo {author} {\bibfnamefont
  {Y.}~\bibnamefont {{Wang}}}, \bibinfo {author} {\bibfnamefont
  {A.}~\bibnamefont {{Zacchei}}}, \bibinfo {author} {\bibfnamefont
  {J.}~\bibnamefont {{Zoubian}}}, \ and\ \bibinfo {author} {\bibfnamefont
  {E.}~\bibnamefont {{Zucca}}},\ }\href {\doibase 10.1051/0004-6361/202038071}
  {\bibfield  {journal} {\bibinfo  {journal} {\aap}\ }\textbf {\bibinfo
  {volume} {642}},\ \bibinfo {eid} {A191} (\bibinfo {year} {2020})},\ \Eprint
  {http://arxiv.org/abs/1910.09273} {arXiv:1910.09273 [astro-ph.CO]}
  \BibitemShut {NoStop}%
\bibitem [{\citenamefont {Ruan}\ \emph {et~al.}(2022)\citenamefont {Ruan},
  \citenamefont {Cuesta-Lazaro}, \citenamefont {Eggemeier}, \citenamefont
  {Hern\'andez-Aguayo}, \citenamefont {Baugh}, \citenamefont {Li},\ and\
  \citenamefont {Prada}}]{Ruan:2021rqv}%
  \BibitemOpen
  \bibfield  {author} {\bibinfo {author} {\bibfnamefont {C.-Z.}\ \bibnamefont
  {Ruan}}, \bibinfo {author} {\bibfnamefont {C.}~\bibnamefont {Cuesta-Lazaro}},
  \bibinfo {author} {\bibfnamefont {A.}~\bibnamefont {Eggemeier}}, \bibinfo
  {author} {\bibfnamefont {C.}~\bibnamefont {Hern\'andez-Aguayo}}, \bibinfo
  {author} {\bibfnamefont {C.~M.}\ \bibnamefont {Baugh}}, \bibinfo {author}
  {\bibfnamefont {B.}~\bibnamefont {Li}}, \ and\ \bibinfo {author}
  {\bibfnamefont {F.}~\bibnamefont {Prada}},\ }\href {\doibase
  10.1093/mnras/stac1345} {\bibfield  {journal} {\bibinfo  {journal} {Mon. Not.
  Roy. Astron. Soc.}\ }\textbf {\bibinfo {volume} {514}},\ \bibinfo {pages}
  {440} (\bibinfo {year} {2022})},\ \Eprint {http://arxiv.org/abs/2110.10033}
  {arXiv:2110.10033 [astro-ph.CO]} \BibitemShut {NoStop}%
\bibitem [{\citenamefont {Thomas}\ \emph {et~al.}(2023)\citenamefont {Thomas},
  \citenamefont {Clifton},\ and\ \citenamefont {Anton}}]{Thomas_2023}%
  \BibitemOpen
  \bibfield  {author} {\bibinfo {author} {\bibfnamefont {D.~B.}\ \bibnamefont
  {Thomas}}, \bibinfo {author} {\bibfnamefont {T.}~\bibnamefont {Clifton}}, \
  and\ \bibinfo {author} {\bibfnamefont {T.}~\bibnamefont {Anton}},\ }\href
  {\doibase 10.1088/1475-7516/2023/04/016} {\bibfield  {journal} {\bibinfo
  {journal} {Journal of Cosmology and Astroparticle Physics}\ }\textbf
  {\bibinfo {volume} {2023}},\ \bibinfo {pages} {016} (\bibinfo {year}
  {2023})}\BibitemShut {NoStop}%
\bibitem [{\citenamefont {Clifton}\ and\ \citenamefont
  {Sanghai}(2019)}]{Clifton_2019}%
  \BibitemOpen
  \bibfield  {author} {\bibinfo {author} {\bibfnamefont {T.}~\bibnamefont
  {Clifton}}\ and\ \bibinfo {author} {\bibfnamefont {V.~A.}\ \bibnamefont
  {Sanghai}},\ }\href {\doibase 10.1103/physrevlett.122.011301} {\bibfield
  {journal} {\bibinfo  {journal} {Physical Review Letters}\ }\textbf {\bibinfo
  {volume} {122}} (\bibinfo {year} {2019}),\
  10.1103/physrevlett.122.011301}\BibitemShut {NoStop}%
\bibitem [{\citenamefont {Sanghai}\ and\ \citenamefont
  {Clifton}(2015)}]{Sanghai_2015}%
  \BibitemOpen
  \bibfield  {author} {\bibinfo {author} {\bibfnamefont {V.~A.}\ \bibnamefont
  {Sanghai}}\ and\ \bibinfo {author} {\bibfnamefont {T.}~\bibnamefont
  {Clifton}},\ }\href {\doibase 10.1103/physrevd.91.103532} {\bibfield
  {journal} {\bibinfo  {journal} {Physical Review D}\ }\textbf {\bibinfo
  {volume} {91}} (\bibinfo {year} {2015}),\
  10.1103/physrevd.91.103532}\BibitemShut {NoStop}%
\bibitem [{\citenamefont {Troxel}\ \emph {et~al.}(2018)\citenamefont {Troxel},
  \citenamefont {MacCrann}, \citenamefont {Zuntz}, \citenamefont {Eifler},
  \citenamefont {Krause}, \citenamefont {Dodelson}, \citenamefont {Gruen},
  \citenamefont {Blazek}, \citenamefont {Friedrich}, \citenamefont {Samuroff},
  \citenamefont {Prat}, \citenamefont {Secco}, \citenamefont {Davis},
  \citenamefont {Fert{\'{e}}}, \citenamefont {DeRose}, \citenamefont {Alarcon},
  \citenamefont {Amara}, \citenamefont {Baxter}, \citenamefont {Becker},
  \citenamefont {Bernstein}, \citenamefont {Bridle}, \citenamefont {Cawthon},
  \citenamefont {Chang}, \citenamefont {Choi}, \citenamefont {Vicente},
  \citenamefont {Drlica-Wagner}, \citenamefont {Elvin-Poole}, \citenamefont
  {Frieman}, \citenamefont {Gatti}, \citenamefont {Hartley}, \citenamefont
  {Honscheid}, \citenamefont {Hoyle}, \citenamefont {Huff}, \citenamefont
  {Huterer}, \citenamefont {Jain}, \citenamefont {Jarvis}, \citenamefont
  {Kacprzak}, \citenamefont {Kirk}, \citenamefont {Kokron}, \citenamefont
  {Krawiec}, \citenamefont {Lahav}, \citenamefont {Liddle}, \citenamefont
  {Peacock}, \citenamefont {Rau}, \citenamefont {Refregier}, \citenamefont
  {Rollins}, \citenamefont {Rozo}, \citenamefont {Rykoff}, \citenamefont
  {S{\'{a}}nchez}, \citenamefont {Sevilla-Noarbe}, \citenamefont {Sheldon},
  \citenamefont {Stebbins}, \citenamefont {Varga}, \citenamefont {Vielzeuf},
  \citenamefont {Wang}, \citenamefont {Wechsler}, \citenamefont {Yanny},
  \citenamefont {Abbott}, \citenamefont {Abdalla}, \citenamefont {Allam},
  \citenamefont {Annis}, \citenamefont {Bechtol}, \citenamefont
  {Benoit-L{\'{e}}vy}, \citenamefont {Bertin}, \citenamefont {Brooks},
  \citenamefont {Buckley-Geer}, \citenamefont {Burke}, \citenamefont {Rosell},
  \citenamefont {Kind}, \citenamefont {Carretero}, \citenamefont {Castander},
  \citenamefont {Crocce}, \citenamefont {Cunha}, \citenamefont {D'Andrea},
  \citenamefont {da~Costa}, \citenamefont {DePoy}, \citenamefont {Desai},
  \citenamefont {Diehl}, \citenamefont {Dietrich}, \citenamefont {Doel},
  \citenamefont {Fernandez}, \citenamefont {Flaugher}, \citenamefont {Fosalba},
  \citenamefont {Garc{\'{\i}}a-Bellido}, \citenamefont {Gaztanaga},
  \citenamefont {Gerdes}, \citenamefont {Giannantonio}, \citenamefont
  {Goldstein}, \citenamefont {Gruendl}, \citenamefont {Gschwend}, \citenamefont
  {Gutierrez}, \citenamefont {James}, \citenamefont {Jeltema}, \citenamefont
  {Johnson}, \citenamefont {Johnson}, \citenamefont {Kent}, \citenamefont
  {Kuehn}, \citenamefont {Kuhlmann}, \citenamefont {Kuropatkin}, \citenamefont
  {Li}, \citenamefont {Lima}, \citenamefont {Lin}, \citenamefont {Maia},
  \citenamefont {March}, \citenamefont {Marshall}, \citenamefont {Martini},
  \citenamefont {Melchior}, \citenamefont {Menanteau}, \citenamefont {Miquel},
  \citenamefont {Mohr}, \citenamefont {Neilsen}, \citenamefont {Nichol},
  \citenamefont {Nord}, \citenamefont {Petravick}, \citenamefont {Plazas},
  \citenamefont {Romer}, \citenamefont {Roodman}, \citenamefont {Sako},
  \citenamefont {Sanchez}, \citenamefont {Scarpine}, \citenamefont {Schindler},
  \citenamefont {Schubnell}, \citenamefont {Smith}, \citenamefont {Smith},
  \citenamefont {Soares-Santos}, \citenamefont {Sobreira}, \citenamefont
  {Suchyta}, \citenamefont {Swanson}, \citenamefont {Tarle}, \citenamefont
  {Thomas}, \citenamefont {Tucker}, \citenamefont {Vikram}, \citenamefont
  {Walker}, \citenamefont {Weller},\ and\ \citenamefont {Zhang}}]{Troxel_2018}%
  \BibitemOpen
  \bibfield  {author} {\bibinfo {author} {\bibfnamefont {M.}~\bibnamefont
  {Troxel}}, \bibinfo {author} {\bibfnamefont {N.}~\bibnamefont {MacCrann}},
  \bibinfo {author} {\bibfnamefont {J.}~\bibnamefont {Zuntz}}, \bibinfo
  {author} {\bibfnamefont {T.}~\bibnamefont {Eifler}}, \bibinfo {author}
  {\bibfnamefont {E.}~\bibnamefont {Krause}}, \bibinfo {author} {\bibfnamefont
  {S.}~\bibnamefont {Dodelson}}, \bibinfo {author} {\bibfnamefont
  {D.}~\bibnamefont {Gruen}}, \bibinfo {author} {\bibfnamefont
  {J.}~\bibnamefont {Blazek}}, \bibinfo {author} {\bibfnamefont
  {O.}~\bibnamefont {Friedrich}}, \bibinfo {author} {\bibfnamefont
  {S.}~\bibnamefont {Samuroff}}, \bibinfo {author} {\bibfnamefont
  {J.}~\bibnamefont {Prat}}, \bibinfo {author} {\bibfnamefont {L.}~\bibnamefont
  {Secco}}, \bibinfo {author} {\bibfnamefont {C.}~\bibnamefont {Davis}},
  \bibinfo {author} {\bibfnamefont {A.}~\bibnamefont {Fert{\'{e}}}}, \bibinfo
  {author} {\bibfnamefont {J.}~\bibnamefont {DeRose}}, \bibinfo {author}
  {\bibfnamefont {A.}~\bibnamefont {Alarcon}}, \bibinfo {author} {\bibfnamefont
  {A.}~\bibnamefont {Amara}}, \bibinfo {author} {\bibfnamefont
  {E.}~\bibnamefont {Baxter}}, \bibinfo {author} {\bibfnamefont
  {M.}~\bibnamefont {Becker}}, \bibinfo {author} {\bibfnamefont
  {G.}~\bibnamefont {Bernstein}}, \bibinfo {author} {\bibfnamefont
  {S.}~\bibnamefont {Bridle}}, \bibinfo {author} {\bibfnamefont
  {R.}~\bibnamefont {Cawthon}}, \bibinfo {author} {\bibfnamefont
  {C.}~\bibnamefont {Chang}}, \bibinfo {author} {\bibfnamefont
  {A.}~\bibnamefont {Choi}}, \bibinfo {author} {\bibfnamefont {J.~D.}\
  \bibnamefont {Vicente}}, \bibinfo {author} {\bibfnamefont {A.}~\bibnamefont
  {Drlica-Wagner}}, \bibinfo {author} {\bibfnamefont {J.}~\bibnamefont
  {Elvin-Poole}}, \bibinfo {author} {\bibfnamefont {J.}~\bibnamefont
  {Frieman}}, \bibinfo {author} {\bibfnamefont {M.}~\bibnamefont {Gatti}},
  \bibinfo {author} {\bibfnamefont {W.}~\bibnamefont {Hartley}}, \bibinfo
  {author} {\bibfnamefont {K.}~\bibnamefont {Honscheid}}, \bibinfo {author}
  {\bibfnamefont {B.}~\bibnamefont {Hoyle}}, \bibinfo {author} {\bibfnamefont
  {E.}~\bibnamefont {Huff}}, \bibinfo {author} {\bibfnamefont {D.}~\bibnamefont
  {Huterer}}, \bibinfo {author} {\bibfnamefont {B.}~\bibnamefont {Jain}},
  \bibinfo {author} {\bibfnamefont {M.}~\bibnamefont {Jarvis}}, \bibinfo
  {author} {\bibfnamefont {T.}~\bibnamefont {Kacprzak}}, \bibinfo {author}
  {\bibfnamefont {D.}~\bibnamefont {Kirk}}, \bibinfo {author} {\bibfnamefont
  {N.}~\bibnamefont {Kokron}}, \bibinfo {author} {\bibfnamefont
  {C.}~\bibnamefont {Krawiec}}, \bibinfo {author} {\bibfnamefont
  {O.}~\bibnamefont {Lahav}}, \bibinfo {author} {\bibfnamefont
  {A.}~\bibnamefont {Liddle}}, \bibinfo {author} {\bibfnamefont
  {J.}~\bibnamefont {Peacock}}, \bibinfo {author} {\bibfnamefont
  {M.}~\bibnamefont {Rau}}, \bibinfo {author} {\bibfnamefont {A.}~\bibnamefont
  {Refregier}}, \bibinfo {author} {\bibfnamefont {R.}~\bibnamefont {Rollins}},
  \bibinfo {author} {\bibfnamefont {E.}~\bibnamefont {Rozo}}, \bibinfo {author}
  {\bibfnamefont {E.}~\bibnamefont {Rykoff}}, \bibinfo {author} {\bibfnamefont
  {C.}~\bibnamefont {S{\'{a}}nchez}}, \bibinfo {author} {\bibfnamefont
  {I.}~\bibnamefont {Sevilla-Noarbe}}, \bibinfo {author} {\bibfnamefont
  {E.}~\bibnamefont {Sheldon}}, \bibinfo {author} {\bibfnamefont
  {A.}~\bibnamefont {Stebbins}}, \bibinfo {author} {\bibfnamefont
  {T.}~\bibnamefont {Varga}}, \bibinfo {author} {\bibfnamefont
  {P.}~\bibnamefont {Vielzeuf}}, \bibinfo {author} {\bibfnamefont
  {M.}~\bibnamefont {Wang}}, \bibinfo {author} {\bibfnamefont {R.}~\bibnamefont
  {Wechsler}}, \bibinfo {author} {\bibfnamefont {B.}~\bibnamefont {Yanny}},
  \bibinfo {author} {\bibfnamefont {T.}~\bibnamefont {Abbott}}, \bibinfo
  {author} {\bibfnamefont {F.}~\bibnamefont {Abdalla}}, \bibinfo {author}
  {\bibfnamefont {S.}~\bibnamefont {Allam}}, \bibinfo {author} {\bibfnamefont
  {J.}~\bibnamefont {Annis}}, \bibinfo {author} {\bibfnamefont
  {K.}~\bibnamefont {Bechtol}}, \bibinfo {author} {\bibfnamefont
  {A.}~\bibnamefont {Benoit-L{\'{e}}vy}}, \bibinfo {author} {\bibfnamefont
  {E.}~\bibnamefont {Bertin}}, \bibinfo {author} {\bibfnamefont
  {D.}~\bibnamefont {Brooks}}, \bibinfo {author} {\bibfnamefont
  {E.}~\bibnamefont {Buckley-Geer}}, \bibinfo {author} {\bibfnamefont
  {D.}~\bibnamefont {Burke}}, \bibinfo {author} {\bibfnamefont {A.~C.}\
  \bibnamefont {Rosell}}, \bibinfo {author} {\bibfnamefont {M.~C.}\
  \bibnamefont {Kind}}, \bibinfo {author} {\bibfnamefont {J.}~\bibnamefont
  {Carretero}}, \bibinfo {author} {\bibfnamefont {F.}~\bibnamefont
  {Castander}}, \bibinfo {author} {\bibfnamefont {M.}~\bibnamefont {Crocce}},
  \bibinfo {author} {\bibfnamefont {C.}~\bibnamefont {Cunha}}, \bibinfo
  {author} {\bibfnamefont {C.}~\bibnamefont {D'Andrea}}, \bibinfo {author}
  {\bibfnamefont {L.}~\bibnamefont {da~Costa}}, \bibinfo {author}
  {\bibfnamefont {D.}~\bibnamefont {DePoy}}, \bibinfo {author} {\bibfnamefont
  {S.}~\bibnamefont {Desai}}, \bibinfo {author} {\bibfnamefont
  {H.}~\bibnamefont {Diehl}}, \bibinfo {author} {\bibfnamefont
  {J.}~\bibnamefont {Dietrich}}, \bibinfo {author} {\bibfnamefont
  {P.}~\bibnamefont {Doel}}, \bibinfo {author} {\bibfnamefont {E.}~\bibnamefont
  {Fernandez}}, \bibinfo {author} {\bibfnamefont {B.}~\bibnamefont {Flaugher}},
  \bibinfo {author} {\bibfnamefont {P.}~\bibnamefont {Fosalba}}, \bibinfo
  {author} {\bibfnamefont {J.}~\bibnamefont {Garc{\'{\i}}a-Bellido}}, \bibinfo
  {author} {\bibfnamefont {E.}~\bibnamefont {Gaztanaga}}, \bibinfo {author}
  {\bibfnamefont {D.}~\bibnamefont {Gerdes}}, \bibinfo {author} {\bibfnamefont
  {T.}~\bibnamefont {Giannantonio}}, \bibinfo {author} {\bibfnamefont
  {D.}~\bibnamefont {Goldstein}}, \bibinfo {author} {\bibfnamefont
  {R.}~\bibnamefont {Gruendl}}, \bibinfo {author} {\bibfnamefont
  {J.}~\bibnamefont {Gschwend}}, \bibinfo {author} {\bibfnamefont
  {G.}~\bibnamefont {Gutierrez}}, \bibinfo {author} {\bibfnamefont
  {D.}~\bibnamefont {James}}, \bibinfo {author} {\bibfnamefont
  {T.}~\bibnamefont {Jeltema}}, \bibinfo {author} {\bibfnamefont
  {M.}~\bibnamefont {Johnson}}, \bibinfo {author} {\bibfnamefont
  {M.}~\bibnamefont {Johnson}}, \bibinfo {author} {\bibfnamefont
  {S.}~\bibnamefont {Kent}}, \bibinfo {author} {\bibfnamefont {K.}~\bibnamefont
  {Kuehn}}, \bibinfo {author} {\bibfnamefont {S.}~\bibnamefont {Kuhlmann}},
  \bibinfo {author} {\bibfnamefont {N.}~\bibnamefont {Kuropatkin}}, \bibinfo
  {author} {\bibfnamefont {T.}~\bibnamefont {Li}}, \bibinfo {author}
  {\bibfnamefont {M.}~\bibnamefont {Lima}}, \bibinfo {author} {\bibfnamefont
  {H.}~\bibnamefont {Lin}}, \bibinfo {author} {\bibfnamefont {M.}~\bibnamefont
  {Maia}}, \bibinfo {author} {\bibfnamefont {M.}~\bibnamefont {March}},
  \bibinfo {author} {\bibfnamefont {J.}~\bibnamefont {Marshall}}, \bibinfo
  {author} {\bibfnamefont {P.}~\bibnamefont {Martini}}, \bibinfo {author}
  {\bibfnamefont {P.}~\bibnamefont {Melchior}}, \bibinfo {author}
  {\bibfnamefont {F.}~\bibnamefont {Menanteau}}, \bibinfo {author}
  {\bibfnamefont {R.}~\bibnamefont {Miquel}}, \bibinfo {author} {\bibfnamefont
  {J.}~\bibnamefont {Mohr}}, \bibinfo {author} {\bibfnamefont {E.}~\bibnamefont
  {Neilsen}}, \bibinfo {author} {\bibfnamefont {R.}~\bibnamefont {Nichol}},
  \bibinfo {author} {\bibfnamefont {B.}~\bibnamefont {Nord}}, \bibinfo {author}
  {\bibfnamefont {D.}~\bibnamefont {Petravick}}, \bibinfo {author}
  {\bibfnamefont {A.}~\bibnamefont {Plazas}}, \bibinfo {author} {\bibfnamefont
  {A.}~\bibnamefont {Romer}}, \bibinfo {author} {\bibfnamefont
  {A.}~\bibnamefont {Roodman}}, \bibinfo {author} {\bibfnamefont
  {M.}~\bibnamefont {Sako}}, \bibinfo {author} {\bibfnamefont {E.}~\bibnamefont
  {Sanchez}}, \bibinfo {author} {\bibfnamefont {V.}~\bibnamefont {Scarpine}},
  \bibinfo {author} {\bibfnamefont {R.}~\bibnamefont {Schindler}}, \bibinfo
  {author} {\bibfnamefont {M.}~\bibnamefont {Schubnell}}, \bibinfo {author}
  {\bibfnamefont {M.}~\bibnamefont {Smith}}, \bibinfo {author} {\bibfnamefont
  {R.}~\bibnamefont {Smith}}, \bibinfo {author} {\bibfnamefont
  {M.}~\bibnamefont {Soares-Santos}}, \bibinfo {author} {\bibfnamefont
  {F.}~\bibnamefont {Sobreira}}, \bibinfo {author} {\bibfnamefont
  {E.}~\bibnamefont {Suchyta}}, \bibinfo {author} {\bibfnamefont
  {M.}~\bibnamefont {Swanson}}, \bibinfo {author} {\bibfnamefont
  {G.}~\bibnamefont {Tarle}}, \bibinfo {author} {\bibfnamefont
  {D.}~\bibnamefont {Thomas}}, \bibinfo {author} {\bibfnamefont
  {D.}~\bibnamefont {Tucker}}, \bibinfo {author} {\bibfnamefont
  {V.}~\bibnamefont {Vikram}}, \bibinfo {author} {\bibfnamefont
  {A.}~\bibnamefont {Walker}}, \bibinfo {author} {\bibfnamefont
  {J.}~\bibnamefont {Weller}}, \ and\ \bibinfo {author} {\bibfnamefont
  {Y.}~\bibnamefont {Zhang}},\ }\href {\doibase 10.1103/physrevd.98.043528}
  {\bibfield  {journal} {\bibinfo  {journal} {Physical Review D}\ }\textbf
  {\bibinfo {volume} {98}} (\bibinfo {year} {2018}),\
  10.1103/physrevd.98.043528}\BibitemShut {NoStop}%
\end{thebibliography}%
\bibliographystyle{apsrev4-1}

\appendix

\section{Spherical Collapse in \texttt{ReACT}}
\label{app:spherical_collapse}
In this appendix, we describe the physics behind the \texttt{ReACT} approach and provide some qualitative justification for the shape of the fitting function in our extended implementation. The \texttt{ReACT} method augments the halo model with the spherical collapse formalism. The modifications to gravity enter via the Poisson equation \eqref{eq:MGParam}. Halo formation is modelled by following the evolution of a spherical top-hat overdensity of radius $R_{\rm TH}$ in terms of the gravitational potential and $\mu(z)$. The equation solved within the code is given by \cite{ref:reactionCataneo}
\begin{equation}\label{eq:SColl}
    y^{\prime\prime} + \frac{H^{\prime}}{H}y^{\prime} - \left(1 + \frac{H^{\prime}}{H}\right)y + \left[\frac{H_0}{H}\right]^2\frac{\Omega_{\rm m, 0}}{2a^3}\mu(z)\delta\left(\frac{a}{a_i} + y\right) = 0
\end{equation}
where y = $R_{\rm TH}/R_i - a/a_i$, $a_i$ is the initial scale factor, $R_i$ is the initial radius of the overdensity and the prime denotes a derivative w.r.t $\ln(a)$.  The initial conditions are set deep into matter domination when an Einstein-de-Sitter Universe is an accurate description. The initial overdensity $\delta_i$ is iteratively adjusted until the collapse condition corresponding to $R_{\rm TH} = 0$ at characterised by the scale factor at the time of collapse $(a_{\rm coll})$ is reached. One can then apply the virial theorem (see appendix A in \cite{ref:ReactTheory} for details) to determine the viral overdensity parameter given by
\begin{equation}
    \Delta_{\rm vir} = [1 + \delta(a_{\rm vir})]\left(\frac{a_{\rm coll}}{a_{\rm vir}}\right)^3\, .
\end{equation}

\begin{figure}
    \centering
    \includegraphics[width=0.85\textwidth]{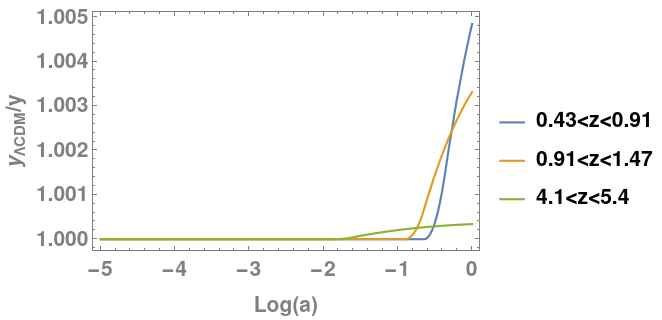}
    \caption{We show the solution to the differential equation \eqref{eq:SColl} for different redshift bins for a constant value of $\mu$. Clearly, the spherical collapse parameter is more dramatically affected when $\mu$ is modified at late times.  }
    \label{fig:SphericalColl}
\end{figure}

The results of the spherical collapse calculation are used to compute a modified Sheth-Tormen mass function given by
\begin{equation}
    n_{\rm vir} = \frac{dn}{d\ln M_{\rm vir}} = \frac{\bar{\rho}_{\rm m, 0}}{M_{\rm vir}}\nu f(\nu) \frac{d \ln\nu}{d\ln M_{\rm vir}}\, ,
\end{equation}
where $\nu = \delta_{\rm c}/\sigma$ is the peak height with $\sigma$ is the variance in the amplitude of the density fluctuations within a top-hat of a fixed radius (typically set to $8\, h^{-1}{\rm Mpc}$). Note that the inclusion of the spherical collapse calculation leads to a different peak height entering the halo model. The concentration-mass relation measured in simulations have a weak negative power law, meaning the largest halos have the smallest $c$. Since larger halos typically form at late times, the spherical collapse results shown in fig.~\ref{fig:SphericalColl} indicate that one would expect that variation in the fitting function for the amplitude $\hat{c} = c/c_{\rm \Lambda CDM}$ of the concentration-mass relation in \texttt{ReACT} will be more dramatic at small redshifts.

Effectively, the change in the redshift at which halo-collapse and virialisation occur is propagated into the mass function. The non-linear matter power spectrum may be computed using the halo model via the sum of the 1-halo and 2-halo terms given by 
\begin{eqnarray}
P_{1h} & = & \int {\rm d}\ln M_{\rm vir}n_{\rm vir}\left(\frac{M_{\rm vir}}{\bar{\rho}_{\rm m, 0}}\right)^2\left|u(k, M_{\rm vir})\right|^2 \, , \\
P_{2h} & = & \left( \int {\rm d}\ln M_{\rm vir}n_{\rm vir}\frac{M_{\rm vir}}{\bar{\rho}_{\rm m, 0}}u(k, M_{\rm vir})b_{\rm L}(M_{\rm vir}) \right)^2 P_{\rm L}(k)  \, ,
\end{eqnarray}
where $u(k, m)$ is the Fourier transform of the density profile, $b_{\rm L}$ is the linear bias parameter and $P_{\rm L}(k)$ is the linear power spectrum. 

\texttt{ReACT} also implements second order perturbation-theory in order to get better accuracy for the specific behaviour seen in $f(R)$ gravity and DGP gravity. However, in our formalism, all modifications to gravity are captured by the parameter $\mu(z)$. Therefore, we switch off the flags that implement these routines within the code (see appendix B in \cite{Srinivasan2021}). 

\section{Sensitivity of analysis to simulation parameters}
In our analysis in this work as well as in \cite{Srinivasan2021}, we ran our $N$-body simulations with Planck-2018 \cite{ref:Planck2018} (TT+TE+low P +BAO) cosmological parameters and fixed random seed for the initial conditions. Since we compute the ratio of matter power spectra with the identical initial conditions, our results are realisation-independent \cite{ref:McDonaldRatio2006}. We verified this by computing the $R(z,k)$ for two of our simulations after averaging over three realisations and found identical results.
We also investigated the effects of varying the standard cosmological parameters. We obtain the same $R(z,k)$ at both $z=0$ and $z=0.5$ when we vary cosmological parameters, to the DES Y1 cosmological parameters \cite{Troxel_2018} and the joint Planck-DES Y1 analysis \cite{ref:Planck2018}. 
These results justify that the results in the main text are not dependent on the realisations that were examined, or on the $\Lambda$ CDM parameters. This ensures that the fitting function has a wide range of validity and is suitable for future data analyses.

\begin{figure}
    \centering
    \includegraphics[width = 0.6\textwidth]{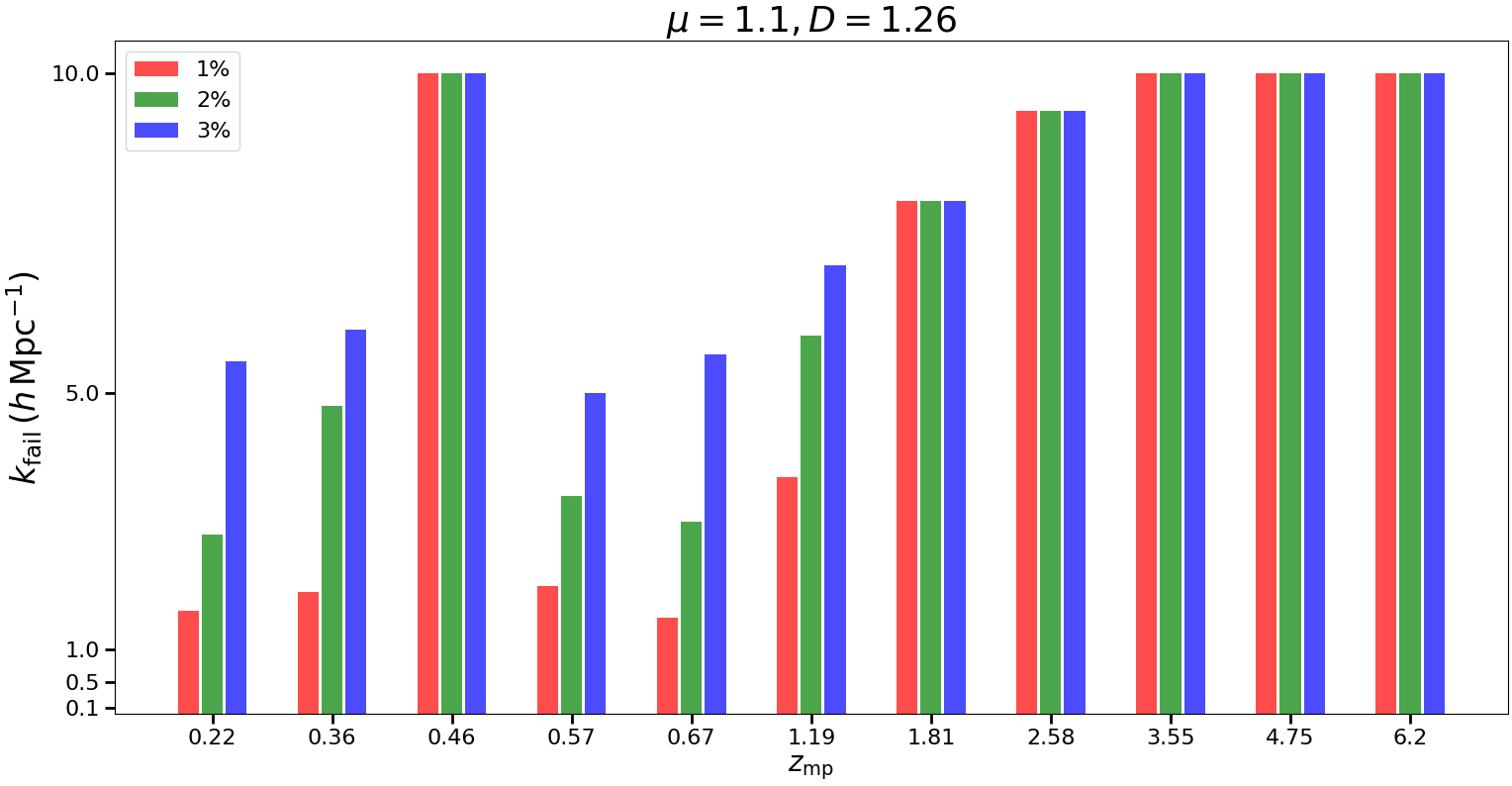}
    \includegraphics[width = 0.6\textwidth]{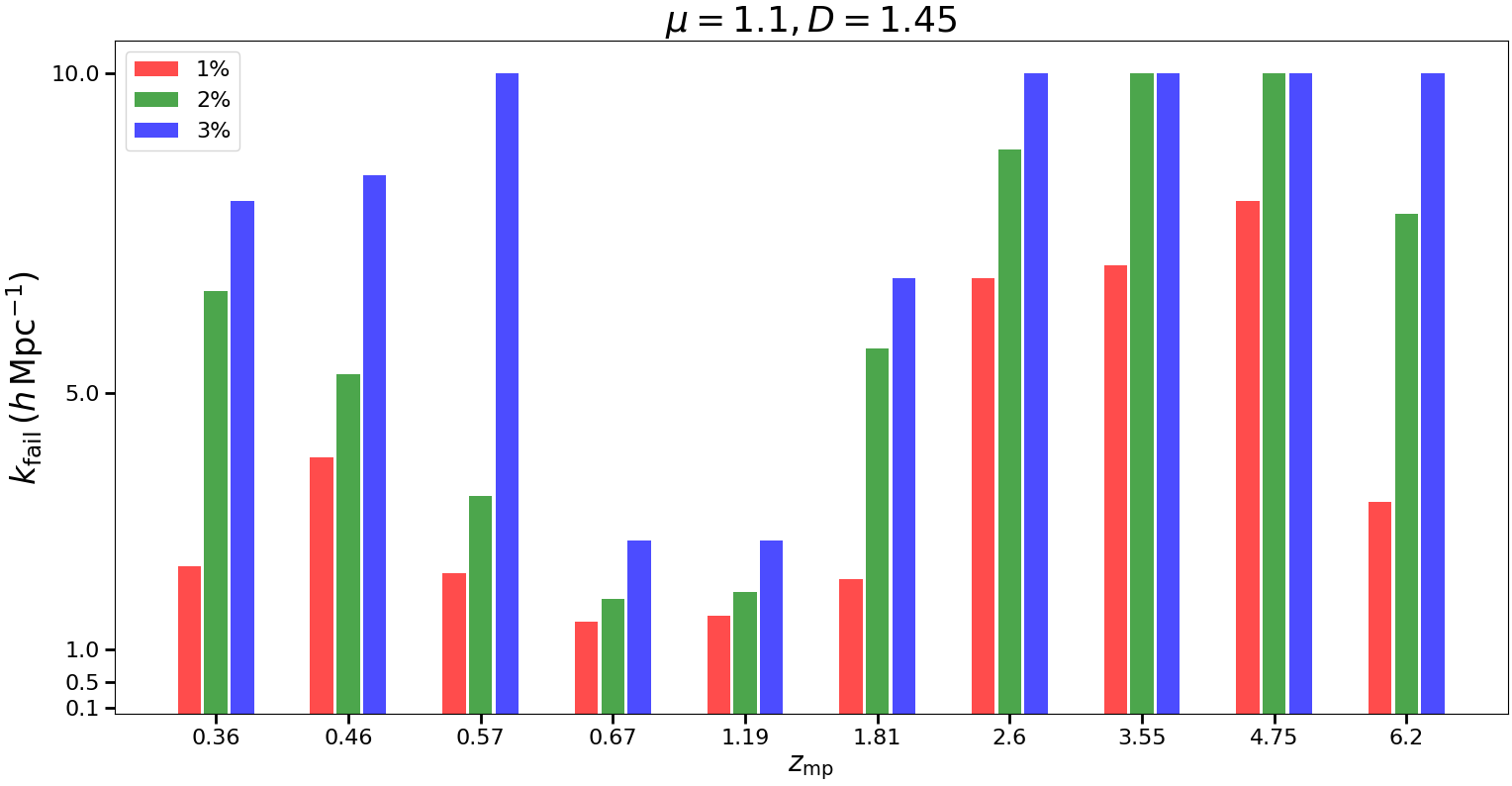}
    \includegraphics[width = 0.6\textwidth]{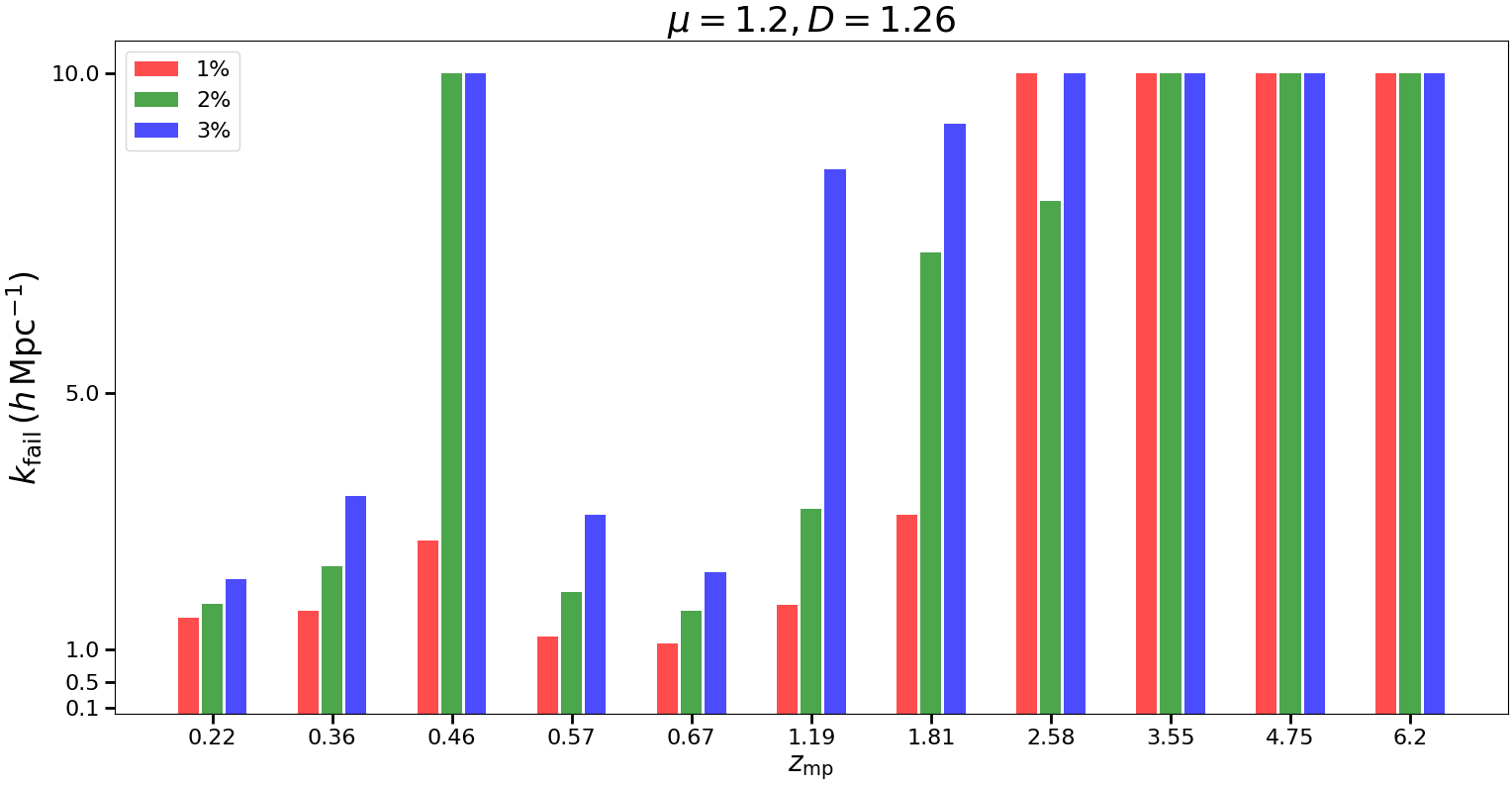}
    \includegraphics[width = 0.6\textwidth]{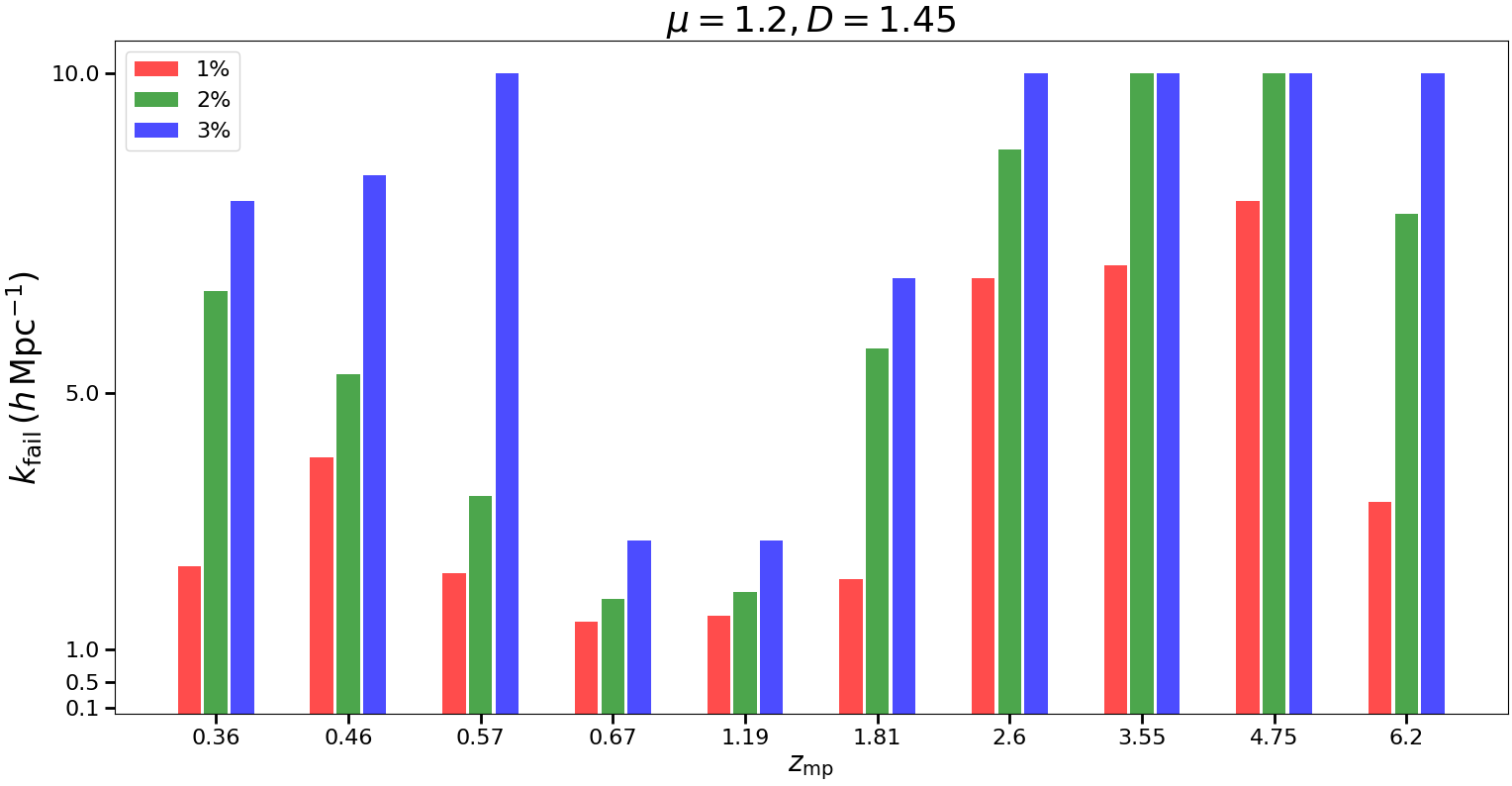}
    \caption{We present the $k_{\rm fail}$ parameter with different accuracy thresholds for our simulations across the our parameter-space of ($\mu, D(z)$). This is particularly relevant for the lensing prediction since we would cut-off $P(k)$ at $k_{\rm fail}$ in the computation of the convergence power spectrum. In particular, our choice of $1\%$ in the main text may be conservative, given that baryonic effects may limit the accuracy of \texttt{ReACT} to $3\%$. We also note that the variation in $k_{\rm fail}$ is substantial between $\mu=1.1$ and $\mu=1.2$, whereas the difference is much less significant between the two cases $D=1.26$ and $D=1.45$. These observations motivate out choice of presenting the corresponding weak-lensing $\ell_{\rm fail}$ as a function of $\mu$ for `conservative' and `realistic' cases, respectively.  }
    \label{fig:kfails}
\end{figure}

We make use of the $k_{\rm fail}$ statistic in order to derive the fitting parameters in our extended implementation of \texttt{ReACT} (see sec.~\ref{subsec:results_conc}). We derive our fitting parameters by imposing 1\% accuracy threshold in computing $k_{\rm fail}$. This might be an overly conservative choice considering that recent release of \texttt{ReACT} that incorporated baryonic modelling \cite{Bose_2021} reported an accuracy of 3\%. In fig.~\ref{fig:kfails}, we show the value of $k_{\rm fail}$ we achieve for 1\%, 2\% and 3\% thresholds, respectively for across the range of our simulation parameters $(\mu, D(z))$. We see that our choice of using $k_{\rm fail}$ means that we are most sensitive to variation in $\mu$ rather than $D(z)$. This motivates our choice of presenting our weak-lensing results for the extreme values of $\mu$ in our simulation suite.

\label{app:kfail_analysis}

\section{Concentration Fitting function}
\label{app:Fitting_Function}
\begin{figure}
    \centering
    \includegraphics[width = 0.6\textwidth]{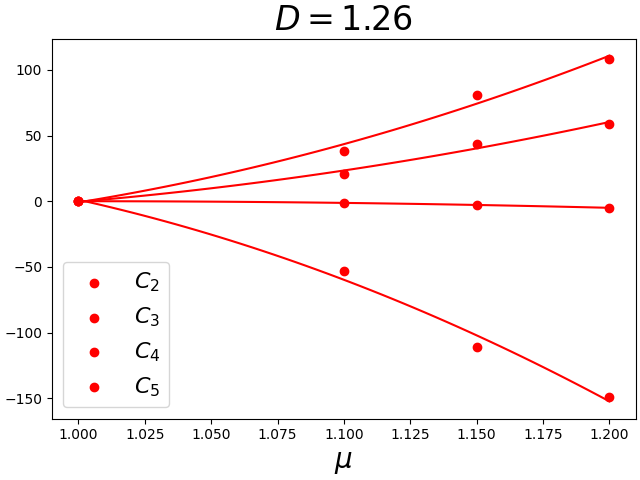}
    \includegraphics[width = 0.6\textwidth]{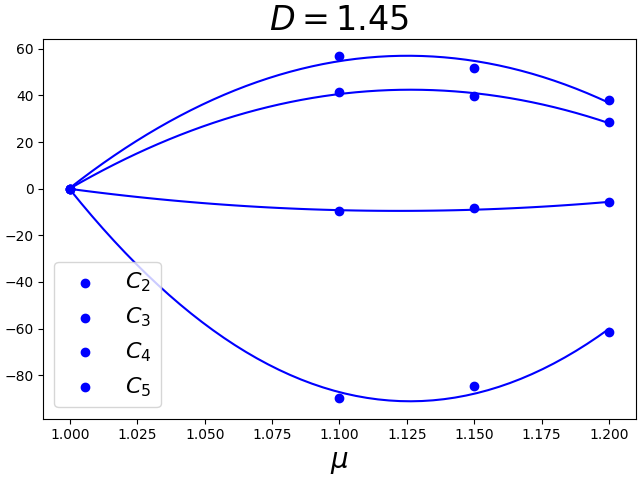}
    \includegraphics[width = 0.6\textwidth]{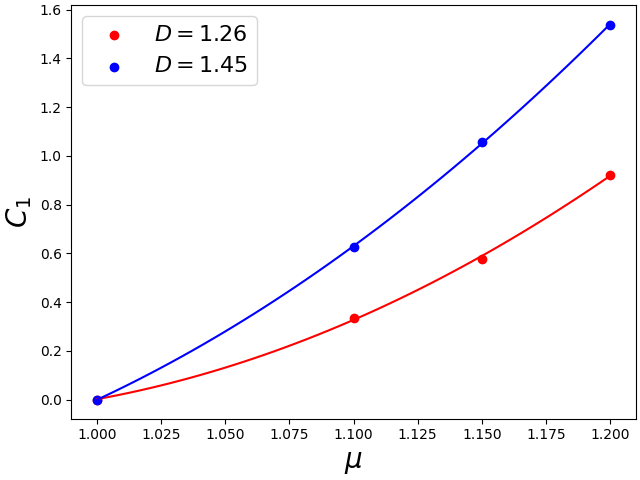}
    \caption{The variation of the fitting parameters as a function of $\mu$ at $z=0$. }
    \label{fig:scaling_plots}
\end{figure}

In the main text, we derive the fitting function for the amplitude of the concentration-mass relation as a function of $\bar{z}$ to be an exponential at $\bar{z}>1$ and a cubic polynomial at $\bar{z}<1$. We noted that the fitting parameters vary linearly as a function of $D$ and quadratically as a function of $\mu$ (see fig.~\ref{fig:scaling_plots}. We thus derived the general form of the fitting parameters at $z_{\rm pk} = 0$. We now give the explicit functional forms we use to construct the fitting parameters at $z_{\rm pk} = 0$.

We derive the fitting parameters $\tilde{C}_i$ as a function of $\mu$ at $D=1.26$ to be $C_i = \gamma_i (\mu-1) + \delta_i (\mu-1)^2$. We show in table \ref{tab:Ci_vals} the values of these parameters for the extreme values of $D(z)$. Note that these have been derived for $\mu>1$, but can be used to to derive the fit for arbitrary $\mu$. We will return to this point shortly. We derive the $\hat{C}_i$ at arbitrary values of $D(z)$ by linearly interpolating between $\tilde{C}_i$ and $\bar{C}_i$ see eq.~\eqref{eq:Cfit} in main text. 

\begin{table}[]
    \centering
    \begin{tabular}{|c|c|c|c|}
    \hline
        $D(z)$ & $C_i$ & $\gamma_i$  & $\delta_i$  \\
        \hline
        \multirow{5}{*}{1.26} & $\tilde{C}_1$   & 1.9 & 1.3  \\
        
        & $\tilde{C}_2$   & -0.6 & -122.9 \\
        
        & $\tilde{C}_3$   & 172.7 & 653.0  \\
        
        & $\tilde{C}_4$   & -450.6 & -1579.0  \\
        
        & $\tilde{C}_5$   & 328.5 & 1145.6  \\
        \hline
        \multirow{5}{*}{1.45} & $\bar{C}_1$  & 4.9 & 14.0  \\
        & $\bar{C}_2$   & -154.7 & 634.0   \\
        & $\bar{C}_3$   & 904.5 & -3610.1   \\
        & $\bar{C}_4$   & -1438.7 & 5700.0   \\
        & $\bar{C}_5$   & 669.6 & -2648.0  \\
        \hline
    \end{tabular}
    \caption{The fitting parameters $\tilde{C}_i$ and $\bar{C}_i$ as a function of $\mu$. We linearly interpolate between these values for the general case of arbitrary $D(z)$.}
    \label{tab:Ci_vals}
\end{table}

\begin{figure}
    \centering
    \includegraphics[width = 0.7\textwidth]{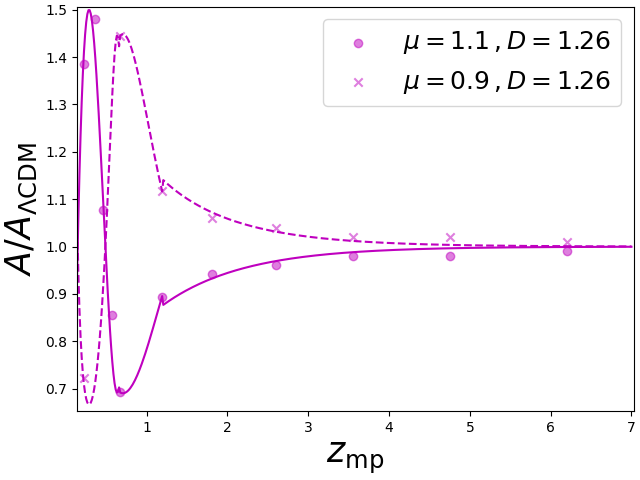}
    \caption{We show the symmetry in the shape of the matter power spectrum across $\mu=1$ is also seen in the fitting function. In the above figure, the data points are derived from the simulations for $\mu=1.1$ (circles) and $\mu=0.9$, (crosses) respectively but we only derive the fit for $\mu=1.1$. We then derive the fit to the $\mu=0.9$ datapoints by taking the reciprocal of the fit the $\mu=1.1$ points. }
    \label{fig:Reflection_fit}
\end{figure}

While this fit is derived for $\mu>1$, we find that there is a symmetry in the shape of the matter power spectrum across $\mu = 1$ (see figs.~\ref{fig:pk} and \ref{fig:pk_transition} and the relevant discussion in the main text). Therefore, $A/A_{\rm \Lambda CDM}$ for $\mu<1$ can be computed by taking the reciprocal of the fit at the reflected value at $\mu>1$. We show an example of this behaviour in fig.~\ref{fig:Reflection_fit}. 

\begin{figure}
    \centering
    \includegraphics[width = 0.7\textwidth]{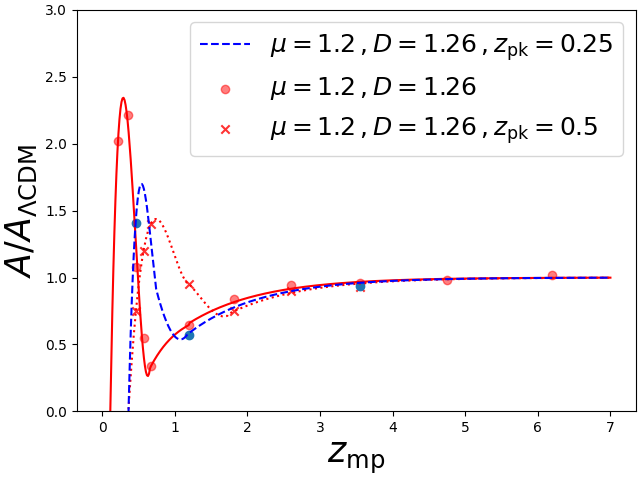}
    \caption{We show the fitting function at $z=0$ and $z_{\rm pk}=0.5$, where the circular and cross data points represent the measurement of the fit at $z_{\rm mp}$ corresponding to the simulations at $z_{\rm pk} = 0$ and $z_{\rm pk} = 0.5$, respectively. We linearly interpolate between the these two curves to obtain the fitting parameters at $z_{\rm pk}=0.25$, and indeed at other redshifts to produce the weak-lensing convergence spectra in the main text (see fig.~\ref{fig:lfail}). In other words, we do not use any simulation data to produce the curve at $z_{\rm pk}=0.25$, and still obtain $\ell_{\rm fail}>800$ consistently across our simulation set as well as agreement with the simulations (blue data points). 
    }
    \label{fig:fit_redshift_vary}
\end{figure}

We find that the general functional form of the fit is preserved at all redshifts. However, we find that the ratio of the amplitudes of  the concentration parameter relative to $\Lambda$CDM $A/A_{\Lambda CDM}$ is inversely proportional to redshift. As a result, at $z_{\rm pk}\geq 1$ we find that vanilla implementation of \texttt{ReACT} (see sec.~\ref{sec:results}) achieves a $k_{\rm fail} \geq 3\,h\,{\rm Mpc}^{-1}$ at an accuracy threshold of 1\%. Therefore, we derive the fitting function at arbitrary redshifts in the range $0\leq z_{\rm pk} \leq 1$ by linearly interpolating between the redshift at which we know the fitting parameters, i.e., $z_{\rm pk}=\{0, 0.5, 1\}$ \footnote{We will be publicly releasing this code with our extended \texttt{ReACT} implementation shortly}. We show an example of this for $D=1.26$ in fig.~\ref{fig:fit_redshift_vary}.

\end{document}